\documentclass[10pt]{iopart}
\usepackage{epsfig}
\usepackage{iopams}
\def\nn{\nonumber}

\def\be{\begin{equation}}
\def\ee{\end{equation}}

\def\apj{Ap. J.}
\def\Phix{\phi}

\eqnobysec

\def\alt{\mathrel{\hbox{\rlap{\hbox{\lower4pt\hbox{$\sim$}}}\hbox{$<$}}}}
\def\agt{\mathrel{\hbox{\rlap{\hbox{\lower4pt\hbox{$\sim$}}}\hbox{$>$}}}}
\begin{document}

\title[Computing inspirals in Kerr in the adiabatic regime]{Computing inspirals in Kerr in the adiabatic regime.  I. The scalar case}

\author{Steve Drasco\dag, \'{E}anna \'{E}.\ Flanagan\dag\ddag, Scott A. Hughes\S
$\|$}

\address{\dag\ Center for Radiophysics and Space Research, Cornell
University, Ithaca, NY 14853}
\address{\ddag\ Laboratory for Elementary Particle Physics, Cornell
University, Ithaca, NY 14853}
\address{\S\ Department of Physics, MIT, 77 Massachusetts Ave.,
Cambridge, MA 02139}
\address{$\|$\ MIT Kavli Institute for Astrophysics
and Space Research, MIT, 77 Massachusetts
Ave., Cambridge, MA 02139}

\date{\today}
%%%%%%%%%%%%%%%%%%%%%%%%%%%%%%%%%%%%%%%%%%%%%%%%%%%%%%%%%%%%%%%%%%%%%%%%%%%%%%%
\begin{abstract}

A key source for LISA will be the inspiral of compact objects
into massive black holes.
Recently Mino has shown that in the adiabatic limit, gravitational
waveforms for these sources
can be computed by using for the radiation reaction force the gradient of one
half the difference between the retarded and advanced metric
perturbations.
Using post-Newtonian expansions, we argue that the
resulting waveforms should be sufficiently accurate for signal detection with LISA.
Data-analysis templates will require higher accuracy, going
beyond adiabaticity; this remains a significant challenge.

We describe an explicit computational procedure for obtaining
waveforms based on Mino's result, for the case of a point particle
coupled to a scalar field.  We derive an explicit expression for the
time-averaged time derivative of the Carter constant, and verify that
the expression correctly predicts that circular orbits remain circular
while evolving under
the influence of radiation reaction.
The derivation uses detailed properties of mode expansions, Green's
functions and bound geodesic orbits in the Kerr spacetime, which we
review in detail.
This paper is about three quarters review and one quarter
new material.  The intent is to
give a complete and self-contained treatment of scalar radiation
reaction in the Kerr spacetime, in a single unified notation, starting
with the Kerr metric, and ending with formulae for the
time evolution of all three constants of the motion that are sufficiently
explicit to be used immediately in a numerical code.

\end{abstract}
%%%%%%%%%%%%%%%%%%%%%%%%%%%%%%%%%%%%%%%%%%%%%%%%%%%%%%%%%%%%%%%%%%%%%%%%%%%%%%%
\maketitle
%%%%%%%%%%%%%%%%%%%%%%%%%%%%%%%%%%%%%%%%%%%%%%%%%%%%%%%%%%%%%%%%%%%%%%%%%%%%%%%%%%%%%%%%%%%%%%%%%%%%%%%%%%%%%%%%%%%%%%%%%%%%%%%%%%%%%%%%%%%%%%%%%%%%%%%%%%%%%%

\section{Summary and overview}

\subsection{Gravitational radiation reaction of point particles}

A key unsolved problem in general relativity is to compute
the gravitational radiation produced by a small object spiralling into
a much larger black hole.  This problem is of direct observational relevance.
Inspirals of compact objects into intermediate
mass black holes ($M \sim 10^2 - 10^3 M_\odot)$
may be observed by LIGO and other ground based interferometers
\cite{2002ApJ...581..438M}; recent observations suggest the existence of
black holes in this mass range \cite{Miller:2003sc,Fiorito:2004qh}.
In addition, a key source for the space-based gravitational wave
detector LISA is the
final epoch of inspiral of a stellar-mass compact object into a
massive ($M \sim 10^6 M_\odot$) black hole at the center of a galaxy.
Gair et. al. \cite{Gair:2004iv} have estimated that LISA should
see over a thousand such inspiral events during its multi-year mission
lifetime, based on Monte Carlo simulations of the dynamics of stellar
cusps by Freitag \cite{2003ApJ...583L..21F}.

Observations of these signals will have several major scientific
payoffs \cite{Cutler:2002me}:

\begin{itemize}

\item From the observed waveform, one can measure the mass
and spin of the central black hole with fractional accuracies of order
$10^{-4}$ \cite{Poisson:1996tc,Barack:2003fp}.
%Observing many
%events will therefore provide a census of the masses and spins of the
%massive central black holes in non-active galactic nuclei like M31 and
%M32.
The spin can provide useful information about the growth
history (mergers versus accretion) of the black hole \cite{Hughes:2002ei}.

\item Likewise, one obtains a census of the inspiralling
objects' masses with precision $\sim 10^{-4}$, teaching us about the
stellar mass function and mass segregation in the central parsec of
galactic nuclei.

\item The measured event rate will teach us about
dynamics in the central parsec of galaxies.

\item The gravitational waves will be wonderful tools for probing the
nature of black holes in galactic nuclei, allowing us to
observationally map, for the first time, the spacetime geometry of a
black hole, and providing a high precision test of general relativity
in the strong field regime \cite{Ryan:1995wh,Ryan:1997hg}.

\end{itemize}

To be in the LISA waveband, the mass $M$ of the black hole must be in
the range $10^5 \, M_\odot \alt M \alt 10^7 \, M_\odot$
\cite{Finn:2000sy}.  The ratio between the mass $\mu \sim (1-100)
M_\odot$ of the compact object and $M$ is therefore in the range
$10^{-7} \alt \mu/M \alt 10^{-3}$.
%The smallness of this mass ratio
%%is the foundation for approximation methods used to compute the
%signal.
These systems spend the last year or so of
their lives in the very relativistic regime close to the black hole
horizon, where the post-Newtonian approximation has completely broken
down, and emit $\sim M/\mu \sim 10^5$ cycles of waveform \cite{Finn:2000sy}.

Realizing the above science goals will require accurate theoretical models
(templates) of the gravitational waves.  This is because the method of matched
filtering will be used both to detect the signals buried in the
detector noise, and to measure the parameters characterizing detected
signals.  The accuracy requirement is roughly that the template should
gain or lose no more than $\sim 1$ cycle of phase
compared to the true waveform over the $\sim 10^5$ cycles of inspiral.
These sources must therefore be modeled with a fractional
accuracy $\sim 10^{-5}$. The past several years have seen a
significant research effort in the relativity community aimed
at providing these accurate templates.

To date, there have been several approaches to generating
waveforms.  The foundation for all the approaches is the fact that, since
$\mu/M \ll 1$, the field of the compact object can be treated as a
linear perturbation to the large black hole's gravitational field.  On
short timescales, the
compact object moves on a geodesic of the Kerr geometry,
characterized by its conserved energy $E$,
$z$-component of angular momentum $L_z$, and Carter
constant $Q$.
Over longer timescales, radiation reaction causes the parameters $E$,
$L_z$ and $Q$ to evolve and the orbit to shrink.

The various approaches are:

\begin{enumerate}

\item {\it Use of post-Newtonian methods:}
Fairly crude waveforms can be obtained using post-Newtonian methods
\cite{Glampedakis:2002cb,Barack:2003fp}.
These have been used to approximately scope out LISA's
ability to detect inspiral events \cite{Gair:2004iv} and
to measure the waveform's parameters \cite{Barack:2003fp}.
However, since the orbital speeds are a substantial fraction of the speed
of light, these waveforms are insufficiently accurate for the eventual
detection and data analysis of real signals.

\item {\it Use of conservation laws:}
In this approach
\cite{Cutler:1994pb,Shibata:1994xk,Glampedakis:2002ya,Hughes:1999bq}
one uses the Teukolsky-Sasaki-Nakamura (TSN) formalism
\cite{Teukolsky:1972my,Teukolsky:1973ha,Sasaki:1981sx} to compute
the fluxes
of energy $E$ and angular momentum $L_z$ to infinity and down the
black hole horizon generated by a
a compact object on a geodesic orbit.
Imposing global conservation of energy and
angular momentum, one infers the rates of change of the orbital energy
and angular momentum.  For certain special classes of orbits (circular
and equatorial orbits) this
provides enough information that one can also infer the rate of change
of the Carter constant $Q$, and thus the inspiralling trajectory.

\item {\it Direct computation of the self-force:}
In this more fundamental approach one computes the self-force or
radiation-reaction force arising
from the interaction of the compact object with its own gravitational
field.
%This force is analogous to the
%Abraham-Lorentz-Dirac force that acts on accelerated electric charges.
A formal expression for this force in a general vacuum spacetime in terms
of the retarded Green's function was computed several years ago
\cite{Mino:1997nk,Quinn:1997am}.
Translating this expression into a practical computational
scheme for Kerr black holes is very difficult and is still in
progress.  Roughly 100 papers
devoted to this problem have appeared in the last few years; see, for
example Poisson \cite{Poisson:2003nc}
for an overview and
references.

\item {\it Time-domain numerical simulations:}
Another technique is to numerically integrate the
Teukolsky equation as a 2+1 PDE in the time domain
\cite{Krivan:1997hc,Burko:2002bt,Scheel:2003vs,Martel:2003,Lopez-Aleman:2003ik,Khanna:2003qv,Pazos-Avalos:2004rp},
and to model the compact object as a
finite-sized source.  This approach faces considerable challenges:
(i) There is a separation of timescales --- the
orbital period is much shorter than the radiation reaction timescale.
(ii) There is a separation of lengthscales --- the compact object is
much smaller than the black hole.  (iii) The self-field of the
small object must be computed with extremely high accuracy, as the
piece of the self-field responsible for the self-force is a tiny
fraction ($\sim \mu/M$) of the divergent self-field.
This approach may eventually be competitive with (ii) and (iii) but is
currently somewhat far from being competitive [flux accuracies are
$\sim 10\%$ as compared to $\sim 10^{-6}$ for (ii)].
\end{enumerate}

\subsection{Radiation reaction in the adiabatic regime}

A key feature of these systems is that they evolve adiabatically: the
radiation reaction timescale $\sim M^2/\mu$ is much longer than the
orbital timescale $\sim M$, by a factor of the inverse $M/\mu$ of the
mass ratio.  This has implications for the nature of the signal.
The self-acceleration ${\vec a}$ of the compact object can be expanded
in powers of the mass ratio as
\begin{equation}
{\vec a} = \frac{\mu}{M} \left[ {\vec a}_{1,{\rm diss}} + {\vec
    a}_{1,{\rm cons}} + \frac{\mu}{M} {\vec a}_{2,{\rm diss}} +
  \frac{\mu}{M} {\vec a}_{2,{\rm cons}} + O \left( \frac{\mu^2}{M^2}
  \right) \right]\;.
\label{eq:selfaccel}
\end{equation}
Here ${\vec a}_{1,{\rm diss}}$ and ${\vec a}_{1,{\rm cons}}$ are the
dissipative and conservative pieces of the leading-order
self-acceleration computed in Refs.\ \cite{Mino:1997nk,Quinn:1997am}.
Similarly, ${\vec a}_{2,{\rm diss}}$ and ${\vec a}_{2,{\rm cons}}$ are
the corresponding pieces of the first correction to the
self-acceleration, which has not yet been computed (although see Ref.\
\cite{Rosenthal:2005ju} for work in this direction).

The effect of the dissipative pieces of the self force
will accumulate secularly, while the effect of the conservative pieces
will not.  Hence the effect of the
dissipative pieces on the phase of the orbit will be larger than that of the
conservative pieces by a factor of the number of cycles of inspiral,
$\sim M/\mu$.  Consider now, for example, the azimuthal phase
$\Phix(t)$ of the orbit.  This can be
expanded in powers of the mass ratio using a two-timescale expansion
as \cite{dfh05,Hinderer}
%\footnote{Here we focus on the
% piece of the phase that grows linearly with time in the absence of
%radiation reaction.  There are also oscillatory terms which are
%unimportant for this discussion; see Eq.\ (3.23) of Ref.\
% \cite{Drasco:2004tv}.}
\begin{equation}
\Phix(t) = \frac{M}{\mu} \left[ \Phix_1(t,t_1) + \frac{\mu}{M} \Phix_2(t,t_1) +
O\left( \frac{\mu^2}{M^2}\right) \right]\;,
\label{eq:emri_phase}
\end{equation}
where the leading-order phase
$\Phix_1(t,t_1)$ is determined by ${\vec a}_{1,{\rm diss}}$,
and the higher order correction $\Phix_2(t,t_1)$ to the phase is
determined by ${\vec a}_{1,{\rm cons}}$ and ${\vec a}_{2,{\rm diss}}$.
Here $t_1$ is a ``slow'' time variable which satisfies $dt_1/dt =
O(\mu/M)$.

We shall call leading-order waveforms containing only terms
corresponding to
$\Phix_1(t,t_1)$ {\it adiabatic waveforms}.  From Eq.\
(\ref{eq:emri_phase}) these waveforms are accurate
to $\sim 1$ cycle over the inspiral (more precisely, the phase error
is independent of $\mu/M$ in the limit $\mu/M \to 0$).
To compute adiabatic waveforms we need only keep the
dissipative piece of the self-force, and we can discard any
conservative pieces.  The conservation-law method discussed above
yields waveforms that are accurate to this leading order, but is limited in that it
cannot be used for generic orbits.

Adiabatic waveforms will likely be sufficiently accurate for detecting
inspiral events with LISA \cite{dfh05}.  The correction $\Phix_2$ to
the phase of the waveform can be estimated using the post-Newtonian
approximation, and amounts to $\alt 1$ cycle over the
inspiral for typical parameter values.
While the post-Newtonian approximation is not strictly valid in the
highly relativistic regime near the horizon of interest here, it
suffices to give some indication of the accuracy of adiabatic
waveforms.  This computation is described in \ref{sec:accuracy}.

Recently Yasushi Mino derived a key result that paves the way for
computations of adiabatic waveforms for generic inspirals in Kerr
\cite{Mino:2003yg}.
Mino showed that the time average of the self-force formula of
\cite{Mino:1997nk,Quinn:1997am} for bound orbits in Kerr yields the
same result as the
gradient of one half the difference between the retarded and advanced
metric perturbations.
This ``half retarded minus half advanced'' prescription is the
standard result for electromagnetic radiation reaction in flat
spacetime due to Dirac \cite{Dirac:1938nz}.
This prescription was also posited, without proof, for scalar,
electromagnetic and gravitational radiation reaction in the Kerr
spacetime by Gal'tsov \cite{Galtsov:1982}.
However, prior to Mino's analysis, the prescription was not generally
thought to be applicable or relevant in Kerr.

As one might expect, knowledge of the infinite-time-averaged self-force is
sufficient to compute the leading-order, adiabatic waveforms
\cite{Mino:2003yg}, although there are subtleties related to the
choice of gauge \cite{2005PThPh.113..733M,2005Minob,2005Minoc}.  The
sufficiency can also be established using a two-timescale expansion
\cite{Hinderer}.

\subsection{Implementing Mino's prescription for generic inspirals}

In this paper we apply Mino's result to compute an explicit expression
for the time-averaged time derivative of the Carter constant.  We
specialize to the case of a particle endowed with a scalar charge,
coupled to a scalar field.  This computation is useful as a warm up
exercise for the more complicated case of a particle emitting
gravitational waves \cite{tensorcase}.

We start in Sec.\ \ref{sec:scalar} by describing the model of a point
particle coupled to a scalar field, and review how the self-force
causes both an acceleration of the particle's motion and also an
evolution of the renormalized rest mass of the particle \cite{Quinn:2000wa}.
In Sec.\ \ref{sec:geosesics} we review the properties of generic bound
geodesic orbits in the Kerr spacetime.  A crucial result we use later
is that the $r$ and $\theta$ motions are periodic when expressed as
functions of a particular time parameter we call Mino time
\cite{Mino:2003yg}, and that
the $t$ and $\phi$ motions consist of linearly growing terms plus
terms that are periodic with the period of the $r$-motion, plus terms
that are periodic with the period of the $\theta$-motion
\cite{Drasco:2004tv}.

Section \ref{sec:minoproof} reviews Mino's derivation of the
half-retarded-minus-half-advanced prescription for radiation reaction
in the adiabatic limit \cite{Mino:2003yg}.  In Sec.\ \ref{sec:modes} we
discuss a convenient basis of modes for solutions of the scalar wave
equation in Kerr, namely the ``in'', ``out'', ``up'' and ``down''
modes used by Chrzanowski \cite{Chrzanowski:1975wv} and Gal'tsov
\cite{Galtsov:1982}.  We also review several key properties of these
modes including symmetry relations and relations between the various
reflection and transmission coefficients that appear in the
definitions of the modes.  Section \ref{sec:retarded} reviews the
standard derivation of the mode expansion of the retarded Green's function,
and Sec.\ \ref{sec:radiative} reviews the derivation of the mode
expansion of the radiative
Green's function given by Gal'tsov \cite{Galtsov:1982}, correcting
several typos in Gal'tsov which are detailed in \ref{sec:typos}.

Next, in Sec.\ \ref{sec:harmonic} we turn to the source term in the
wave equation for the scalar field.  We review the derivation of
Drasco and Hughes of the harmonic decomposition of this source term in
terms of a discrete sum over frequencies in which harmonics of three
different fundamental frequencies occur \cite{Drasco:2004tv}.
We derive expressions for the mode coefficients in the expansion of
the retarded field near future null infinity and near the future event
horizon.  These coefficients are expressed as integrals over a torus in
phase space, which is filled ergodically by
the geodesic motion.

Section \ref{sec:Edot} combines the results of the preceding sections
to derive expressions for the time-averaged rates of change of two of the
conserved quantities of geodesic motion, namely the energy $E$ and the angular
momentum $L_z$.  The expressions are derived from the radiative self
force, following Gal'tsov \cite{Galtsov:1982}.  Gal'tsov also shows that
identical expressions are obtained by using the fluxes of energy and
angular momentum to infinity and down the black hole horizon.
[This result has recently been independently derived for
circular, equatorial orbits in Ref.\ \protect{\cite{Gralla:2005et}}.]
We also show that the time-averaged rate of change of the renormalized
  rest mass of the particle vanishes.

In section \ref{sec:Kdot} we derive an expression for the
time-averaged rate of change of the Carter constant, using the
radiative self-force and the mode expansion of the radiative
Green's function.  This expression [Eq.\ (\ref{eq:dKdt}) below] is the
main new result in this paper.  It involves two new amplitudes that
are computed in terms of integrals over the torus in phase space,
just as for the amplitudes appearing in the energy and angular
momentum fluxes.  Finally, in Sec.\ \ref{sec:circular} we show
that our result correctly predicts the known result that circular
orbits remain circular while evolving under the influence of radiation reaction.
This prediction serves as a check of our result.

As apparent from the above summary, about 25 percent of this paper is
new material, and the remaining 75 percent is review.
The intent is to
give a complete and self-contained treatment of scalar radiation
reaction in the Kerr spacetime, in a single unified notation, starting
with the Kerr metric, and ending with formulae for the
evolution of all three constants of the motion that are sufficiently
explicit to be used immediately in a numerical code.

\section{The scalar field model}
\label{sec:scalar}

We consider a point particle of scalar charge $q$
coupled to a scalar field $\Phi$.  We denote by $\mu_0$ the bare
rest mass of the particle, to be distinguished from a renormalized
rest mass $\mu$ which will occur below.
The particle moves in a spacetime
with metric $g_{\alpha\beta}$ which is fixed; we neglect the gravitational
waves generated by the particle.  The worldline of the particle is
$x^\alpha = z^\alpha(\tau)$, where $\tau$ is proper time.
The action is taken to be
\be
S = - \frac{1}{2} \int d^4 x \sqrt{-g} (\nabla \Phi)^2 - \int d\tau \left\{\mu_0 - q \Phi[z(\tau)] \right\}.
\label{eq:action}
\ee
Varying this action with respect to the worldline yields the equation
of motion
\begin{equation}
(\mu_0 - q \Phi) a^\alpha = q (g^{\alpha\beta} + u^\alpha u^\beta) \nabla_\beta \Phi,
\label{eq:acc00}
\end{equation}
where $u^\alpha$ is the 4-velocity and $a^\alpha$ is the 4-acceleration.
Following Poisson \cite{Poisson:2003nc}, we define the renormalized mass
$\mu$ by\footnote{If we endow the particle with an electric charge $e$
and add the electromagnetic coupling term $e \int d\tau u^\alpha
A_\alpha$ to the action (\protect{\ref{eq:action}}), the equation of motion
(\ref{eq:acc}) becomes
$\mu a^\alpha = q (g^{\alpha\beta} + u^\alpha u^\beta) \nabla_\beta
\Phi + e F^{\alpha\beta} u_\beta$, where $F^{\alpha\beta}$ is the
Faraday tensor.  This shows that the renormalized mass
(\ref{eq:mudef}) is the mass that would be measured by coupling to
other fields.}
\be
\mu(\tau) = \mu_0 - q \Phi[z(\tau)].
\label{eq:mudef}
\ee
The equation of motion (\ref{eq:acc00}) can then be written as
\begin{equation}
a^\alpha = \frac{q}{\mu} (g^{\alpha\beta} + u^\alpha u^\beta) \nabla_\beta \Phi.
\label{eq:acc}
\end{equation}
It is also useful to rewrite the definition (\ref{eq:mudef}) of the
renormalized mass in terms of a differential equation for its evolution with
time:
\be
\frac{d \mu}{d\tau}(\tau) = - q u^\alpha \nabla_\alpha \Phi.
\label{eq:dmudt0}
\ee
The phenomenon of evolution of renormalized rest-mass is discussed
further in Refs.\ \cite{Burko:2002ge,Burko:2002gf,Haas:2004kw}.
Equations (\ref{eq:acc}) and (\ref{eq:dmudt0}) can also
be combined to give the expression
\be
f^\alpha = u^\beta \nabla_\beta ( \mu u^\alpha) = q \nabla^\alpha \Phi.
\label{eq:tsf}
\ee
for the total self force.  The
components of $f^\alpha$ parallel to and perpendicular to the four
velocity yield the quantity $d\mu/d\tau$ and $\mu$ times the self-acceleration,
respectively.

Varying the action (\ref{eq:action}) with respect to the field $\Phi$
gives the equation of motion
\be
\Box \Phi(x) = {\cal T}(x),
\label{eq:wave}
\ee
where the scalar source is
\be
{\cal T}(x) = - q \int_{-\infty}^\infty d \tau \delta^{(4)}[x,z(\tau)].
\label{eq:sourcedef}
\ee
Here
\be
\delta^{(4)}(x,x') = \delta^{(4)}(x - x')/\sqrt{-g}
\label{eq:deltafndef}
\ee
is the
generalized Dirac delta function, where $\delta^{(4)}(x) =
\delta(x^0) \delta(x^1) \delta(x^2) \delta(x^3)$.
We will assume that there is no
incoming scalar radiation, so that the physical solution of the wave
equation (\ref{eq:wave}) is the retarded solution
\be
\fl
\Phi_{\rm ret}(x) = \int d^4 x' \sqrt{-g(x')} G_{\rm ret}(x,x') {\cal
  T}(x') = - q \int d\tau G_{\rm ret}[x,z(\tau)].
\label{eq:Phiret}
\ee
Here $G_{\rm ret}(x,x')$ is the retarded Green's function for the
scalar wave equation (\ref{eq:wave}).

The retarded field (\ref{eq:Phiret})
must of course be regularized before
being inserted into the expression (\ref{eq:acc}) for the
self-acceleration and into the expression (\ref{eq:dmudt0}) for $d\mu/d\tau$, or
else divergent results will be obtained.  The appropriate regularization
prescription has been derived by Quinn \cite{Quinn:2000wa}.
The regularized self-acceleration at the point $z^\alpha(\tau)$
is \cite{Quinn:2000wa}
\be
\fl
a^\alpha(\tau) = \frac{q^2}{\mu} (g^{\alpha\beta} + u^\alpha
u^\beta ) \left[ \frac{1}{6} R_{\beta\gamma} u^\gamma -
    \lim_{\varepsilon \to 0} \int_{-\infty}^{\tau -\varepsilon} d\tau'
    \nabla_\beta G_{\rm ret}[z(\tau),z(\tau')] \right],
\label{eq:selfa1}
\ee
and the regularized expression for $d\mu/d\tau$ is
\be
\frac{d \mu}{d \tau}(\tau) = -\frac{q^2}{12} R + q^2
\lim_{\varepsilon \to 0} \int_{-\infty}^{\tau -\varepsilon} d\tau'
    u^\alpha \nabla_\alpha G_{\rm ret}[z(\tau),z(\tau')].
\label{eq:dmudt1}
\ee

As discussed in the introduction, we specialize in this paper to bound
motion about a Kerr black hole of
mass $M$ with $M \gg \mu$.  In this case the terms in Eqs.\
(\ref{eq:selfa1}) and (\ref{eq:dmudt1}) involving the Ricci tensor
$R_{\alpha\beta}$ vanish.
Also, as long as the orbit is not very close to the innermost stable orbit,
the evolution of the orbit is adiabatic: the orbital evolution
timescale $\sim M^2/\mu$ is much longer than the orbital timescale
$\sim M$.  This adiabaticity allows a significant simplification of
the formulae (\ref{eq:selfa1}) and (\ref{eq:dmudt1})
for the self-acceleration and for the rate of change of mass
$d\mu/d\tau$, as shown by Mino
\cite{Mino:2003yg}.
%For gravitational radiation reaction, the regularized
%self-acceleration expression that corresponds to Eq.\ (\ref{eq:selfa1})
%was derived by Mino, Sasaki and Tanaka \cite{Mino:1997nk} and by
%Quinn and Wald \cite{Quinn:1997am}.
%Mino showed that in the adiabatic limit in Kerr, the
%the Mino-Sasaki-Tanaka-Quinn-Wald expression reduces
%to the gradient of one half the difference
%between the retarded and advanced metric perturbations \cite{Mino:2003yg}.
Namely, these formulae
%The result is that the expressions
%(\ref{eq:selfa1}) and (\ref{eq:dmudt1}) for the self-acceleration
%$a^\alpha$ and the rate of change of mass $d\mu/d\tau$
reduce to the
simple expressions (\ref{eq:acc}) and (\ref{eq:dmudt0}), but
with $\Phi$ replaced
by the
radiative field
\begin{equation}
\Phi_{\rm rad} = \frac{1}{2} ( \Phi_{\rm ret} - \Phi_{\rm adv}).
\end{equation}
Here $\Phi_{\rm adv}$ is the advanced solution of Eq.\
(\ref{eq:wave}).
In Sec.\ \ref{sec:minoproof} below we review Mino's argument,
specialized to the scalar case.

To summarize, the starting point for our analysis is the
self-acceleration expression
\be
a^\alpha = \frac{q}{\mu} (g^{\alpha\beta} + u^\alpha u^\beta)
\nabla_\beta \Phi_{\rm rad}.
\label{eq:selfacc0}
\ee
together with the expression for the evolution of rest mass
\be
\frac{d \mu}{d \tau}(\tau) = - q u^\alpha \nabla_\alpha \Phi_{\rm
  rad}.
\label{eq:dmudt2}
\ee
Equations (\ref{eq:selfacc0}) and (\ref{eq:dmudt2}) are equivalent to
the equation
\be
f^\alpha = u^\beta \nabla_\beta ( \mu u^\alpha) = q \nabla^\alpha
\Phi_{\rm rad}.
\label{eq:tsf1}
\ee
for the total self-force.

\section{Generic bound geodesics in the Kerr spacetime}
\label{sec:geosesics}

This section summarizes the notation and results of Drasco \& Hughes
\cite{Drasco:2004tv}, and also some of the results of Mino \cite{Mino:2003yg}, for
generic bound geodesic orbits in Kerr.

\subsection{The Kerr spacetime}
\label{sec:kerr}

In Boyer-Lindquist coordinates $(t,r,\theta,\phi)$, the Kerr metric is
\begin{eqnarray}
\fl
ds^2 = - \left(1 - \frac{2 M r}{\Sigma}\right) dt^2 - \frac{4 a \sin^2 \theta M r
}{\Sigma} dt d\phi + ( \varpi^4 - \Delta a^2 \sin^2 \theta)
\frac{\sin^2 \theta}{\Sigma} d\phi^2 \nonumber \\
\lo + \Sigma d\theta^2 +
\frac{\Sigma}{\Delta} dr^2.
\label{eq:kerrmetric}
\end{eqnarray}
Here
\begin{eqnarray}
\Sigma & \equiv & r^2 + a^2 \cos^2 \theta,  \\
\Delta & \equiv & r^2 + a^2 - 2 M r, \\
\varpi & \equiv & \sqrt{r^2 + a^2},
\end{eqnarray}
and $M$, $a$ are the black hole mass and spin parameter.
Throughout the rest of this paper we use units in which $M=1$, for
simplicity.  The
square root of the determinant of the metric is
\be
\sqrt{-g} = \Sigma \sin \theta,
\label{eq:detg}
\ee
and the wave operator is given by
\begin{eqnarray}
\fl
\Sigma \Box \Phi = - \left[ \frac{\varpi^4}{\Delta} - a^2 \sin^2
  \theta \right] \Phi_{,tt} - \frac{4 a r}{\Delta} \Phi_{,t\phi} +
\left( \frac{1}{\sin^2 \theta} - \frac{a^2}{\Delta} \right)
\Phi_{,\phi\phi} \nonumber \\
\lo + (\Delta \Phi_{,r})_{,r} + \frac{1}{\sin \theta}
(\sin \theta \Phi_{,\theta})_{,\theta}.
\label{eq:waveoperator}
\end{eqnarray}
The differential operator that appears on the right hand side of this
equation is the Teukolsky differential operator
\cite{Teukolsky:1973ha} for spin $s=0$.

We will use later the Kinnersley null tetrad ${\vec l}$, ${\vec n}$,
${\vec m}$, ${\vec m}^*$, which is given by
\be
{\vec l} = \frac{\varpi^2}{\Delta} \partial_t + \partial_r +
\frac{a}{\Delta} \partial_\phi,
\label{eq:vecldef}
\ee
\be
{\vec n} = \frac{\varpi^2}{2 \Sigma} \partial_t - \frac{\Delta}{2
  \Sigma} \partial_r + \frac{a}{2 \Sigma} \partial_\phi,
\label{eq:vecndef}
\ee
and
\be
{\vec m} = \frac{1}{\sqrt{2}(r + i a \cos\theta)} \left( i a
\sin\theta \partial_t + \partial_\theta + \frac{i}{\sin\theta}
\partial_\phi \right).
\ee
The corresponding one-forms are
\be
{\bf l} = - dt + a \sin^2 \theta d\phi + \frac{\Sigma}{\Delta} dr,
\ee
\be
{\bf n} = - \frac{\Delta}{2 \Sigma} dt + \frac{a \Delta \sin^2
  \theta}{2 \Sigma} d\phi - \frac{1}{2} dr,
\ee
and
\be
{\bf m} =  \frac{1}{\sqrt{2}(r + i a \cos\theta)} \left( - i a
\sin\theta dt + \Sigma d\theta + i \varpi^2 \sin \theta d\phi \right).
\ee
The basis vectors obey the orthonormality relations ${\vec l} \cdot {\vec n} = -1$
and ${\vec m} \cdot {\vec m}^* = 1$ while all other inner products
vanish.  The metric can be written in terms of the basis one-forms as
\be
g_{\alpha\beta} = -2 l_{(\alpha} n_{\beta)} + 2 m_{(\alpha} m_{\beta)}^*.
\label{eq:gabformula}
\ee

\subsection{Constants of the motion}

We define the conserved energy per unit rest mass $\mu$
\be
E = - {\vec u} \cdot \frac{\partial}{\partial t},
\label{eq:Edef}
\ee
the conserved $z$-component of angular momentum divided by $\mu M$
\be
L_z = {\vec u} \cdot \frac{\partial}{\partial \phi},
\label{eq:Lzdef}
\ee
and Carter constant divided by $\mu^2 M^2$
\be
Q = u_\theta^2 - a^2 \cos^2 \theta E^2 +  \cot^2 \theta L_z^2 + a^2
\cos^2 \theta.
\label{eq:Qdef}
\ee
[From now on we will for simplicity call these dimensionless quantities ``energy'',
  ``angular momentum'' and ``Carter constant''.]
The geodesic equations can then be written in the form \cite{Drasco:2004tv}
\begin{eqnarray}
\fl
\left(\frac{dr}{d\lambda}\right)^2 = \left[E(r^2+a^2)
- a L_z\right]^2- \Delta\left[r^2 + (L_z - a E)^2 +
Q\right]
\equiv V_r(r)\;,
\label{eq:rdot}\\
\fl
\left(\frac{d\theta}{d\lambda}\right)^2 = Q - \cot^2\theta L_z^2
-a^2\cos^2\theta(1 - E^2)
\equiv V_\theta(\theta)\;,
\label{eq:thetadot}\\
\fl
\frac{d\phi}{d\lambda} =
\csc^2\theta L_z + aE\left(\frac{r^2+a^2}{\Delta} - 1\right) -
\frac{a^2L_z}{\Delta}
\equiv V_\phi(r,\theta)\;,
\label{eq:phidot}\\
\fl
\frac{dt}{d\lambda} =
E\left[\frac{(r^2+a^2)^2}{\Delta} - a^2\sin^2\theta\right] +
aL_z\left(1 - \frac{r^2+a^2}{\Delta}\right)
\equiv V_t(r,\theta)\;.
\label{eq:tdot}
\end{eqnarray}
Here $\lambda$ is the Mino time parameter \cite{Mino:2003yg}, related to
proper time $\tau$ by
\be
d\lambda = \frac{1}{\Sigma} d\tau.
\label{eq:Minotime}
\ee
Also these equations define the potentials $V_r(r)$,
$V_\theta(\theta)$, $V_\phi(r,\theta)$ and $V_t(r,\theta)$
\footnote{These quantities were denoted $R(r)$, $\Theta(\theta)$,
$\Phi(r,\theta)$ and $T(r,\theta)$ in Ref.\ \protect{\cite{Drasco:2004tv}}.  We no
not use this notation here since it would clash with the functions
$R$, $\Theta$ and $\Phi$ defined in Eq.\ (\protect{\ref{eq:ansatz}}) below.}.

Sometimes it will be convenient to use
instead of the Carter constant $Q$ the quantity
\be
K = Q + (L_z - a E)^2.
\ee
For convenience we will also call this quantity the ``Carter constant''.
In the Schwarzschild limit $Q$ and $K$ are given by $Q =
L_x^2 + L_y^2$ and $K = L_x^2 + L_y^2 + L_z^2$.
The quantity $K$ can be written as
\be
K = K^{\alpha\beta} u_\alpha u_\beta,
\label{eq:Kdef}
\ee
where $K^{\alpha\beta}$ is the Killing tensor
\be
K^{\alpha\beta} = 2 \Sigma m^{(\alpha} {\bar m}^{\beta)} - a^2 \cos^2
\theta g^{\alpha\beta}.
\ee
Using the identity (\ref{eq:gabformula}) this can also be written as
\be
K^{\alpha\beta} = 2 \Sigma l^{(\alpha} n^{\beta)} + r^2
g^{\alpha\beta}.
\label{eq:Kalphabeta}
\ee
Using the formulae (\ref{eq:vecldef}) and (\ref{eq:vecndef}) for the
null vectors ${\vec l}$ and ${\vec n}$ together with the definitions
(\ref{eq:Edef}) and (\ref{eq:Lzdef}) of $E$ and $L_z$, we obtain from
Eq.\ (\ref{eq:Kalphabeta}) the following formula for $K$:
\be
K = \frac{1}{\Delta} (\varpi^2 E - a L_z)^2 - \Delta u_r^2 - r^2.
\label{eq:Kformula1}
\ee
Solving this for $u_r$ gives
\be
u_r = \pm \sqrt{ \frac{1}{\Delta^2} ( \varpi^2 E - a L_z)^2 -
\frac{r^2}{\Delta} - \frac{K}{\Delta} };
\label{eq:urformula}
\ee
this formula will be useful later.

\subsection{Parameterization of solutions}
\label{sec:geodesicparameters}

Following Mino \cite{Mino:2003yg}, we parameterize any geodesic by seven
parameters:
\be
E,L_z,Q,\lambda_{r0},\lambda_{\theta0},t_0,\phi_0.
\ee
Here $t_0$ and $\phi_0$ are the values of $t(\lambda)$ and
$\phi(\lambda)$ at $\lambda = 0$.  The quantity $\lambda_{r0}$ is the
value of $\lambda$ nearest to $\lambda=0$ for which $r(\lambda) =
r_{\rm min}$, where $r_{\rm min}$ is the minimum value of $r$ attained
on the geodesic.  Similarly $\lambda_{\theta0}$ is the
value of $\lambda$ nearest to $\lambda=0$ for which $\theta(\lambda) =
\theta_{\rm min}$, where $\theta_{\rm min}$ is the minimum value of
$\theta$ attained on the geodesic.
This parameterization is degenerate because of the freedom to
reparametrize the geodesic via $\lambda \to \lambda + \Delta
\lambda$.  We discuss this degeneracy further in Sec.\
\ref{sec:degeneracy}.

Frequently in this paper we will focus on the {\it fiducial
geodesic} associated with the constants $E$, $L_z$ and $Q$, namely
the geodesic with
\be
\lambda_{r0} = \lambda_{\theta0} = t_0 = \phi_0 =0.
\ee

\subsection{Motions in $r$ and $\theta$}

It follows from the geodesic equations (\ref{eq:rdot}) and
(\ref{eq:thetadot}) that the functions
$r(\lambda)$ and $\theta(\lambda)$ are periodic.  We denote the
periods by $\Lambda_r$ and $\Lambda_\theta$, respectively, so
\be
r(\lambda + \Lambda_r) = r(\lambda), \ \ \ \ \ \
\theta(\lambda + \Lambda_\theta) = \theta(\lambda).
\label{eq:periodic}
\ee
Using the initial condition $r(\lambda_{r0}) = r_{\rm min}$ we can
write the solution $r(\lambda)$ to Eq.\ (\ref{eq:rdot}) explicitly as
\be
r(\lambda) = {\hat r}(\lambda - \lambda_{r0}),
\label{eq:rmotion}
\ee
where the function ${\hat r}(\lambda)$ is defined by
\be
\int_{r_{\rm min}}^{{\hat r}(\lambda)} \frac{dr}{\pm \sqrt{V_r(r)}} =
\lambda.
\label{eq:hatrdef}
\ee
Similarly we can
write the solution $\theta(\lambda)$ to Eq.\ (\ref{eq:thetadot}) explicitly as
\be
\theta(\lambda) = {\hat \theta}(\lambda - \lambda_{\theta0}),
\label{eq:thetamotion}
\ee
where the function ${\hat \theta}(\lambda)$ is defined by
\be
\int_{\theta_{\rm min}}^{{\hat \theta}(\lambda)} \frac{d\theta}{\pm \sqrt{V_\theta(\theta)}} = \lambda.
\label{eq:hatthetadef}
\ee
The functions ${\hat r}(\lambda)$ and ${\hat \theta}(\lambda)$ are
just the $r$ and $\theta$ motions for the fiducial geodesic.

\subsection{Motion in $t$}

Next, the function $V_t(r,\theta)$ that appears on the right hand side
of Eq.\ (\ref{eq:tdot}) is a sum of a function of $r$ and a function
of $\theta$:
\be
V_t(r,\theta) = V_{tr}(r) + V_{t\theta}(\theta),
\ee
where
$V_{tr}(r)  = E \varpi^4/\Delta  + aL_z(1 - \varpi^2/\Delta)$ and
$V_{t\theta}(\theta) = - a^2 E \sin^2 \theta$.
Therefore using $t(0) = t_0$ we obtain
\be
t(\lambda) = t_0 + \int_0^\lambda d\lambda' \left\{ V_{tr}[r(\lambda')] +
  V_{t\theta}[\theta(\lambda')] \right\}.
\label{eq:tmotion0}
\ee
Next we define the averaged value $\langle V_{tr} \rangle$ of $V_{tr}$ to be
\be
\langle V_{tr} \rangle = \frac{1}{\Lambda_r} \int_0^{\Lambda_r} d\lambda
\
V_{tr}[r(\lambda)] = \frac{1}{\Lambda_r} \int_0^{\Lambda_r} d\lambda\
V_{tr}[{\hat r}(\lambda)].
\ee
Here the second equality follows from the representation (\ref{eq:rmotion}) of
the $r$ motion together with the periodicity condition (\ref{eq:periodic}).
Similarly we define
\be
\langle V_{t\theta} \rangle = \frac{1}{\Lambda_\theta} \int_0^{\Lambda_\theta} d\lambda
\
V_{t\theta}[\theta(\lambda)] = \frac{1}{\Lambda_\theta} \int_0^{\Lambda_\theta} d\lambda\
V_{t\theta}[{\hat \theta}(\lambda)].
\ee
Inserting these definitions into Eq.\ (\ref{eq:tmotion0})
allows us to write $t(\lambda)$ as a sum of a linear term and terms
that are periodic \cite{Drasco:2004tv}:
\begin{eqnarray}
\label{eq:tmotion00}
t(\lambda) &=& t_0 + \Gamma \lambda + \Delta t_r(\lambda) + \Delta
t_\theta(\lambda)  \\
&\equiv& t_0 + \Gamma \lambda + \Delta t(\lambda).
\label{eq:tmotion1}
\end{eqnarray}
Here we have defined the constant
\be
\Gamma = \langle V_{tr} \rangle + \langle V_{t\theta} \rangle,
\label{eq:Gammadef}
\ee
and the functions
\be
\Delta t_r(\lambda) = \int_0^\lambda d\lambda' \bigg\{
V_{tr}[r(\lambda')] - \langle V_{tr} \rangle \bigg\},
\label{eq:deltatrdef}
\ee
\be
\Delta t_\theta(\lambda) = \int_0^\lambda d\lambda' \bigg\{
V_{t\theta}[\theta(\lambda')] - \langle V_{t\theta} \rangle \bigg\}.
\label{eq:deltatthetadef}
\ee
The key property of these functions is that they are periodic:
\be
\Delta t_r(\lambda + \Lambda_r) = \Delta t_r(\lambda),
\ee
\be
\Delta t_\theta(\lambda + \Lambda_\theta) = \Delta t_\theta(\lambda);
\ee
this follows from the definitions (\ref{eq:deltatrdef}) and
(\ref{eq:deltatthetadef}) together with the periodicity condition
(\ref{eq:periodic}).  We can exhibit the dependence of these
functions on the parameters $\lambda_{r0}$ and $\lambda_{\theta0}$
by substituting the formulae (\ref{eq:rmotion}) and
(\ref{eq:thetamotion}) for $r(\lambda)$ and $\theta(\lambda)$ into
Eqs.\ (\ref{eq:deltatrdef}) and
(\ref{eq:deltatthetadef}).  The result is
\be
\Delta t_r(\lambda) = {\hat t}_r(\lambda - \lambda_{r0}) - {\hat
  t}_r(-\lambda_{r0}),
\label{eq:deltatrformula}
\ee
\be
\Delta t_\theta(\lambda) = {\hat t}_\theta(\lambda - \lambda_{\theta0}) - {\hat
  t}_\theta(-\lambda_{\theta0}),
\label{eq:deltatthetaformula}
\ee
where the functions ${\hat t}_r(\lambda)$ and ${\hat t}_\theta(\lambda)$ are defined by
\be
{\hat t}_r(\lambda) = \int_0^\lambda d\lambda' \bigg\{
V_{tr}[{\hat r}(\lambda')] - \langle V_{tr} \rangle \bigg\},
\label{eq:hattrdef}
\ee
\be
{\hat t}_\theta(\lambda) = \int_0^\lambda d\lambda' \bigg\{
V_{t\theta}[{\hat \theta}(\lambda')] - \langle V_{t\theta} \rangle \bigg\}.
\label{eq:hattthetadef}
\ee

\subsection{Motion in $\phi$}

The motion in $\phi$ can be analyzed in exactly the same way as the
motion in $t$.  First, the function $V_\phi(r,\theta)$ that appears on
the right hand side of Eq.\ (\ref{eq:phidot}) is a sum of a function
of $r$ and a function of $\theta$:
\be
V_\phi(r,\theta) = V_{\phi r}(r) + V_{\phi\theta}(\theta),
\ee
where
$V_{\phi r}(r)=a E (\varpi^2/\Delta - 1) - a^2 L_z/\Delta$ and
$V_{\phi\theta}(\theta) = \csc^2\theta L_z$.  Therefore using $\phi(0)
= \phi_0$ we obtain
\be
\phi(\lambda) = \phi_0 + \int_0^\lambda d\lambda' \left\{ V_{\phi r}[r(\lambda')] +
  V_{\phi\theta}[\theta(\lambda')] \right\}.
\label{eq:phimotion0}
\ee
Next we define the averaged value $\langle V_{\phi r} \rangle$ of
$V_{\phi r}$ to be
\be
\langle V_{\phi r} \rangle = \frac{1}{\Lambda_r} \int_0^{\Lambda_r} d\lambda
\
V_{\phi r}[r(\lambda)] = \frac{1}{\Lambda_r} \int_0^{\Lambda_r} d\lambda\
V_{\phi r}[{\hat r}(\lambda)].
\ee
Here the second equality follows from the representation (\ref{eq:rmotion}) of
the $r$ motion together with the periodicity condition (\ref{eq:periodic}).
Similarly we define
\be
\langle V_{\phi\theta} \rangle = \frac{1}{\Lambda_\theta}
\int_0^{\Lambda_\theta} d\lambda
\
V_{\phi\theta}[\theta(\lambda)] = \frac{1}{\Lambda_\theta} \int_0^{\Lambda_\theta} d\lambda\
V_{\phi\theta}[{\hat \theta}(\lambda)].
\ee
Inserting these into Eq.\ (\ref{eq:phimotion0}) and using $\phi(0) = \phi_0$
allows us to write $\phi(\lambda)$ as a sum of a linear term and terms
that are periodic \cite{Drasco:2004tv}:
\begin{eqnarray}
\label{eq:phimotion00}
\phi(\lambda) &=& \phi_0 + \Upsilon_\phi \lambda + \Delta \phi_r(\lambda) + \Delta
\phi_\theta(\lambda)  \\
&\equiv& \phi_0 + \Upsilon_\phi \lambda + \Delta \phi(\lambda).
\label{eq:phimotion1}
\end{eqnarray}
Here we have defined the constant
\be
\Upsilon_\phi = \langle V_{\phi r} \rangle + \langle V_{\phi\theta} \rangle,
\label{eq:Upsilonphidef}
\ee
and the functions
\be
\Delta \phi_r(\lambda) = \int_0^\lambda d\lambda' \bigg\{
V_{\phi r}[r(\lambda')] - \langle V_{\phi r} \rangle \bigg\},
\label{eq:deltaphirdef}
\ee
\be
\Delta \phi_\theta(\lambda) = \int_0^\lambda d\lambda' \bigg\{
V_{\phi\theta}[\theta(\lambda')] - \langle V_{\phi\theta} \rangle \bigg\}.
\label{eq:deltaphithetadef}
\ee
The key property of these functions is that they are periodic:
\be
\Delta \phi_r(\lambda + \Lambda_r) = \Delta \phi_r(\lambda),
\ee
\be
\Delta \phi_\theta(\lambda + \Lambda_\theta) = \Delta \phi_\theta(\lambda);
\ee
this follows from the definitions (\ref{eq:deltaphirdef}) and
(\ref{eq:deltaphithetadef}) together with the periodicity condition
(\ref{eq:periodic}).  We can exhibit the dependence of these
functions on the parameters $\lambda_{r0}$ and $\lambda_{\theta0}$
by substituting the formulae (\ref{eq:rmotion}) and
(\ref{eq:thetamotion}) for $r(\lambda)$ and $\theta(\lambda)$ into
Eqs.\ (\ref{eq:deltaphirdef}) and
(\ref{eq:deltaphithetadef}).  The result is
\be
\Delta \phi_r(\lambda) = {\hat \phi}_r(\lambda - \lambda_{r0}) - {\hat
  \phi}_r(-\lambda_{r0}),
\label{eq:deltaphirformula}
\ee
\be
\Delta \phi_\theta(\lambda) = {\hat \phi}_\theta(\lambda -
\lambda_{\theta0}) - {\hat \phi}_\theta(-\lambda_{\theta0}),
\label{eq:deltaphithetaformula}
\ee
where the functions ${\hat \phi}_r$ and ${\hat \phi}_\theta$ are defined by
\be
{\hat \phi}_r(\lambda) = \int_0^\lambda d\lambda' \bigg\{
V_{\phi r}[{\hat r}(\lambda')] - \langle V_{\phi r} \rangle \bigg\},
\label{eq:hatphirdef}
\ee
\be
{\hat \phi}_\theta(\lambda) = \int_0^\lambda d\lambda' \bigg\{
V_{\phi\theta}[{\hat \theta}(\lambda')] - \langle V_{\phi\theta}
\rangle \bigg\}.
\label{eq:hatphithetadef}
\ee

\subsection{Re-parameterization freedom}
\label{sec:degeneracy}

Not all of the parameters $E$, $L_z$, $Q$, $\lambda_{r0}$,
$\lambda_{\theta0}$, $t_0$ and $\phi_0$ that characterize the geodesic
are independent.  This is because of the freedom to change the
dependent variable $\lambda$ via $\lambda \to {\tilde \lambda} = \lambda +
\Delta \lambda$.  Under this change of variable the parameters
$\lambda_{r0}$ and $\lambda_{\theta0}$
transform as
\be
\lambda_{r0} \to {\tilde \lambda}_{r0} = \lambda_{r0} + \Delta
\lambda,
\label{eq:t1}
\ee
\be
\lambda_{\theta0} \to {\tilde \lambda}_{\theta0} = \lambda_{\theta0} +
\Delta \lambda.
\label{eq:t2}
\ee
We can compute how the parameters $t_0$ and $\phi_0$ transform as
follows.  Combining Eqs.\ (\ref{eq:tmotion1}),
(\ref{eq:deltatrformula}) and (\ref{eq:deltatthetaformula})
gives the following formula for the $t$ motion:
\be
\fl
t = t_0 + \Gamma \lambda
+ {\hat t}_r(\lambda - \lambda_{r0}) - {\hat t}_r(-\lambda_{r0})
+ {\hat t}_\theta(\lambda - \lambda_{\theta0}) - {\hat
  t}_\theta(-\lambda_{\theta0}).
\ee
Rewriting the right hand side in terms of ${\tilde \lambda}$, ${\tilde \lambda}_{r0}$
and ${\tilde \lambda}_{\theta0}$ yields
\begin{eqnarray}
\fl
t \ = t_0 + \Gamma {\tilde \lambda} - \Gamma \Delta \lambda
+ {\hat t}_r({\tilde \lambda} - {\tilde \lambda}_{r0}) - {\hat t}_r(-\lambda_{r0})
+ {\hat t}_\theta({\tilde \lambda} - {\tilde \lambda}_{\theta0}) - {\hat
  t}_\theta(-\lambda_{\theta0}) \nn \\
\fl
\ \ \ \equiv {\tilde t}_0 + \Gamma {\tilde \lambda}
+ {\hat t}_r({\tilde \lambda} - {\tilde \lambda}_{r0}) - {\hat
  t}_r(-{\tilde \lambda}_{r0})
+ {\hat t}_\theta({\tilde \lambda} - {\tilde \lambda}_{\theta0}) - {\hat
  t}_\theta(-{\tilde \lambda}_{\theta0}).
\end{eqnarray}
Comparing the first and second lines here allows us to read off the
value of ${\tilde t}_0$:
\be
\fl
{\tilde t}_0 = t_0 - \Gamma \Delta \lambda
+ {\hat  t}_r(-\lambda_{r0} - \Delta \lambda) - {\hat
  t}_r(-\lambda_{r0})
+ {\hat  t}_\theta(-\lambda_{\theta0} - \Delta \lambda) - {\hat t}_\theta(-\lambda_{\theta0}).
\label{eq:t3}
\ee
Similarly we obtain
\be
\fl
\phi_0 \to {\tilde \phi}_0 = \phi_0 - \Upsilon_\phi \Delta \lambda
+ {\hat  \phi}_r(-\lambda_{r0} - \Delta \lambda) - {\hat
  \phi}_r(-\lambda_{r0})
+ {\hat  \phi}_\theta(-\lambda_{\theta0} - \Delta \lambda) - {\hat
  \phi}_\theta(-\lambda_{\theta0}).
\label{eq:t4}
\ee
In Sec.\ \ref{sec:dep} below we will explicitly show that all of the amplitudes
and waveforms we compute are invariant under these transformations.

\subsection{Averages over Mino time in terms of angular variables}

We will often encounter functions $F_\theta(\lambda)$ of Mino time
$\lambda$ that are periodic with period $\Lambda_\theta$.  For such
functions the average over Mino time is given
by
\be
\langle F_\theta \rangle_\lambda = \frac{1}{\Lambda_\theta} \int_0^{\Lambda_\theta}
F_\theta(\lambda) d \lambda.
\label{eq:thetaaverage}
\ee
In order to compute such averages it will be convenient to
change the variable of integration from $\lambda$ to a new variable
$\chi$.
This variable is a generalization of the Newtonian true anomaly and is defined as
follows \cite{Wilkins:1972rs,Hughes:1999bq}.
The potential $V_\theta(\theta)$
defined by Eq.\ (\ref{eq:thetadot}) can be written in terms of the
variable
\be
z \equiv \cos^2 \theta
\ee
as
\be
V_\theta(z) = \frac{\beta (z - z_+) (z - z_-)}{1 - z}.
\label{eq:Thetaformula}
\ee
Here $\beta = a^2(1-E^2)$ and $z_-$ and $z_+$ are the two zeros of
$V_\theta(z)$.  They are
ordered such that $0 \le z_- \le 1 \le z_+$.  The variable $\chi$ is
defined by
\be
\cos^2 {\hat \theta}(\lambda) = z_- \cos^2 \chi,
\ee
together with the requirement that $\chi$ increases monotonically as
$\lambda$ increases.
From the definition (\ref{eq:hatthetadef}) of ${\hat \theta}(\lambda)$
together with the formula (\ref{eq:Thetaformula}) for $V_\theta(z)$ we
get
\be
\frac{d\chi}{d\lambda} = \sqrt{\beta (z_+  - z_- \cos^2 \chi)}.
\ee
Therefore the average (\ref{eq:thetaaverage}) can be written as
\be
\langle F_\theta \rangle_\lambda = \frac{1}{\Lambda_\theta} \int_0^{2
  \pi} d\chi
\frac{F_\theta[\lambda(\chi)] }{\sqrt{\beta(z_+ - z_- \cos^2 \chi)}}.
\label{eq:thetaaverage1}
\ee
%Using $\langle 1 \rangle  = 1$ gives an expression for
%$\Lambda_\theta$:
%\be
%\Lambda_\theta = \frac{4}{\sqrt{\beta z_+}} \int_0^{\pi/2}
%\frac{1}{\sqrt{1 - (z_-/z_+) \cos^2 \chi}} = \frac{4}{\sqrt{\beta z_+}}
%  K(\sqrt{z_-/z_+}),
%\ee
%where $K$ is the complete elliptic integral of the first kind.  Using
%this in Eq.\ (\ref{eq:thetaaverage1}) gives
%\be
%\langle F_\theta \rangle_\lambda = \frac{1}{4 K(\sqrt{z_-/z_+})} \int_0^{2
%  \pi} d\chi
%\frac{F_\theta[\lambda(\chi)] }{\sqrt{1 - (z_-/z_+) \cos^2 \chi}}.
%\label{eq:thetaaverage2}
%\ee

A similar analysis applies to averages of functions $F_r(\lambda)$
that are periodic with period $\Lambda_r$ \cite{Drasco:2004tv}.  The
average over $\lambda$ of such a function is
\be
\langle F_r \rangle_\lambda = \frac{1}{\Lambda_r} \int_0^{\Lambda_r}
F_r(\lambda) d \lambda.
\label{eq:raverage}
\ee
Now the
potential  $V_r(r)$ in
Eq.\ (\ref{eq:rdot}) can be written as \cite{Drasco:2004tv}
\be
V_r(r) = (1-E^2)(r_1-r)(r-r_2)(r-r_3)(r-r_4),
\ee
where $r_1$, $r_2$, $r_3$ and $r_4$ are the four roots of the quartic
$V_r(r)=0$, ordered such that $r_4 \le r_3 \le r_2 \le r_1$.
For stable orbits the motion takes place in $r_2 \le r \le r_1$.
The orbital eccentricity $\varepsilon$ and semi-latus rectum $p$ are
defined by
\be
r_1 = \frac{p}{1-\varepsilon}
\ee
and
\be
r_2 = \frac{p}{1 + \varepsilon}.
\ee
We also define the parameters $p_3$ and $p_4$ by
\be
r_3 = \frac{p_3 }{1 - \varepsilon}, \ \ \ \ \
r_4 = \frac{p_4 }{1 + \varepsilon}.
\ee
Then following Ref.\ \cite{Cutler:1994pb}
we define the parameter $\psi$ by
\be
{\hat r}(\lambda) = \frac{p }{1 + \varepsilon \cos \psi}.
\label{eq:psidef}
\ee
It follows from these definitions that \cite{Drasco:2004tv}
\be
\frac{d \psi}{d \lambda} = {\cal P}(\psi),
\ee
where
\be
\fl
{\cal P}(\psi) = \frac{\sqrt{1-E^2}}{1 - \varepsilon^2}
\left[ p(1-\varepsilon) - p_3(1 + \varepsilon \cos \psi) \right]^{1/2}
\left[ p(1+\varepsilon) - p_4(1 + \varepsilon \cos \psi) \right]^{1/2}.
\ee
The average (\ref{eq:raverage}) over $\lambda$ is therefore
\be
\langle F_r \rangle_\lambda = \frac{1}{\Lambda_r} \int_0^{2 \pi} d\psi
\frac{F_r(\lambda)}{{\cal P}(\psi)} = \frac{
\int_0^{2 \pi} d\psi F_r(\lambda)/ {\cal P}(\psi)}
{\int_0^{2 \pi} d\psi / {\cal P}(\psi)}.
\label{eq:raverage1}
\ee

Note that the viewpoint on the new variables $\chi$,$\psi$ adopted
here is slightly different to that in \cite{Drasco:2004tv}: Here they are
defined in terms of the fiducial motions ${\hat r}(\lambda)$ and
${\hat \theta}(\lambda)$ instead of the actual motions $r(\lambda)$
and $\theta(\lambda)$.  The new viewpoint facilitates the computation
of amplitudes and fluxes for arbitrary geodesics; if one is working
with the fiducial geodesic the distinction is unimportant.

\subsubsection{Bi-periodic functions}

Finally suppose we have a function $F(\lambda_r,\lambda_\theta)$ of
two parameters $\lambda_r$ and $\lambda_\theta$ which is biperiodic:
\begin{eqnarray}
F(\lambda_r + \Lambda_r, \lambda_\theta) = F(\lambda_r,\lambda_\theta)
\nn \\
F(\lambda_r , \lambda_\theta + \Lambda_\theta) =
F(\lambda_r,\lambda_\theta).
\end{eqnarray}
The average value of this function is
\be
\langle F \rangle_\lambda = \frac{1}{\Lambda_r \Lambda_\theta}
\int_0^{\Lambda_r} d\lambda_r \, \int_0^{\Lambda_\theta}
d\lambda_\theta F(\lambda_r,\lambda_\theta).
\ee
By combining the results (\ref{eq:thetaaverage1}) and
(\ref{eq:raverage1}) we can write this average as a double integral
over the new variables $\chi$ and $\psi$:
\be
\langle F \rangle_\lambda = \frac{1}{\Lambda_r \Lambda_\theta}
\int_0^{2 \pi} d\chi \int_0^{2 \pi} d\psi
\frac{F[\lambda_r(\psi),\lambda_\theta(\chi)]}{ \sqrt{ \beta( z_+ -
    z_- \cos^2 \chi) } {\cal P}(\psi)}.
\label{eq:averageidentity}
\ee
This formula will be used in later sections.

\section{Mino's derivation of half retarded minus half advanced prescription}
\label{sec:minoproof}

In this section we review the proof by Mino \cite{Mino:2003yg}
that for computing radiation reaction in the adiabatic limit in Kerr,
one can use the ``half retarded minus half advanced'' prescription.
We specialize to the scalar case.

We start by defining some notation for self-forces.
Suppose that we have a particle with scalar charge $q$ at a point ${\cal P}$ in a
spacetime $M$ with metric $g_{\alpha\beta}$.  Suppose that the 4-velocity
of the particle at ${\cal P}$ is $u^\alpha$, and that we are given a
solution $\Phi$ (not necessarily the retarded solution) of the wave
equation (\ref{eq:wave}) for the scalar field, for which the source is
a delta function on the geodesic determined by ${\cal P}$ and
$u^\alpha$.  The self-force on the particle is then some
functional of ${\cal P}$, $u^\alpha$, $g_{\alpha\beta}$ and $\Phi$,
which we write as
\be
f^\alpha\left[ {\cal P}, u^\alpha, g_{\alpha\beta}, \Phi \right].
\ee
Here we suppress the trivial dependence on $q$.
Note that this functional does not depend on a choice of time
orientation for the manifold, and also it is invariant under $u^\alpha
\to - u^\alpha$.

Next, we define the retarded self-force as
\be
f^\alpha_{\rm ret}\left[ {\cal P}, u^\alpha, g_{\alpha\beta} \right]
= f^\alpha\left[ {\cal P}, u^\alpha, g_{\alpha\beta},
  \Phi_{\rm ret} \right],
\label{eq:selfret}
\ee
where $\Phi_{\rm ret}$ is the retarded solution to the wave equation
(\ref{eq:wave}) using the time orientation that is determined by
demanding that $u^\alpha$ be future directed.  Similarly we define the
advanced self force by
\be
f^\alpha_{\rm adv}\left[ {\cal P}, u^\alpha, g_{\alpha\beta} \right]
= f^\alpha\left[ {\cal P}, u^\alpha, g_{\alpha\beta},
  \Phi_{\rm adv} \right],
\ee
where $\Phi_{\rm adv}$ is the advanced solution.
It follows from these definitions that
\be
f^\alpha_{\rm ret}\left[ {\cal P}, -u^\alpha, g_{\alpha\beta} \right]
= f^\alpha_{\rm adv}\left[ {\cal P}, u^\alpha, g_{\alpha\beta}
  \right].
\label{eq:flip}
\ee

\subsection{Properties of the self-force}

The derivation of the half-retarded-minus-half-advanced prescription
in the adiabatic limit
rests on two properties of the self-force.  The first
property is the fact that the self-force can be computed by
subtracting from the divergent field $\Phi$ a locally constructed
singular field $\Phi_{\rm sing}$ that depends only on the spacetime geometry in
the vicinity of the point ${\cal P}$ and on the four velocity at
${\cal P}$.  This property is implicit in the work of Quinn \cite{Quinn:2000wa}
and Quinn and Wald \cite{Quinn:1997am}, and was proved explicitly by
Detweiler and Whiting \cite{Detweiler:2002mi}.  We can write this
property as
\be
f^\alpha\left[ {\cal P}, u^\alpha, g_{\alpha\beta}, \Phi \right] = q
\nabla^\alpha \left( \Phi - \Phi_{\rm sing}[{\cal
    P},u^\alpha,g_{\alpha\beta}] \right).
\label{eq:property0}
\ee

The second property is a covariance property.   Suppose that $\psi$ is
a diffeomorphism from the manifold $M$ to another manifold $N$ that
takes ${\cal P}$ to $\psi({\cal P})$.  We denote by $\psi^*$ the
natural mapping of tensors over the tangent space at ${\cal P}$ to
tensors over the tangent space at $\psi({\cal P})$.  We also denote by
$\psi^*$ the associated natural mapping of tensor fields on $M$ to
tensor fields on $N$.  Then the covariance property is
\be
f^\alpha[\psi({\cal P}), \psi^* u^\alpha, \psi^* g_{\alpha\beta},
  \psi^* \Phi] = \psi^* f^\alpha[{\cal P}, u^\alpha, g_{\alpha\beta},
  \Phi].
\ee
This expresses the fact that the self-force does not depend on
quantities such as a choice of coordinates.
It follows from the definition (\ref{eq:selfret}) that the retarded
self-force $f^\alpha_{\rm ret}$ satisfies a similar covariance relation:
\be
f^\alpha_{\rm ret}[\psi({\cal P}), \psi^* u^\alpha, \psi^* g_{\alpha\beta}
 ] = \psi^* f^\alpha_{\rm ret}[{\cal P}, u^\alpha, g_{\alpha\beta}].
\label{eq:covariance}
\ee

\subsection{Property of generic bound geodesics in Kerr}

Next we review a property of bound geodesics in Kerr upon which the
proof depends.  This is the fact that
for generic bound geodesics there exists isometries $\psi$
of the Kerr spacetime of the form
\be
t \to 2 t_1 - t, \ \ \ \ \phi \to 2 \phi_1 - \phi,
\label{eq:isometry}
\ee
where $t_1$ and $\phi_1$ are constants,
which come arbitrarily close to mapping the  geodesic onto itself.

To see this, note that if a geodesic is mapped onto itself by the
mapping (\ref{eq:isometry}), then the point
$[t,r(t),\theta(t),\phi(t)]$ is mapped onto
$[2 t_1 - t, r(t), \theta(t),2 \phi_1 -
  \phi(t)]$,
which must equal $[t',r(t'),\theta(t'),\phi(t')]$
where $t' = 2 t_1 -t$.
If we specialize to $t = t_1$ then we see that $\phi(t_1) =
\phi_1$.  We denote this point by ${\cal Q} =
(t_1,r_1,\theta_1,\phi_1)$.  Next, since the geodesic is determined
by ${\cal Q}$ and by the four-velocity $u^\alpha$ at ${\cal
  Q}$, the geodesic is mapped onto itself by $\psi$ if and only if
$\psi^* u^\alpha = - u^\alpha$ at ${\cal Q}$.  This will be true if
and only
if the $r$ and $\theta$ components of $u^\alpha$ vanish.  Thus, a
geodesic will be invariant under a map of the form (\ref{eq:isometry})
if and only if it contains a point ${\cal Q}$ which is simultaneously
a turning point of the $r$ and $\theta$ motions.

Consider now a generic bound geodesic.  Such a geodesic can be
 characterized by the parameters $E$, $L_z$, $Q$,
$\lambda_{r0}$, $\lambda_{\theta0}$, $t_0$, $\phi_0$
in the notation of Sec.\ \ref{sec:geodesicparameters}.
The turning points of the $r$ motion occur at
$\lambda = \lambda_{r,n} \equiv \lambda_{r0} + n \Lambda_r/2$, where
 $n$ is an integer.
Similarly the turning points of the $\theta$ motion occur at
$\lambda = \lambda_{\theta,m} \equiv \lambda_{\theta0} + m
 \Lambda_\theta/2$, where $m$ is an
 integer.  Since generically $\Lambda_r$ and $\Lambda_\theta$ will be
 incommensurate, for any $\varepsilon>0$ we can find values of the
 integers $m$ and $n$ so that
\be
|\lambda_{r,n} - \lambda_{\theta,m}| \le \varepsilon.
\ee
Thus we can find a point on the geodesic that is arbitrarily to close
to being a turning point for both the $r$ and $\theta$ motions, and hence
there is an isometry of the form (\ref{eq:isometry})
that comes arbitrarily close to mapping the geodesic onto itself.

Below we will simply assume that the isometry does map the geodesic
onto itself; the error associated with this assumption can be made
arbitrarily small in the adiabatic limit\footnote{Note however that
real inspirals will consist of finite curves which are approximately
locally geodesic; the corresponding error for such curves will
contribute to the post-adiabatic correction to the inspiral, i.e., to
the second term in Eq.\ (\protect{\ref{eq:emri_phase}}).}.

\subsection{Property of the adiabatic limit}

The self-acceleration $a^\alpha$ at a point ${\cal P}$ on a geodesic
has three independent components, since $a^\alpha u_\alpha =0$.  These
three components are determined by the time derivatives of the three
conserved quantities
$$
{dE \over dt},\ \ \
{dL_z \over dt},\ \ \
{dK \over dt}.
$$
In order to compute the evolution of the orbit in the adiabatic limit,
it is sufficient to know the time averaged quantities
\be
\left< {dE \over dt} \right>_t,\ \ \
\left< {dL_z \over dt} \right>_t,\ \ \
\left< {dK \over dt} \right>_t,
\label{eq:timeaverages}
\ee
where
\be
\left< {dE \over dt} \right>_t \equiv \lim_{T \to \infty} {1 \over 2
T} \int_{-T}^T dt \, {dE \over dt}(t).
\ee
This was shown in Ref.\ \cite{Mino:2003yg}
(although see Refs.\ \cite{2005PThPh.113..733M,2005Minob,2005Minoc}
for caveats related to gauge issues), and is
also discussed in detail using a two-timescale expansion in Ref.\
\cite{Hinderer}.

\subsection{Derivation}

We now combine the various results discussed above to derive the half
retarded minus half advanced prescription.
We parameterize the geodesic as $x^\alpha = z^\alpha(t)$, and denote
by
\be
f^{\alpha}_{\rm ret}(t) = f_{\rm ret}^\alpha[ z^\alpha(t),
  u^\alpha(t), g_{\alpha\beta} ]
\ee
the retarded self force at $z^\alpha(t)$.
The key idea is to use the covariance property (\ref{eq:covariance}),
with $\psi$ taken to be the isometry associated with the geodesic.
If we take ${\cal P}$ to be $z^\alpha(t)$, then the
left hand side of (\ref{eq:covariance}) evaluates to
\begin{eqnarray}
\fl
f_{\rm ret}^\alpha[ z^\alpha(2 t_1 -t), - u^\alpha(2 t_1 - t),
  g_{\alpha\beta}] &=& f_{\rm adv}^\alpha[ z^\alpha(2 t_1 -t), u^\alpha(2 t_1 - t),
  g_{\alpha\beta}] \nn \\
&=&  f_{\rm adv}^\alpha(2 t_1 - t).
\end{eqnarray}
Here we have used the property (\ref{eq:flip}) and the fact that
$\psi^* g_{\alpha\beta} = g_{\alpha\beta}$.
This gives
\be
f_{\rm adv}^\alpha(2 t_1 -t ) = \psi^* f_{\rm ret}^\alpha(t).
\label{eq:flip1}
\ee

Now let ${\cal E}$ denote one of the three conserved quantities $E$,
$L_z$ or $K$.  The time derivative of ${\cal E}$ depends linearly on
the self force, so we can write
\be
{d {\cal E} \over d t}(t) = \Xi_\alpha(t) f^\alpha_{\rm ret}(t)
\label{eq:dcalEdt000}
\ee
for some $\Xi_\alpha$.
Below we will compute $\Xi_\alpha$ explicitly for the three different
conserved quantities and derive the key property that
\be
\psi^* \Xi_\alpha(t) = - \Xi_\alpha(2 t_1 -t).
\label{eq:flip2}
\ee
Now acting on Eq. (\ref{eq:dcalEdt000}) with $\psi^*$ and using the
relations (\ref{eq:flip1}) and (\ref{eq:flip2})
yields
\be
(\Xi_\alpha f^\alpha_{\rm ret})(t) = - (\Xi_\alpha f^\alpha_{\rm
  adv})(2t_1 -t).
\ee
Taking a time average gives
\be
\left< \Xi_\alpha f_{\rm ret}^\alpha \right>_t = -
\left< \Xi_\alpha f_{\rm adv}^\alpha \right>_t,
\ee
and hence
\be
\left< { d {\cal E} \over dt } \right>_t =
\left< \Xi_\alpha f_{\rm ret}^\alpha \right>_t =
{1 \over 2} \left<
\Xi_\alpha ( f^\alpha_{\rm ret} - f^\alpha_{\rm adv}) \right>_t.
\ee
Finally we use the explicit formula (\ref{eq:property0}) for the self force.  The
contributions from the locally constructed singular field $\Phi_{\rm
  sing}$ cancel, and we obtain
\be
\left< { d {\cal E} \over dt } \right>_t =
{q \over 2} \left<
\Xi_\alpha \nabla^\alpha ( \Phi_{\rm ret} - \Phi_{\rm adv}) \right>_t.
\ee
This can be written as
\be
\left< { d {\cal E} \over dt } \right>_t = \left<
\Xi_\alpha f_{\rm rad}^\alpha \right>_t
\ee
where
\be
f_{\rm rad}^\alpha = {q \over 2} \nabla^\alpha ( \Phi_{\rm ret} -
\Phi_{\rm adv} )
\ee
is the self-force one obtains from the half retarded minus half
advanced prescription, cf.\ Eqs.\ (\ref{eq:selfacc0}) and
(\ref{eq:dmudt2}) above.
What we have shown is that although the quantity $f_{\rm rad}^\alpha$
differs from the true self-force
$f_{\rm ret}^\alpha$, the difference averages to zero when one
takes a time average over the entire geodesic.

It remains to derive the formula (\ref{eq:flip2}).
For energy and angular momentum, we can write the conserved quantity as
the inner product of a Killing vector $\xi^\alpha$ with the 4-velocity
of the particle:
\be
{\cal E} = \xi_\alpha u^\alpha.
\ee
Here ${\vec \xi} = - \partial/\partial t$ for ${\cal E} = E$, and
${\vec \xi} = \partial /\partial \phi$ for ${\cal E} = L_z$.
Taking a derivative with respect to proper time $\tau$ gives
\be
\frac{d {\cal E}}{d \tau} = u^\beta \nabla_\beta ( \xi_\alpha
u^\alpha) = (\nabla_\beta \xi_\alpha) u^\beta u^\alpha + \xi_\alpha
(u^\beta \nabla_\beta u^\alpha) = \xi_\alpha a^\alpha,
\ee
where $a^\alpha$ is the 4-acceleration.  Here we have used the fact
that ${\vec \xi}$ is a Killing vector so $\nabla_{(\alpha}
\xi_{\beta)} =0$.  Next we use the geodesic equations (\ref{eq:tdot})
and (\ref{eq:Minotime}) to
translate from a proper time derivative to coordinate time derivative,
and we use the relation between self-force and self-acceleration given
by Eqs.\ (\ref{eq:acc}) and (\ref{eq:tsf}).  Using the definition
(\ref{eq:dcalEdt000}) of $\Xi_\alpha$ now gives
\be
\Xi_\alpha = \frac{\Sigma(r)}{\mu V_t(r,\theta)} (\delta^\beta_\alpha +
u_\alpha u^\beta) \xi_\beta.
\ee
The formula (\ref{eq:flip2}) now follows from $\psi^* u^\alpha(t) = -
u^\alpha(2 t_1 -t)$ and $\psi^* \xi_\alpha = - \xi_\alpha$.

Turn now to the Carter constant.
Taking a time derivative of the expression (\ref{eq:Kdef}) for $K$ gives
\begin{eqnarray}
\frac{d K}{d \tau} &=& u^\gamma \nabla_\gamma (K_{\alpha\beta} u^\alpha u^\beta) =
\nabla_{(\gamma} K_{\alpha\beta)} u^\gamma u^\alpha u^\beta + 2 K_{\alpha\beta} u^\alpha u^\gamma \nabla_\gamma u^\beta \nn \\
\mbox{} &=& 2 K_{\alpha\beta} u^\alpha a^\beta.
\end{eqnarray}
Here we have used the Killing tensor equation $\nabla_{(\gamma} K_{\alpha\beta)} =0$.
Proceeding as before now gives the expression
\be
\Xi_\alpha = \frac{2 \Sigma(r)}{\mu V_t(r,\theta)} (\delta^\beta_\alpha +
u_\alpha u^\beta) K_{\beta\gamma} u^\gamma.
\ee
The formula (\ref{eq:flip2}) now follows from the relations $\psi^* u^\alpha(t) = -
u^\alpha(2 t_1 -t)$ and $\psi^* K_{\alpha\beta} = K_{\alpha\beta}$.

\section{Separation of variables and basis of modes}
\label{sec:modes}

\subsection{Separation of variables for scalar wave equation}

Separation of variables for the scalar wave equation in Kerr was first
carried out by Dieter Brill and others in 1972 \cite{Brill:1972xj},
and subsequently generalized to higher spins by Teukolsky
\cite{Teukolsky:1972my,Teukolsky:1973ha}.  We substitute the ansatz
\be
\Phi(t,r,\theta,\phi) = R(r) \Theta(\theta) e^{i m \phi} e^{-i \omega
  t}
\label{eq:ansatz}
\ee
into the homogeneous version of Eq.\ (\ref{eq:wave}), and make use of the
expression (\ref{eq:waveoperator}) for the wave operator.  The result
is the two equations
\be
\frac{1}{\sin \theta} (\sin \theta \Theta_{,\theta})_{,\theta} -
\frac{m^2}{\sin^2 \theta} \Theta + a^2 \omega^2 \cos^2\theta \Theta =
- \lambda \Theta,
\label{eq:Thetaeqn}
\ee
and
\be
( \Delta R_{,r})_{,r} + \frac{1}{\Delta} \left( \omega^2 \varpi^4 - 4
r a m \omega + m^2 a^2 \right) R = (\lambda + \omega^2 a^2) R.
\label{eq:Reqn}
\ee
Here $\lambda$ is the separation constant
We label the successive
eigenvalues $\lambda$ of Eq.\ (\ref{eq:Thetaeqn}) by an integer $l$,
with $l = |m|,|m|+1,|m|+2, \ldots$, and we denote these eigenvalues by
$\lambda_{\omega lm}$.
In the Schwarzschild limit $a=0$ we have $\lambda_{\omega lm} = l(l+1)$.
Since the differential equation
(\ref{eq:Thetaeqn}) does not depend on the sign of $\omega$ or the
sign of $m$, we have
\be
\lambda_{-\omega l-m} = \lambda_{\omega lm},
\label{eq:lambdaidentity}
\ee
a property which will be used later.
We denote by $\Theta_{\omega lm}(\theta)$ the corresponding solutions to
Eq.\ (\ref{eq:Thetaeqn}).  The functions
\be
S_{\omega lm}(\theta,\phi) \equiv e^{i m \phi} \Theta_{\omega lm}(\theta)
\label{eq:Somegalmdef}
\ee
are the spheroidal harmonics.
These harmonics are orthogonal on the sphere, and we can choose them
to be orthonormal like the spherical harmonics:
\be
\int d^2 \Omega \ S_{\omega lm}(\theta,\phi)^* S_{\omega
  l'm'}(\theta,\phi) = \delta_{l l'} \delta_{m m'}.
\label{eq:orthonormal}
\ee
Following Gal'tsov \cite{Galtsov:1982} we also choose the phases of the
spheroidal harmonics to satisfy
\be
(P S_{\omega lm})(\theta,\phi) \equiv S_{\omega lm}(\pi - \theta, \pi +
\phi) = (-1)^l S_{\omega lm}(\theta,\phi)
\label{eq:Zparity}
\ee
and
\be
S_{-\omega l -m}(\theta,\phi)^* = (-1)^m S_{\omega lm}(\theta,\phi).
\label{eq:Zparity1}
\ee
Here $P$ is the parity operator.

The radial equation (\ref{eq:Reqn}) can now be simplified by
defining
\be
u(r) = \varpi R(r),
\label{eq:udef}
\ee
and by using the tortoise coordinate
$r^*$ defined by
\be
dr^* = (\varpi^2/\Delta) dr.
\label{eq:rstardef0}
\ee
An explicit
expression for $r^*$ is
\be
r^* = r + \frac{2 r_+}{ r_+ - r_-}\ln \frac{r-r_+}{ 2 }
 - \frac{2 r_-}{ r_+ - r_-}\ln \frac{r-r_-}{ 2 },
\label{eq:rstardef}
\ee
where $r_\pm = 1 \pm \sqrt{1^2 - a^2}$ are the two roots of $\Delta(r)
=0$.  [Recall that we are using units in which the mass $M$ of the
black hole is unity.]
The resulting simplified radial equation is
\be
\frac{d^2 u}{d r^{*\,2}}(r^*) + V_{\omega lm}(r^*) u(r^*)  = 0,
\label{eq:diffsimple}
\ee
where the potential is
\begin{eqnarray}
\fl
V_{\omega lm} = - \frac{\Delta}{\varpi^6}(3r^2 - 4  r + a^2) + \frac{3 r^2
  \Delta^2}{\varpi^8} + \omega^2
\nonumber \\
+ \frac{1}{\varpi^4} \left[ m^2 a^2 -
  4  r a m \omega - \Delta (\lambda_{\omega l m} + \omega^2 a^2 )
  \right].
\label{eq:potential}
\end{eqnarray}
From Eq.\ (\ref{eq:lambdaidentity}) it follows that this potential satisfies
\be
V_{-\omega l-m}(r) = V_{\omega lm}(r),
\label{eq:Videntity}
\ee
an identity which will be used later.

\subsection{Basis of modes}

In this section we follow the treatment of Chrzanowski
\cite{Chrzanowski:1975wv} as modified slightly by Gal'tsov \cite{Galtsov:1982}.
We follow Gal'tsov's notation except that we omit
his subscripts $s$ denoting spin, since we have specialized to $s=0$.

We start by analyzing the asymptotic behavior of the potential
(\ref{eq:potential}) as $r^* \to -\infty$ and as $r^* \to \infty$.
We consider first the limit $r^* \to -\infty$, the past and future
event horizons.  In this limit $\Delta \to 0$ and $r \to r_+ = 1 +
\sqrt{1 - a^2}$.
We find $V_{\omega lm}(r) \to V_{\omega lm}(r_+) = p_{m\omega}^2$,
where
\be
p_{m\omega} = \omega - m \omega_+,
\label{eq:kdef}
\ee
and
\be
\omega_+ = \frac{a}{2  r_+}
\ee
is the angular velocity of rotation of the horizon.  Therefore the
solutions of the radial equation near the horizons are of the form
\be
u(r) \propto e^{\pm i p_{m\omega} r^*}.
\ee
In the limit of $r^* \to \infty$ (past and future
null infinity), $V_{\omega lm}(r^*) \to \omega^2$, so the solutions are of the
form
\be
u(r) \propto e^{\pm i \omega r^*}.
\ee

\subsubsection{``In'', ``up'', ``out'' and ``down'' modes}

We now define, following Gal'tsov, the following solution
\be
u_{\omega lm}^{\rm in}(r^*) = \alpha_{\omega lm}
\left\{ \begin{array}{ll} \tau_{\omega lm} |p_{m\omega}|^{-1/2} e^{- i p_{m\omega} r^*},
  & \mbox{ $r^* \to -\infty$}\\
|\omega|^{-1/2} \left[ e^{-i \omega r^*} + \sigma_{\omega lm} e^{i
      \omega r^*} \right],   & \mbox{
        $r^* \to \infty$.}\\ \end{array} \right.
\label{eq:uindef}
\ee
This equation serves to define the mode as well as the complex
transmission and reflection coefficients $\tau_{\omega lm}$ and
$\sigma_{\omega lm}$.  The coefficient $\alpha_{\omega lm}$ is a
normalization constant whose value is arbitrary; we will discuss a
convenient choice of $\alpha_{\omega lm}$ later.
We will often denote the set of subscripts $\omega lm$ collectively by
$\Lambda$: thus $\sigma_\Lambda$, $\tau_\Lambda$,
and sometimes we will omit the subscript and use simply $\sigma$,
$\tau$.

The ``in'' mode (\ref{eq:uindef}) is a mixture of outgoing and ingoing
components at past and future null infinity, since the mode function
is multiplied by $e^{-i \omega t}$.  At the past and future event
horizons, the mode is purely ingoing when the sign of $p_{m\omega}$ is the same
as the sign of $\omega$.  However, from the definition (\ref{eq:kdef})
of $p_{m\omega}$ we see that the sign of $\omega p_{m\omega}$ can be negative;
this occurs
for superradiant
modes.  Thus, at the future event horizon the ``in'' modes can be
either ingoing or outgoing.

\begin{figure}
\begin{center}
\epsfig{file=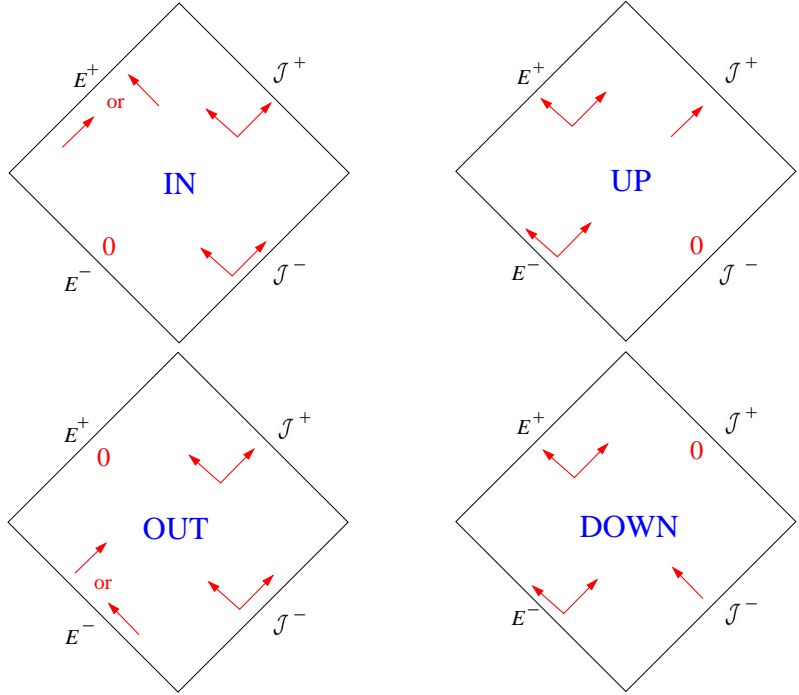,angle=0,width=10.5cm}
\caption{An illustration of the various types of modes in black hole spacetimes.
Here ${\cal J}^-$ denotes past null infinity, ${\cal J}^+$ future null
infinity, $E^-$ the past event horizon, and $E^+$ the future event horizon.
The four panels give the behavior of the four different modes ``in'',
``out'', ``up'' or ``down'' as indicated.
A zero indicates the mode vanishes at the indicated boundary.  Two
arrows indicates that the
mode consists of a mixture of ingoing and outgoing radiation at that
boundary.  Two arrows with an ``OR'' means that the mode is either
purely ingoing or purely outgoing at that boundary, depending on the
relative sign of $p_{m\omega}$ and $\omega$.
The ``in'' modes vanish on the past event horizon, and the ``up''
modes vanish on past null infinity.  Thus the ``in'' and ``up'' modes
together form a complete basis of modes.  Similarly the ``down'' and
``out'' modes together form a complete basis of modes.}
\label{fig:modes}
\end{center}
\end{figure}

The crucial feature of the ``in'' modes is that they vanish on the
past event horizon.  This feature will be used later in constructing
the various Green's functions.  A more precise statement of the result
is that a solution $\Phi$ of the wave equation which is a
linear combination of ``in'' modes with coefficients $c_{\omega lm}$,
such that the coefficients depend smoothly on $\omega$ (a reasonable
requirement), must vanish at the past event horizon.  To see this, note
from Eqs.\ (\ref{eq:ansatz}) and (\ref{eq:udef})
that the solution can be written as
\be
\Phi(t,r,\theta,\phi) = \int_{-\infty}^\infty d\omega \sum_{l=0}^\infty
\sum_{m=-l}^l e^{-i \omega t} c_{\omega lm} S_{\omega lm}(\theta,\phi)
\frac{u_{\omega lm}^{\rm in}(r^*)}{\varpi(r^*)}.
\ee
We now insert the asymptotic form (\ref{eq:uindef}) of the mode
function near the horizon, and we use the definition (\ref{eq:kdef})
of $p_{m\omega}$.  This gives
\begin{eqnarray}
\fl
\Phi(t,r,\theta,\phi) =
\frac{1}{\varpi} \int_{-\infty}^\infty d\omega \sum_{lm} e^{-i \omega (t + r^*)}
c_{\omega lm} S_{\omega lm}(\theta,\phi) \alpha_{\omega lm}
\tau_{\omega lm} |p_{m\omega}|^{-1/2} e^{i m \omega_+ r^*}  \\
\fl
\ \ \ \ \ \ \ \ \ \ \ \ \ \
\mbox{} \equiv \frac{1}{\varpi} \sum_{lm} G_{lm}(t+r^*;\theta,\phi)
e^{i m \omega_+ r^*}.
\label{eq:Gdef}
\end{eqnarray}
Now all of the quantities that depend on $\omega$ in the integrand are
smooth functions of $\omega$.  Since Fourier transforms of smooth
functions go to zero at infinity,
it follows that the function $G_{lm}(v;\theta,\phi)$ defined by
Eq.\ (\ref{eq:Gdef}) satisfies $G_{lm} \to 0$ as $v \to -\infty$,
where $v=t+r^*$.  Thus,
$\Phi$ will vanish as $v\to -\infty$, on the past event horizon.

Another important property of the ``in'' modes is the identity
\be
u^{\rm in}_{-\omega l-m}(r^*) = u^{\rm in}_{\omega lm}(r^*)^*.
\label{eq:uinidentity}
\ee
To derive this, note that the potential $V_{\omega lm}$ is real and
also satisfies the identity (\ref{eq:Videntity}).  It follows that the
function $u^{\rm in}_{-\omega l-m}(r^*)^*$ satisfies the same differential
equation (\ref{eq:diffsimple}) as $u^{\rm in}_{\omega lm}(r)$, and has
the same asymptotic behavior as $r^* \to \infty$ if we choose the
normalization constants to satisfy
\be
\alpha_{-\omega l-m} = \alpha_{\omega lm}^*.
\label{eq:alphacondt}
\ee
The result (\ref{eq:uinidentity}) now follows.  In particular the
identity (\ref{eq:uinidentity})
implies that
\be
\sigma_{-\omega l-m} = \sigma_{\omega lm}^*
\ee
and
\be
\tau_{-\omega l-m} = \tau_{\omega lm}^*.
\label{eq:tauidentity}
\ee

We next define the ``up'' modes
\be
\fl
u_{\omega lm}^{\rm up}(r^*) = \beta_{\omega lm}
\left\{ \begin{array}{ll}  |p_{m\omega}|^{-1/2} \frac{\omega p_{m\omega}}{|\omega p_{m\omega}|}
  \left[ \mu_{\omega lm} e^{ i p_{m\omega} r^*} + \nu_{\omega lm} e^{-i p_{m\omega} r^*} \right],
  & \mbox{ $r^* \to -\infty$}\\
|\omega|^{-1/2} e^{i \omega r^*},   & \mbox{
        $r^* \to \infty$.}\\ \end{array} \right.
\label{eq:uupdef}
\ee
As before this equation defines both the mode as well as the complex
coefficients $\mu_{\omega lm}$ and
$\nu_{\omega lm}$.  The coefficient $\beta_{\omega lm}$ is a
normalization constant whose value is arbitrary; we will discuss a
convenient choice of $\beta_{\omega lm}$ later.
The ``up'' modes are a mixture of ingoing and outgoing components at the
past and future event horizons.  At future null infinity, the mode is
purely outgoing.  By using an argument similar to that given above,
one can show that these modes vanish at past null infinity.
Also we can argue as before that
\be
u_{-\omega l-m}^{\rm up}(r^*) = u_{\omega lm}^{\rm up}(r^*)^*,
\label{eq:uupidentity}
\ee
\be
\mu_{-\omega l-m} = \mu_{\omega lm}^*,
\ee
and
\be
\nu_{-\omega l-m} = \nu_{\omega lm}^*,
\ee
as long as we choose the normalization constants to satisfy
\be
\beta_{-\omega l-m} = \beta_{\omega lm}^*.
\label{eq:betacondt}
\ee

Since the ``in'' modes vanish at the past event horizon, and the ``up'' modes
vanish at past null infinity, the two sets of modes are orthogonal to
one another and together form a complete basis of modes.
%They are
%related to the ``H'' and ``$\infty$'' modes of Hughes \cite{Hughes:1999bq} by
%\be
%R^H_{\omega lm}(r) \propto \frac{ u^{\rm in}_{\omega lm}}{\varpi}
%\ee
%and
%\be
%R^\infty_{\omega lm}(r) \propto \frac{ u^{\rm up}_{\omega lm}}{\varpi}.
%\ee

Next we note that $(u^{\rm in}_{\omega lm})^*$ is a solution of the
differential equation (\ref{eq:diffsimple}) since the potential
(\ref{eq:potential}) is real.  We denote this quantity by
\be
u^{\rm out}_{\omega lm} = (u^{\rm in}_{\omega lm})^*.
\label{eq:uoutdef}
\ee
Similarly we define
\be
u^{\rm down}_{\omega lm} = (u^{\rm up}_{\omega lm})^*.
\label{eq:udowndef}
\ee
These ``down'' and ``out'' modes together form a basis, as
the ``out'' modes vanish on the future event horizon and the ``down''
modes vanish at future null infinity.
The properties of all of these modes are summarized in Fig. \ref{fig:modes}.

\subsubsection{Relations between scattering and transmission
  coefficients}

The four sets of constants $\sigma_{\omega lm}$, $\tau_{\omega lm}$,
$\mu_{\omega lm}$ and $\nu_{\omega lm}$ are not all independent.
One can derive relations between these coefficients by using the fact
that the Wronskian
\be
W(u_1,u_2) = u_1 \frac{d u_2}{d r^*} - \frac{d u_1}{d r^*} u_2.
\label{eq:wronskiandef}
\ee
is conserved for any two solutions $u_1$
and $u_2$ of the homogeneous equation (\ref{eq:diffsimple}).
By evaluating $W(u^{\rm up}_{\omega lm},u^{\rm in}_{\omega lm})$
at $r^* = -\infty$ and at $r^* = \infty$ using
the asymptotic relations (\ref{eq:uindef}) and (\ref{eq:uupdef}) and
equating the results, we obtain
\be
\tau_{\omega lm} \mu_{\omega lm}=1.
\label{eq:widentity1}
\ee
A similar calculation with the modes $u^{\rm up}_{\omega lm}$ and
$u^{\rm out}_{\omega lm}$ yields
\be
\tau_{\omega lm}^{*} \nu_{\omega lm} = - \sigma_{\omega lm}^*.
\label{eq:widentity2}
\ee
Finally, using the modes $u^{\rm in}_{\omega lm}$ and
$u^{\rm out}_{\omega lm}$ yields
\be
| \sigma_{\omega lm}|^2 + \frac{\omega p_{m\omega}}{|\omega
p_{m\omega}|} | \tau_{\omega lm}|^2 =1.
\label{eq:widentity3}
\ee

\subsubsection{Relations between mode functions}

The ``down'' and ``out'' modes can be expressed as linear combinations
of the the ``in'' and ``up'' modes, since the latter form a basis of
modes.  Using the asymptotic forms (\ref{eq:uindef}) and (\ref{eq:uupdef}) of the modes
at $r^* \to \infty$ together with the definition (\ref{eq:udowndef})
allows us to identity the coefficients for $u^{\rm down}_{\omega lm}$, giving
\be
u^{\rm down}_{\omega lm}(r) = \frac{\beta_{\omega lm}^*}{\alpha_{\omega lm}}
u^{\rm in}_{\omega lm}(r) - \sigma_{\omega lm}
\frac{\beta_{\omega lm}^*}{\beta_{\omega lm}} u^{\rm up}_{\omega
  lm}(r).
\label{eq:udownformula}
\ee
A similar computation using the definition (\ref{eq:uoutdef}) and
using the asymptotic forms at $r^* \to -\infty$ gives
\be
u^{\rm out}_{\omega lm}(r) = \frac{\omega p_{m\omega}}{|\omega p_{m\omega}|}
\frac{\tau_{\omega lm}^*}{\mu_{\omega lm}} \frac{\alpha^*_{\omega
    lm}}{\beta_{\omega lm}} u^{\rm up}_{\omega lm}(r)
- \frac{\tau_{\omega lm}^* \nu_{\omega lm}}{\mu_{\omega lm}
  \tau_{\omega lm}} \frac{\alpha_{\omega lm}^*}{\alpha_{\omega lm}}
u^{\rm in}_{\omega lm}(r).
\ee
This can be simplified using the relations (\ref{eq:widentity1}) and
(\ref{eq:widentity2}) to give
\be
u^{\rm out}_{\omega lm}(r) = \frac{\omega p_{m\omega}}{|\omega p_{m\omega}|}
\frac{\alpha^*_{\omega lm}}{\beta_{\omega lm}} | \tau_{\omega lm}|^2
u^{\rm up}_{\omega lm}(r)
+ \frac{\alpha_{\omega lm}^*}{\alpha_{\omega lm}} \sigma_{\omega lm}^*
u^{\rm in}_{\omega lm}(r).
\label{eq:uoutformula}
\ee

\subsubsection{Complete mode functions}

Finally, following Gal'tsov \cite{Galtsov:1982}, we define the complete
mode functions
\be
\pi^{\rm in}_{\omega lm}(t,r,\theta,\phi) = \frac{2}{\varpi} e^{-i \omega t}
S_{\omega lm}(\theta,\phi) u^{\rm in}_{\omega lm}(r^*),
\label{eq:piindef}
\ee
\be
\pi^{\rm up}_{\omega lm}(t,r,\theta,\phi) = \frac{2}{\varpi} e^{-i \omega t}
S_{\omega lm}(\theta,\phi) u^{\rm up}_{\omega lm}(r^*),
\label{eq:piupdef}
\ee
\be
\pi^{\rm out}_{\omega lm}(t,r,\theta,\phi) = \frac{2}{\varpi} e^{-i \omega t}
S_{\omega lm}(\theta,\phi) u^{\rm out}_{\omega lm}(r^*),
\label{eq:pioutdef}
\ee
and
\be
\pi^{\rm down}_{\omega lm}(t,r,\theta,\phi) = \frac{2}{\varpi} e^{-i \omega t}
S_{\omega lm}(\theta,\phi) u^{\rm down}_{\omega lm}(r^*).
\label{eq:pidowndef}
\ee
Gal'tsov actually includes a factor of $1 + p P$ in these definitions,
where $P$ is the parity operator defined by
\be
(Pf)(t,r,\theta,\phi) = f(t,r,\pi-\theta,\pi+\phi)
\ee
for any function $f$ and $p = \pm 1$ is an additional mode index.
However when acting on
the scalar modes considered in this paper, we have from Eq.\
(\ref{eq:Zparity}) that
\be
1 + pP = 1 + p (-1)^l = 2 \delta_{p,(-1)^l}.
\ee
Thus the $p$ index is redundant in the scalar case; we have simply
dropped it and replaced the factor of $1+pP$ by $2$.

\section{Construction of the retarded Green's function}
\label{sec:retarded}

\subsection{Formulae for retarded Green's function}
\label{sec:retardedformula}

The retarded Green's function $G_{\rm ret}(x,x')$ is defined such that if $\Phi$
obeys the scalar wave equation with source ${\cal T}$
\be
\Box \Phi = {\cal T},
\ee
then the retarded solution is
\be
\Phi_{\rm ret}(x) = \int d^4 x' \sqrt{-g(x')} G_{\rm ret}(x,x') {\cal T}(x').
\label{eq:Gretdef}
\ee
The expression for the retarded Green's function in terms of the modes
defined in the last section is\footnote{This expression agrees with
Eq.\ (2.17) of Gal'tsov \protect{\cite{Galtsov:1982}},
if we assume that the normalization constants obey
$
\alpha_{\omega lm} \beta_{\omega lm} = - 1/(16 \pi).
$
However Eq.\ (2.25) of Gal'tsov
says $\alpha_{\omega lm} \beta_{\omega lm} = 1/4$; thus we disagree
with Gal'tsov by a factor of $-4 \pi$.  For this reason we do not
assume any values for the normalization constants $\alpha_{\omega lm}$
and $\beta_{\omega lm}$ in our computations, but leave them
unevaluated.}
\begin{eqnarray}
\fl
G_{\rm ret}(x,x') = \frac{1}{16 \pi i} \int_{-\infty}^\infty d\omega
\sum_{l=0}^\infty \sum_{m=-l}^l \frac{\omega}{|\omega|}
\frac{1}{\alpha_{\omega lm} \beta_{\omega lm}} \nonumber \\
\lo \times
\left[
\pi^{\rm up}_{\omega lm}(x) \pi^{\rm out}_{\omega lm}(x')^* \theta(r-r') +
\pi^{\rm in}_{\omega lm}(x) \pi^{\rm down}_{\omega lm}(x')^* \theta(r'-r)
\right].
\label{eq:Gretformula0}
\end{eqnarray}
Here $\theta(x)$ is the step function, defined to be $+1$ for $x \ge
0$ and $0$ otherwise.
Note that the expression (\ref{eq:Gretformula0}) is
independent of the values chosen for the normalization
constants $\alpha_{\omega lm}$ and $\beta_{\omega lm}$, since the
factor of $1/\alpha$ cancels a factor of $\alpha$
present in the definition (\ref{eq:uindef}) of the ``in'' modes, and
similarly for $\beta$ and the ``up'' modes.

The expression (\ref{eq:Gretformula0})
can be expanded into a more
explicit form by using the definitions (\ref{eq:piindef}) --
(\ref{eq:pidowndef}) of the complete mode functions $\pi_{\omega
  lm}(t,r,\theta,\phi)$ in terms of the radial mode functions
$u_{\omega lm}(r)$, together with the definitions (\ref{eq:uoutdef})
and (\ref{eq:udowndef}) of the ``out'' and ``down'' modes.
This gives
\begin{eqnarray}
\fl
G_{\rm ret}(t,r,\theta,\phi;t',r',\theta',\phi') = \frac{1}{4 \pi i}
\int_{-\infty}^\infty d\omega e^{-i \omega (t - t')}
\sum_{lm} S_{\omega lm}(\theta,\phi) S_{\omega
lm}(\theta',\phi')^*
\frac{\omega}{|\omega|}
\nn \\
\fl \ \ \ \ \ \   \times
\frac{1}{\varpi \varpi' \alpha_{\omega lm} \beta_{\omega lm}}
\left[ u^{\rm up}_{\omega lm}(r) u^{\rm in}_{\omega lm}(r') \theta(r-r') +
u^{\rm in}_{\omega lm}(r) u^{\rm up}_{\omega lm}(r') \theta(r'-r)
\right].
\label{eq:Gretformula1}
\end{eqnarray}
In this section we review the standard derivation of the formula
(\ref{eq:Gretformula1}).

\subsection{Derivation}

The key idea behind the derivation is the following.  Suppose that the
source ${\cal T}(x)$ is non-zero only in the finite range of values of $r$
\be
r_{\rm min} \le r \le r_{\rm max}.
\ee
Then, the retarded solution $\Phi_{\rm ret}(x)$ will be a solution of the
homogeneous equation in the regions $r < r_{\rm min}$ and $r > r_{\rm
max}$.  Now, the retarded solution is determined uniquely by the
condition that it vanish on the past event horizon and on past null
infinity.  This property will be guaranteed if:
\begin{enumerate}
\item When we expand $\Phi_{\rm ret}$ in the region $r < r_{\rm min}$ on the
basis of solutions $\pi^{\rm in}_{\omega lm}(x)$ and $\pi^{\rm
up}_{\omega lm}$ of the homogeneous equation, only the ``in'' modes
contribute.  Then, since the ``in'' modes vanish on the past event
horizon, $\Phi_{\rm ret}$ must also vanish on the past event horizon.

\item When we expand $\Phi_{\rm ret}$ in the region $r > r_{\rm max}$ on the
basis of solutions $\pi^{\rm in}_{\omega lm}(x)$ and $\pi^{\rm
up}_{\omega lm}$, only the ``up'' modes
contribute.  Then, since the ``up'' modes vanish on past null
infinity, $\Phi_{\rm ret}$ must also vanish on past null infinity.

\end{enumerate}

We start by defining the Fourier transformed quantities
\be
{\tilde {\cal T}}(\omega,r,\theta,\phi) = \int_{-\infty}^\infty dt e^{i
\omega t} {\cal T}(t,r,\theta,\phi)
\ee
and
\be
{\tilde \Phi}(\omega,r,\theta,\phi) = \int_{-\infty}^\infty dt e^{i
\omega t} \Phi(t,r,\theta,\phi).
\ee
[For convenience we drop the subscript ``ret'' on $\Phi$ in the
  remainder of this section.]
We make the following ansatz for the Green's function:
\be
\fl
G_{\rm ret}(t,r,\theta,\phi;t',r',\theta',\phi') =
\int_{-\infty}^\infty \frac{d \omega}{2 \pi} e^{-i \omega (t - t')}
{\tilde G}_{\rm ret}(r,\theta,\phi;r',\theta',\phi';\omega).
\label{eq:ansatz0}
\ee
Inserting these definitions into the defining relation
(\ref{eq:Gretdef}) and using the formula (\ref{eq:detg}) for $\sqrt{-g}$
gives
\be
\fl
{\tilde \Phi}(\omega,r,\theta,\phi) = \int_0^\infty dr' \int d^2
\Omega' \Sigma(r',\theta') \, G_{\rm
ret}(r,\theta,\phi;r',\theta',\phi';\omega) {\tilde
{\cal T}}(\omega,r',\theta',\phi').
\label{eq:step1}
\ee

Next, we decompose the quantities ${\tilde \Phi}$ and $\Sigma {\tilde
{\cal T}}$ on the basis of spheroidal harmonics:
\be
{\tilde \Phi}(\omega,r,\theta,\phi) = \sum_{lm} S_{\omega
lm}(\theta,\phi) R_{\omega lm}(r)
\label{eq:Philmdef}
\ee
and
\be
\Sigma(r,\theta) {\tilde {\cal T}}(\omega,r,\theta,\phi) = r^2 \sum_{lm} S_{\omega
lm}(\theta,\phi) {\tilde {\cal T}}_{\omega lm}(r)
\label{eq:Tlmdef}
\ee
We include the factor of $\Sigma$ on the left hand side of Eq.\
(\ref{eq:Tlmdef}) because of the appearance of a factor of $\Sigma$ in
the integrand of Eq.\ (\ref{eq:step1}).  The factor of $r^2$ on the
right hand side of Eq.\ (\ref{eq:Tlmdef}) is so that the coefficients
${\tilde {\cal T}}_{\omega lm}$ reduce to the conventional spherical harmonic
coefficients in the Schwarzschild limit $a =0$.
From the orthogonality relation (\ref{eq:orthonormal}) the inverse
transformations are
\be
R_{\omega lm}(r) = \int d^2 \Omega \, S_{\omega
lm}(\theta,\phi)^* \, {\tilde \Phi}(\omega,r,\theta,\phi)
\ee
and
\be
r^2 {\tilde {\cal T}}_{\omega lm}(r) = \int d^2 \Omega \, S_{\omega
lm}(\theta,\phi)^* \, \Sigma(r,\theta) {\tilde {\cal T}}(\omega,r,\theta,\phi).
\label{eq:inverse2}
\ee

Next, we insert the decompositions (\ref{eq:Philmdef}) and
(\ref{eq:Tlmdef}) of ${\tilde \Phi}$ and ${\tilde {\cal T}}$ into the
Fourier transform of the differential equation (\ref{eq:wave}) using
Eq.\ (\ref{eq:waveoperator}).  This gives
\be
\frac{d^2 u_{\omega lm}}{d r^{*\,2}} + V_{\omega lm}(r^*) u_{\omega lm}(r^*) =
s_{\omega lm},
\label{eq:ulmeqn}
\ee
where $u_{\omega lm}(r) = \varpi(r) R_{\omega lm}(r)$, the potential
$V_{\omega lm}$ is given by Eq.\ (\ref{eq:potential}), and
the source term is
\be
s_{\omega lm} = \frac{r^2 \Delta}{\varpi^3} {\tilde {\cal T}}_{\omega lm}.
\label{eq:slmdef}
\ee
We denote by $G_{\omega lm}(r^*,r^{*\,\prime})$ the relevant
Green's function for the differential equation
(\ref{eq:ulmeqn}):
\be
u_{\omega lm}(r^*) = \int_{-\infty}^\infty dr^{*\,\prime} G_{\omega
lm}(r^*,r^{*\,\prime}) s_{\omega lm}(r^{*\,\prime}).
\label{eq:Gomegalmdef}
\ee
We will derive an explicit formula for $G_{\omega
lm}(r^*,r^{*\,\prime})$ shortly.
First, we note that the Fourier-transformed retarded Green's function
${\tilde G}_{\rm ret}(r,\theta,\phi;r',\theta',\phi';\omega)$ can be
expressed in terms of $G_{\omega lm}$ via the formula
\be
\fl
{\tilde G}_{\rm ret}(r,\theta,\phi;r',\theta',\phi';\omega) =
\sum_{lm} S_{\omega lm}(\theta,\phi) S_{\omega lm}(\theta',\phi')^* \,
\frac{G_{\omega lm}(r^*,r^{*\,\prime})}{\varpi \varpi'}.
\label{eq:ansatz1}
\ee
To see this, insert the ansatz (\ref{eq:ansatz1}) into the
relation (\ref{eq:step1}) and simplify using the definition
(\ref{eq:inverse2}) of ${\tilde {\cal T}}_{\omega lm}$.  This gives
\be
\fl
{\tilde \Phi}(\omega,r,\theta,\phi) = \sum_{lm} \int_0^\infty dr'
S_{\omega lm}(\theta,\phi) \frac{r^{'\,2}}{\varpi \varpi'} G_{\omega
lm}(r^*,r^{*\,\prime}) {\tilde {\cal T}}_{\omega lm}(r').
\ee
Comparing this with the definition (\ref{eq:Philmdef}) of $R_{\omega
lm}$ and simplifying using the relations (\ref{eq:udef}),
(\ref{eq:rstardef0}) and (\ref{eq:slmdef}) finally yields the formula
(\ref{eq:Gomegalmdef}).   Hence the ansatz (\ref{eq:ansatz1}) is correct.

Finally we turn to the derivation of the formula for the
Green's function $G_{\omega lm}$ for the differential equation
(\ref{eq:ulmeqn}).  From the discussion at the start of this section,
the relevant boundary conditions to impose are that
\be
G_{\omega lm}(r^*,r^{*\,\prime}) \propto u^{\rm in}_{\omega lm}(r^*), \
\ \ r^* \to -\infty
\label{eq:b1}
\ee
and
\be
G_{\omega lm}(r^*,r^{*\,\prime}) \propto u^{\rm up}_{\omega lm}(r^*), \
\ \ r^* \to \infty.
\label{eq:b2}
\ee
Consider now the expression
\begin{eqnarray}
\fl
G_{\omega lm}(r^*,r^{*\,\prime}) = \frac{1}{W(u^{\rm in}_{\omega lm},
u^{\rm up}_{\omega lm})} \bigg[
u^{\rm up}_{\omega lm}(r) u^{\rm in}_{\omega lm}(r^\prime) \theta(r-r')
\nonumber \\
\lo +
u^{\rm in}_{\omega lm}(r) u^{\rm up}_{\omega lm}(r^\prime)
\theta(r-r') \bigg],
\label{eq:ansatz2}
\end{eqnarray}
where $W$ is the conserved Wronskian (\ref{eq:wronskiandef}).
This expression satisfies the boundary conditions (\ref{eq:b1}) and
(\ref{eq:b2}).  Also one can directly verify that it satisfies the
differential equation (\ref{eq:ulmeqn}) with the source replaced by
$\delta(r^* - r^{*\,\prime})$, using the fact that the ``in'' and ``up''
modes satisfy the homogeneous version of the differential equation.
This establishes the formula (\ref{eq:ansatz2}).

Next, we compute the Wronskian
$W(u^{\rm in}_{\omega lm}, u^{\rm up}_{\omega lm})$ using the
asymptotic expressions (\ref{eq:uindef}) and (\ref{eq:uupdef}) for the
mode functions for $r^* \to \infty$.  This gives
\be
W(u^{\rm in}_{\omega lm}, u^{\rm up}_{\omega lm}) = 2 i \alpha_{\omega
lm} \beta_{\omega lm} \frac{\omega}{|\omega|}.
\ee
Inserting this into Eq.\ (\ref{eq:ansatz2}) and then into Eqs.\
(\ref{eq:ansatz1}) and (\ref{eq:ansatz0}) finally yields the formula
(\ref{eq:Gretformula1}).

\section{Construction of the radiative Green's function}
\label{sec:radiative}

\subsection{Formulae for radiative Green's function}

Using the retarded Green's function $G_{\rm ret}(x,x')$ discussed in the last section
we can construct the retarded solution $\Phi_{\rm ret}(x)$ of the wave
equation (\ref{eq:wave}).  Similarly, we can construct the advanced solution
$\Phi_{\rm adv}(x)$ using the advanced Green's function $G_{\rm
  adv}(x,x')$; this is the unique solution which vanishes on the
future event horizon and at future null infinity.
One half the retarded solution minus one half the advanced solution
gives the radiative solution:
\be
\Phi_{\rm rad}(x) = \frac{1}{2} \left[ \Phi_{\rm ret}(x) - \Phi_{\rm
    adv}(x) \right].
\ee
Clearly the radiative solution is given in terms of a radiative Green's
function
\be
\Phi_{\rm rad}(x) = \int d^4 x' \sqrt{-g(x')} G_{\rm rad}(x,x') {\cal T}(x'),
\ee
where
\be
G_{\rm rad}(x,x') = \frac{1}{2} \left[ G_{\rm ret}(x,x') - G_{\rm
    adv}(x,x') \right].
\label{eq:Graddef}
\ee
The expression for the radiative Green's function in terms of the modes
defined in Sec.\ \ref{sec:modes} is
\begin{eqnarray}
\fl
G_{\rm rad}(x,x') = \frac{1}{32 \pi i} \int_{-\infty}^\infty d\omega
\sum_{l=0}^\infty \sum_{m=-l}^l \frac{\omega}{|\omega|}
\bigg[
\frac{1}{|\alpha_{\omega lm}|^2}
\pi^{\rm out}_{\omega lm}(x) \pi^{\rm out}_{\omega
    lm}(x')^* \nonumber \\
\lo
 +\frac{\omega p_{m\omega}}{|\omega p_{m\omega}|}
\frac{|\tau_{\omega lm}|^2}{|\beta_{\omega lm}|^2}
\pi^{\rm down}_{\omega lm}(x) \pi^{\rm down}_{\omega lm}(x')^*
\bigg].
\label{eq:Gradformula0}
\end{eqnarray}
Note that this expression
actually independent of the values chosen for the normalization
constants $\alpha_{\omega lm}$ and $\beta_{\omega lm}$, since the
factor of $1/|\alpha|^2$ cancels factors of $\alpha$
present in the definition (\ref{eq:uindef}) and (\ref{eq:uoutdef}) of
the ``out'' modes, and similarly for $\beta$ and the ``down'' modes.

The expression (\ref{eq:Gradformula0})
can be expanded into a more
explicit form by using the definitions (\ref{eq:piindef}) --
(\ref{eq:pidowndef}) of the complete mode functions $\pi_{\omega
  lm}(t,r,\theta,\phi)$ in terms of the radial mode functions
$u_{\omega lm}(r)$.
This gives
\begin{eqnarray}
\fl
G_{\rm rad}(t,r,\theta,\phi;t',r',\theta',\phi') = \frac{1}{8 \pi i}
\int_{-\infty}^\infty d\omega e^{-i \omega (t - t')}
\sum_{lm} S_{\omega lm}(\theta,\phi) S_{\omega
lm}(\theta',\phi')^*
\frac{\omega}{|\omega|}
\frac{1}{\varpi \varpi'} \nn \\
\fl \ \ \  \   \times
\left[
\frac{1}{|\alpha_{\omega lm}|^2}
u^{\rm out}_{\omega lm}(r) u^{\rm out}_{\omega
    lm}(r')^* +\frac{\omega p_{m\omega}}{|\omega p_{m\omega}|}
\frac{|\tau_{\omega lm}|^2}{|\beta_{\omega lm}|^2}
u^{\rm down}_{\omega lm}(r) u^{\rm down}_{\omega lm}(r')^*
\right].
\label{eq:Gradformula1}
\end{eqnarray}
In this section we review the derivation of the formula
(\ref{eq:Gradformula1}) given in Gal'tsov \cite{Galtsov:1982}.

\subsection{Derivation}

We start by deriving the identity
\be
G_{\rm adv}(x,x') = G_{\rm ret}(x',x).
\label{eq:identitya}
\ee
For any two functions $\Phi(x)$ and $\Psi(x)$ we have the identity
\be
\nabla^a ( \Phi \nabla_a \Psi - \Psi \nabla_a \Phi) = \Phi \Box \Psi - \Psi \Box \Phi.
\ee
Integrating this identity over a spacetime region $V$ and using Stokes theorem gives
\be
\int_{\partial V} (\Phi \nabla^a \Psi - \Psi \nabla^a \Phi) d^3 \Sigma_a = \int_V d^4 x \sqrt{-g} ( \Phi \Box \Psi - \Psi \Box \Phi).
\label{eq:identity0}
\ee
If we choose $V$ to be the region of spacetime outside the black hole,
then the boundary $\partial V$ of $V$ consists of the past and future
event horizons, and also past and future null infinity.

We now fix two points $x_1$ and $x_2$ in $V$, and we choose
\be
\Phi(x) = G_{\rm ret}(x,x_1)
\ee
and
\be
\Psi(x) = G_{\rm adv}(x,x_2).
\ee
Then $\Box \Phi = \delta^{(4)}(x,x_1)$ and $\Box \Psi =
\delta^{(4)}(x,x_2)$, where $\delta^{(4)}$ is the covariant delta
function (\ref{eq:deltafndef}).  The right hand side of Eq.\
(\ref{eq:identity0}) therefore evaluates to
\be
\Phi(x_2) - \Psi(x_1) = G_{\rm ret}(x_2,x_1) - G_{\rm adv}(x_1,x_2).
\label{eq:rhsid}
\ee
Consider now the left hand side of Eq.\ (\ref{eq:identity0}).  Since
$\Phi$ vanishes on the past event horizon and at past null infinity,
and $\Psi$ vanishes on the future event horizon and at future null
infinity, there is no portion of $\partial V$ where both $\Phi$ and
$\Psi$ are nonzero.  Therefore the left hand side vanishes, so the
expression (\ref{eq:rhsid}) must vanish, proving the identity
(\ref{eq:identitya}).

We now insert the identity (\ref{eq:identitya}) into the definition
(\ref{eq:Graddef}) of $G_{\rm rad}$, and use the explicit expansion
(\ref{eq:Gretformula1}) of $G_{\rm ret}$.
We also specialize to
\be
\alpha_{\omega lm} = \beta_{\omega lm} =1;
\label{eq:normalizationchoice}
\ee
this causes no loss in generality since the final result
(\ref{eq:Gradformula1}) is
independent of the values of $\alpha_{\omega lm}$ and $\beta_{\omega
lm}$.  This gives
\begin{eqnarray}
\fl
G_{\rm rad}(x,x') &=&
\frac{1}{8 \pi i}
\int_{-\infty}^\infty d\omega e^{-i \omega (t - t')}
\sum_{lm} S_{\omega lm}(\theta,\phi) S_{\omega
lm}(\theta',\phi')^*
\frac{\omega}{|\omega|}
\frac{1}{\varpi \varpi'} \nn \\
\mbox{} && \times
\left[ u^{\rm up}_{\omega lm}(r) u^{\rm in}_{\omega lm}(r') \theta(r-r') +
u^{\rm in}_{\omega lm}(r) u^{\rm up}_{\omega lm}(r') \theta(r'-r)
\right] \nn \\
\mbox{} && - \frac{1}{8 \pi i}
\int_{-\infty}^\infty d\omega e^{i \omega (t - t')}
\sum_{lm} S_{\omega lm}(\theta',\phi') S_{\omega
lm}(\theta,\phi)^*
\frac{\omega}{|\omega|}
\frac{1}{\varpi \varpi'} \nn \\
\mbox{} && \times
\left[ u^{\rm up}_{\omega lm}(r') u^{\rm in}_{\omega lm}(r) \theta(r'-r) +
u^{\rm in}_{\omega lm}(r') u^{\rm up}_{\omega lm}(r) \theta(r-r')
\right],
\label{eq:Gradformula3}
\end{eqnarray}
where $x =(t,r,\theta,\phi)$ and $x'=(t',r',\theta',\phi')$.
In the second sum we relabel $\omega \to -\omega$ and $m \to -m$, and
we use the formula (\ref{eq:Zparity1}).  This gives
\begin{eqnarray}
\fl
G_{\rm rad}(x,x') &=&
\frac{1}{8 \pi i}
\int_{-\infty}^\infty d\omega e^{-i \omega (t - t')}
\sum_{lm} S_{\omega lm}(\theta,\phi) S_{\omega
lm}(\theta',\phi')^*
\frac{\omega}{|\omega|}
\frac{1}{\varpi \varpi'} \nn \\
\mbox{} && \times
\bigg\{ \left[ u^{\rm up}_{\omega lm}(r) u^{\rm in}_{\omega lm}(r')
+ u^{\rm up}_{-\omega l-m}(r) u^{\rm in}_{-\omega l-m}(r') \right]
\theta(r-r') \nn \\
\mbox{} &&
+\left[ u^{\rm in}_{\omega lm}(r) u^{\rm up}_{\omega lm}(r')
+ u^{\rm in}_{-\omega l-m}(r) u^{\rm up}_{-\omega l-m}(r')
\right] \theta(r'-r)
\bigg\}.
\label{eq:Gradformula4}
\end{eqnarray}
Next, since our choice (\ref{eq:normalizationchoice}) of the
normalization constants satisfies the conditions (\ref{eq:alphacondt}) and
(\ref{eq:betacondt}), we can use the identities (\ref{eq:uinidentity})
and (\ref{eq:uupidentity}).
This yields
\begin{eqnarray}
\fl
G_{\rm rad}(x,x') &=&
\frac{1}{8 \pi i}
\int_{-\infty}^\infty d\omega e^{-i \omega (t - t')}
\sum_{lm} S_{\omega lm}(\theta,\phi) S_{\omega
lm}(\theta',\phi')^*
\frac{\omega}{|\omega|}
\frac{1}{\varpi \varpi'} \nn \\
\mbox{} && \times
\bigg\{ \left[ u^{\rm up}_{\omega lm}(r) u^{\rm in}_{\omega lm}(r')
+ u^{\rm up}_{\omega lm}(r)^* u^{\rm in}_{\omega lm}(r')^* \right]
\theta(r-r') \nn \\
\mbox{} &&
+\left[ u^{\rm in}_{\omega lm}(r) u^{\rm up}_{\omega lm}(r')
+ u^{\rm in}_{\omega lm}(r)^* u^{\rm up}_{\omega lm}(r')^*
\right] \theta(r'-r)
\bigg\}.
\label{eq:Gradformula5}
\end{eqnarray}

Consider now the coefficient of $\theta(r-r')$ inside the curly
brackets in Eq.\ (\ref{eq:Gradformula5}).  We denote this quantity by
$H(r,r')$:
\be
H(r,r') = u^{\rm up}(r) u^{\rm in}(r')
+ u^{\rm up}(r)^* u^{\rm in}(r')^* .
\label{eq:Hdef}
\ee
Here and in the next few paragraphs we omit for simplicity the
subscripts $\omega lm$.  Next we use the formula
(\ref{eq:uoutformula}) for the ``out'' modes in terms of the ``in''
and ``up'' modes, together with our choice
(\ref{eq:normalizationchoice}) of normalization constants, to obtain
\be
u^{\rm up}(r) = \frac{\kappa}{|\tau|^2} \left[ u^{\rm out}(r) - \sigma^*
  u^{\rm in}(r) \right].
\label{eq:uupformula1}
\ee
Here
\be
\kappa \equiv \omega p_{m\omega} / |\omega p_{m\omega}|
\ee
is a variable which is $+1$ for normal modes and $-1$ for superradiant
modes.  Inserting this into Eq.\ (\ref{eq:Hdef}) and expanding yields
\begin{eqnarray}
\fl
H(r,r') = \frac{\kappa}{|\tau|^2} \bigg[ u^{\rm in}(r)^* u^{\rm in}(r') +
  u^{\rm in}(r) u^{\rm in}(r')^*
- \sigma^* u^{\rm in}(r) u^{\rm in}(r') \nonumber \\
\lo - \sigma u^{\rm in}(r)^*
u^{\rm in}(r')^* \bigg].
\label{eq:step3}
\end{eqnarray}
Next, by using Eq.\ (\ref{eq:uupformula1}) to evaluate the quantity
$u^{\rm up}(r)^* u^{\rm up}(r')$ we obtain the identity
\begin{eqnarray}
\fl
|\tau|^4 u^{\rm up}(r)^* u^{\rm up}(r') - u^{\rm in}(r) u^{\rm
  in}(r')^* - | \sigma |^2 u^{\rm in}(r)^* u^{\rm in}(r') =
- \sigma^* u^{\rm in}(r) u^{\rm in}(r') \nonumber \\
\lo - \sigma u^{\rm in}(r)^* u^{\rm in}(r')^*.
\end{eqnarray}
We now use this identity to eliminate the last two terms inside the
square brackets in Eq.\ (\ref{eq:step3}), and simplify using the
identity (\ref{eq:widentity3}).  This gives
\be
H(r,r') = u^{\rm in}(r)^* u^{\rm in}(r') + \kappa |\tau|^2 u^{\rm up}(r)^*
u^{\rm up}(r').
\label{eq:Hformula}
\ee
Now the right hand side of Eq.\ (\ref{eq:Hformula}) is explicitly
invariant under the combined transformations of interchanging $r$ and
$r'$ and taking the complex conjugate.  However, from the definition
(\ref{eq:Hdef}) of $H(r,r')$, the left hand side is invariant under
complex conjugation.  It follows that both sides of Eq.\
(\ref{eq:Hformula}) are real and also symmetric under interchange of
$r$ and $r'$:
\be
H(r,r') = H(r',r).
\label{eq:Hsym}
\ee

Next, the quantity inside the curly brackets in the expression
(\ref{eq:Gradformula5}) for $G_{\rm rad}$ is
\be
H(r,r') \theta(r-r') + H(r',r) \theta(r'-r).
\ee
Using the symmetry property (\ref{eq:Hsym}) together with
$\theta(r-r') + \theta(r'-r) =1$, this can be written simply as
$H(r,r')$.  Therefore we can replace the expression in curly brackets
in (\ref{eq:Gradformula5}) with the expression (\ref{eq:Hformula}) for
$H(r,r')$.  This gives
\begin{eqnarray}
\fl
G_{\rm rad}(x,x') &=& \frac{1}{8 \pi i}
\int_{-\infty}^\infty d\omega e^{-i \omega (t - t')}
\sum_{lm} S_{\omega lm}(\theta,\phi) S_{\omega
lm}(\theta',\phi')^*
\frac{\omega}{|\omega|}
\frac{1}{\varpi \varpi'} \nn \\
\mbox{} && \times
\left[
u^{\rm in}_{\omega lm}(r)^* u^{\rm in}_{\omega
    lm}(r') +\frac{\omega p_{m\omega}}{|\omega p_{m\omega}|}
|\tau_{\omega lm}|^2
u^{\rm up}_{\omega lm}(r)^* u^{\rm up}_{\omega lm}(r')
\right].
\label{eq:Gradformula6}
\end{eqnarray}
Finally, rewriting this in terms of the ``down'' and ``out'' modes
using the definitions (\ref{eq:uoutdef}) and (\ref{eq:udowndef}), and
reinserting the appropriate factors of the normalization constants
yields the formula (\ref{eq:Gradformula1}).

\section{Harmonic decomposition of the scalar source}
\label{sec:harmonic}

\subsection{The retarded and radiative fields}

Using the expression (\ref{eq:Gretformula0}) for the retarded Green's
function together with the integral expression (\ref{eq:Gretdef}),
we can compute the retarded field $\Phi_{\rm ret}(x)$ generated by the source
${\cal T}(x)$.  For the case we are interested in, ${\cal T}(x)$ will be nonzero
only in a finite range of values of $r$ of the form
\be
r_{\rm min} \le r \le r_{\rm max}.
\ee
For $r > r_{\rm max}$ only the first term in the square brackets in
Eq.\ (\ref{eq:Gretformula0}) will contribute, and the function
$\theta(r-r')$ will always be $1$.  This gives
\be
\fl
\Phi_{\rm ret}(x) =
\frac{1}{16 \pi i} \int_{-\infty}^\infty d\omega
\sum_{l=0}^\infty \sum_{m=-l}^l \frac{\omega}{|\omega|}
\frac{1}{\alpha_{\omega lm} \beta_{\omega lm}} Z^{\rm out}_{\omega lm} \pi^{\rm
  up}_{\omega lm}(x), \ \ \ \ r \ge r_{\rm max},
\ee
where the coefficients $Z^{\rm out}_{\omega lm}$ are given by the inner
product of the ``out'' modes with the source:
\be
Z^{\rm out}_{\omega lm} \equiv \left(
\pi^{\rm out}_{\omega lm}, {\cal T} \right).
\label{eq:Zoutdef}
\ee
Here following Gal'tsov \cite{Galtsov:1982} we have defined the inner
product of any functions $\Phi(x)$ and
$\Psi(x)$ on spacetime by
\be
\left( \Phi, \Psi \right) \equiv \int d^4 x \sqrt{-g(x)} \Phi(x)^*
\Psi(x).
\label{eq:innerproductdef}
\ee
Similarly for $r < r_{\rm min}$ we obtain
\be
\fl
\Phi_{\rm ret}(x) =
\frac{1}{16 \pi i} \int_{-\infty}^\infty d\omega
\sum_{l=0}^\infty \sum_{m=-l}^l \frac{\omega}{|\omega|}
\frac{1}{\alpha_{\omega lm} \beta_{\omega lm}} Z^{\rm down}_{\omega lm} \pi^{\rm
  in}_{\omega lm}(x), \ \ \ \ r \le r_{\rm min},
\ee
where
\be
Z^{\rm down}_{\omega lm} \equiv \left(
\pi^{\rm down}_{\omega lm}, {\cal T} \right).
\label{eq:Zdowndef}
\ee
Finally, the expression (\ref{eq:Gradformula0}) for the radiative
Green's function together with the definitions (\ref{eq:Zoutdef}) and
(\ref{eq:Zdowndef}) give the following expression for the radiative field
\begin{eqnarray}
\fl
\Phi_{\rm rad}(x) = \frac{1}{32 \pi i} \int_{-\infty}^\infty d\omega
\sum_{l=0}^\infty \sum_{m=-l}^l \frac{\omega}{|\omega|}
\bigg[
\frac{1}{|\alpha_{\omega lm}|^2}
Z^{\rm out}_{\omega lm} \pi^{\rm out}_{\omega lm}(x)
\nonumber \\
\lo
+\frac{\omega p_{m\omega}}{|\omega p_{m\omega}|}
\frac{|\tau_{\omega lm}|^2}{|\beta_{\omega lm}|^2}
Z^{\rm down}_{\omega lm} \pi^{\rm down}_{\omega lm}(x)
\bigg].
\label{eq:Phiradformula0}
\end{eqnarray}
All of these expressions depend on the amplitudes $Z^{\rm out}_{\omega
  lm}$ and $Z^{\rm down}_{\omega lm}$.

\subsection{Harmonic decomposition of amplitudes}

For the case considered here where the source ${\cal T}(x)$ is a point
particle on a bound geodesic orbit, the amplitudes $Z^{\rm
out}_{\omega lm}$ and $Z^{\rm down}_{\omega lm}$ can be expressed
as discrete sums over delta functions \cite{Drasco:2004tv,drasconotes}:
\be
Z_{\omega lm}^{\rm out,down} = \sum_{k=-\infty}^\infty
\sum_{n=-\infty}^\infty \, Z^{\rm out,down}_{lmkn} \, \delta(\omega - \omega_{mkn}).
\label{eq:harmonicdecomposition}
\ee
Here
\be
\omega_{mkn} = m \Omega_\phi + k \Omega_\theta + n \Omega_r,
\label{eq:omegamkndef}
\ee
\be
\Omega_\phi = \frac{\Upsilon_\phi}{\Gamma},\ \ \ \
\Omega_\theta = \frac{\Upsilon_\theta}{\Gamma},\ \ \ \
\Omega_r = \frac{\Upsilon_r}{\Gamma},\ \ \ \
\ee
\be
\Upsilon_\theta = \frac{2 \pi}{\Lambda_\theta},\ \ \ \
\Upsilon_r = \frac{2 \pi}{\Lambda_r},\ \ \ \
\label{eq:Upsilondefs}
\ee
and $\Gamma$ and $\Upsilon_\phi$ are defined by Eqs.\ (\ref{eq:Gammadef}) and
(\ref{eq:Upsilonphidef}).  The formula for the coefficients $Z^{\rm
  out}_{lmkn}$ and $Z^{\rm down}_{lmkn}$ is
\begin{eqnarray}
\fl
Z^{\rm out,down}_{lmkn} = - \frac{4 \pi q}{\Gamma \Lambda_r
  \Lambda_\theta}
e^{-i m \phi_0} e^{i \omega_{mkn} t_0}
\int_0^{\Lambda_r} d\lambda_r \, \int_0^{\Lambda_\theta}
d\lambda_\theta \ e^{i(k \Upsilon_\theta \lambda_\theta + n \Upsilon_r
  \lambda_r)} \nonumber \\
\lo
\times \frac{\Sigma[r(\lambda_r),\theta(\lambda_\theta)]}{\varpi[r(\lambda_r)]}
\Theta_{\omega_{mkn}lm}[\theta(\lambda_\theta)]^*
e^{- i m \Delta \phi_r(\lambda_r)} e^{-i m \Delta \phi_\theta(\lambda_\theta)}
\nn \\  \lo \times
e^{i \omega_{mkn} \Delta t_r(\lambda_r)} e^{i
  \omega_{mkn} \Delta t_\theta(\lambda_\theta)} u^{\rm
  out,down}_{\omega_{mkn}lm}[r(\lambda_r)]^*.
\label{eq:Zlmknformula}
\end{eqnarray}
Here the functions $\Delta t_r$, $\Delta t_\theta$, $\Delta \phi_r$
and $\Delta \phi_\theta$ are defined in Eqs.\
(\ref{eq:deltatrdef}),
(\ref{eq:deltatthetadef}),
(\ref{eq:deltaphirdef}), and
(\ref{eq:deltaphithetadef}).  Also the function $\Theta_{\omega
  lm}(\theta)$ is defined by the decomposition
of the spheroidal harmonic $S_{\omega lm}$:
\be
S_{\omega lm}(\theta,\phi) = \Theta_{\omega lm}(\theta) e^{i m \phi},
\label{eq:Sdecompose}
\ee
cf.\ Eq.\ (\ref{eq:Somegalmdef}) above.
For evaluating the double integral (\ref{eq:Zlmknformula}) it is
convenient to use the identity (\ref{eq:averageidentity}) that
expresses the integral in terms of the variables $\chi$ and $\psi$.

In this section we review the derivation of the harmonic decomposition
(\ref{eq:harmonicdecomposition}) given by Drasco and Hughes
\cite{Drasco:2004tv,drasconotes}, adapted to the scalar case, and
generalized from fiducial geodesics to arbitrary bound geodesics.
We consider only the ``out'' amplitude $Z^{\rm
  out}_{\omega lm}$; the ``down'' case is exactly analogous.

\subsection{Derivation}

We start by inserting the definition (\ref{eq:sourcedef}) of the source
${\cal T}(x)$ into the definition of $Z_{\omega lm}^{\rm out}$ given by Eqs.\
(\ref{eq:Zoutdef}) and (\ref{eq:innerproductdef}).
This gives an expression consisting of an integral along the geodesic
of the mode function:
\be
Z_{\omega lm}^{\rm out} = - q \int_{-\infty}^\infty d\tau \, \pi^{\rm
  out}_{\omega lm}[ t(\tau),r(\tau),\theta(\tau),\phi(\tau)]^*.
\label{eq:Zlmout1}
\ee
Inserting the expression (\ref{eq:pioutdef}) for the mode
function $\pi^{\rm out}_{\omega lm}(t,r,\theta,\phi)$ in terms of
the radial mode function $u^{\rm out}_{\omega lm}(r)$, and using the
decomposition (\ref{eq:Sdecompose}) of the spheroidal harmonic $S_{\omega
lm}(\theta,\phi)$ now gives
\be
\fl
Z_{\omega lm}^{\rm out} = - 2 q \int_{-\infty}^\infty d\tau \,
\Theta_{\omega lm}[\theta(\tau)]^* e^{-i m \phi(\tau)} e^{i \omega
  t(\tau)} \frac{1}{\varpi[r(\tau)]} u_{\omega lm}^{\rm out}[r(\tau)]^*.
\ee
We next change the variable of integration from proper time $\tau$ to
Mino time $\lambda$ using Eq.\ (\ref{eq:Minotime}), and we
use the expressions (\ref{eq:tmotion00}) and (\ref{eq:phimotion00}) for
the functions $t(\lambda)$ and $\phi(\lambda)$.  This gives
\begin{eqnarray}
\fl
Z_{\omega lm}^{\rm out} = - 2 q e^{-i m \phi_0} e^{ i \omega t_0}
\int_{-\infty}^\infty d\lambda \,
e^{i \lambda (\Gamma \omega - m \Upsilon_\phi)}
\Sigma[r(\lambda),\theta(\lambda)]
\Theta_{\omega lm}[\theta(\lambda)]^*
\nn \\
\lo
\times
e^{-i m \Delta  \phi_r(\lambda)}
e^{-i m \Delta  \phi_\theta(\lambda)}
e^{i \omega \Delta t_r(\lambda)}
e^{i \omega \Delta t_\theta (\lambda)}
\frac{1}{\varpi[r(\lambda)]} u_{\omega lm}^{\rm out}[r(\lambda)]^*.
\label{eq:Zoutformula1}
\end{eqnarray}

We now define the function of two variables
\begin{eqnarray}
\fl
J_{\omega lm}(\lambda_r,\lambda_\theta)
&=& - 2 q e^{-i m \phi_0} e^{i \omega t_0} \
\Sigma[r(\lambda_r),\theta(\lambda_\theta)]
\Theta_{\omega lm}[\theta(\lambda_\theta)]^*
e^{-i m \Delta  \phi_r(\lambda_r)}
\nn \\ \mbox{} &\times&
e^{-i m \Delta  \phi_\theta(\lambda_\theta)}
e^{i \omega \Delta t_r(\lambda_r)}
e^{i \omega \Delta t_\theta (\lambda_\theta)}
\frac{1}{\varpi[r(\lambda_r)]} u_{\omega lm}^{\rm out}[r(\lambda_r)]^*.
\label{eq:Jomegalmdef}
\end{eqnarray}
This function has two key properties.  First, when evaluated at
$\lambda_r = \lambda_\theta = \lambda$, it coincides with the
integrand of Eq.\ (\ref{eq:Zoutformula1}), up to the first exponential factor:
\be
Z_{\omega lm}^{\rm out} = \int_{-\infty}^\infty d\lambda \,
J_{\omega lm}(\lambda,\lambda)
e^{i \lambda (\Gamma \omega - m \Upsilon_\phi)}.
\label{eq:Zoutformula2}
\ee
Second, the function $J_{\omega lm}$ is biperiodic:
\begin{eqnarray}
J_{\omega lm}(\lambda_r + \Lambda_r , \lambda_\theta) &=& J_{\omega
  lm}(\lambda_r , \lambda_\theta), \nn \\
\mbox{}
J_{\omega lm}(\lambda_r , \lambda_\theta + \Lambda_\theta) &=& J_{\omega
  lm}(\lambda_r , \lambda_\theta).
\end{eqnarray}
This follows from the fact that the functions $r(\lambda)$, $\Delta
t_r(\lambda)$ and $\Delta \phi_r(\lambda)$ are all periodic with
period $\Lambda_r$, and the functions $\theta(\lambda)$, $\Delta
t_\theta(\lambda)$ and $\Delta \phi_\theta(\lambda)$ are all periodic with
period $\Lambda_\theta$, cf.\ Sec.\ \ref{sec:geosesics} above.
Standard properties of biperiodic functions now imply that $J_{\omega
  lm}$ can be written in terms of a double Fourier series:
\be
J_{\omega lm}(\lambda_r,\lambda_\theta) =
\sum_{k=-\infty}^\infty
\sum_{n=-\infty}^\infty
J_{\omega lmkn} e^{-i (k \Upsilon_\theta \lambda_\theta + n \Upsilon_r
  \lambda_r)},
\label{eq:doubleFourierseries}
\ee
where $\Upsilon_\theta$ and $\Upsilon_r$ are given by Eq.\
(\ref{eq:Upsilondefs}), and where the coefficients $J_{\omega lmkn}$
are given by
\be
J_{\omega lmkn} = \frac{1}{\Lambda_r \Lambda_\theta}
\int_0^{\Lambda_r} d\lambda_r
\int_0^{\Lambda_\theta} d\lambda_\theta
\,
e^{i (k \Upsilon_\theta \lambda_\theta + n \Upsilon_r \lambda_r)}
J_{\omega lm}(\lambda_r,\lambda_\theta).
\label{eq:Jomegalmkndef}
\ee

We now insert the Fourier series (\ref{eq:doubleFourierseries}) for
$J_{\omega lm}$, evaluated at $\lambda_r = \lambda_\theta = \lambda$,
into the expression (\ref{eq:Zoutformula2}) for $Z^{\rm out}_{\omega lm}$.
This gives
\begin{eqnarray}
Z^{\rm out}_{\omega lm} &=& \sum_{kn} \int_{-\infty}^\infty d\lambda
e^{i \lambda ( \Gamma \omega - m \Upsilon_\phi - k \Upsilon_\theta - n
  \Upsilon_r) } J_{\omega lmkn} \nn \\
\mbox{} &=& \sum_{kn} \frac{2 \pi}{\Gamma} \delta(\omega -
\omega_{mkn}) J_{\omega lmkn},
\end{eqnarray}
where we have used the definition (\ref{eq:omegamkndef}) of
$\omega_{mkn}$.  Now comparing with Eq.\
(\ref{eq:harmonicdecomposition}) we can read off that
\be
Z^{\rm out}_{lmkn} = \frac{2 \pi}{\Gamma} J_{\omega_{mkn} lmkn}.
\label{eq:Zlmknans}
\ee
Combining this with the definitions (\ref{eq:Jomegalmdef}) and
(\ref{eq:Jomegalmkndef}) evaluated at $\omega = \omega_{mkn}$ finally
gives the formula (\ref{eq:Zlmknformula}).

Note that it follows from the harmonic decomposition
(\ref{eq:harmonicdecomposition}) that for geodesic sources, the
continuous frequency $\omega$ and the discrete indices $l$, $m$ are
replaced with the four discrete indices $k$,$n$,$l$, and $m$.  In this
context the operation
\be
\omega \to -\omega, \ \ \ \ \ m \to - m, \ \ \ \ l \to l
\ee
associated with the symmetries
(\ref{eq:lambdaidentity}), (\ref{eq:Zparity1}), (\ref{eq:Videntity}),
(\ref{eq:uinidentity}) -- (\ref{eq:tauidentity}) and
(\ref{eq:uupidentity}) -- (\ref{eq:betacondt}) is replaced by the
operation
\be
k \to -k, \ \ \ \ \ n \to -n, \ \ \ \ \ m \to -m, \ \ \ \ \ l \to l.
\ee

\subsection{Dependence of amplitudes on parameters of geodesic}
\label{sec:dep}

The amplitude $Z^{\rm out}_{lmkn}$ is a function of the parameters
characterizing the geodesic, namely $E$, $L_z$, $Q$, $t_0$, $\phi_0$,
$\lambda_{r0}$, and $\lambda_{\theta0}$, cf.\ Sec.\
\ref{sec:geodesicparameters} above.  We write this dependence as
\be
Z^{\rm out}_{lmkn} = Z^{\rm
  out}_{lmkn}(E,L_z,Q,t_0,\phi_0,\lambda_{r0},\lambda_{\theta0}).
\ee
We now specialize to the fiducial geodesic associated with
the constants $E$, $L_z$ and $Q$, i.e. the geodesic for which
$t_0=\phi_0 = \lambda_{r0} = \lambda_{\theta0} =0$.
For this case we can simplify the formula
(\ref{eq:Zlmknformula}) by setting $t_0$ and $\phi_0$ to zero, by
replacing the motions $r(\lambda)$ and $\theta(\lambda)$ with the
fiducial motions ${\hat r}(\lambda)$ and ${\hat \theta}(\lambda)$
defined by Eqs.\ (\ref{eq:hatrdef}) and (\ref{eq:hatthetadef}),
and by replacing the functions $\Delta t_r$, $\Delta t_\theta$,
$\Delta \phi_r$ and $\Delta \phi_\theta$ with the functions ${\hat
  t}_r$, ${\hat t}_\theta$, ${\hat \phi}_r$, and ${\hat \phi}_\theta$.
defined by Eqs.\ (\ref{eq:hattrdef}), (\ref{eq:hattthetadef}),
(\ref{eq:hatphirdef}), and (\ref{eq:hatphithetadef}).
This yields
\begin{eqnarray}
\fl
Z^{\rm out}_{lmkn}(E,L_z,Q,0,0,0,0) = - \frac{4 \pi q}{\Gamma \Lambda_r
  \Lambda_\theta}
\int_0^{\Lambda_r} d\lambda_r \, \int_0^{\Lambda_\theta}
d\lambda_\theta \ e^{i(k \Upsilon_\theta \lambda_\theta + n \Upsilon_r
  \lambda_r)}
\nn \\ \lo \times
 \frac{\Sigma[{\hat r}(\lambda_r),{\hat
  \theta}(\lambda_\theta)]}{\varpi[{\hat r}(\lambda_r)]}
\Theta_{\omega_{mkn}lm}[{\hat \theta}(\lambda_\theta)]^*
e^{- i m {\hat \phi}_r(\lambda_r)}
\nn \\ \lo \times
e^{-i m {\hat \phi}_\theta(\lambda_\theta)}
e^{i \omega_{mkn} {\hat t}_r(\lambda_r)} e^{i
  \omega_{mkn} {\hat t}_\theta(\lambda_\theta)} u^{\rm
  out}_{\omega_{mkn}lm}[{\hat r}(\lambda_r)]^*.
\label{eq:Zlmknformulafiducial}
\end{eqnarray}
For more general geodesics, the amplitude $Z_{lmkn}^{\rm out}$ depends
on the parameters $t_0$,
$\phi_0$, $\lambda_{r0}$ and $\lambda_{\theta0}$ only through an
overall phase:
\be
\fl
Z^{\rm out}_{lmkn}(E,L_z,Q,t_0,\phi_0,\lambda_{r0},\lambda_{\theta0})
= e^{i \chi_{lmkn}(t_0,\phi_0,\lambda_{r0},\lambda_{\theta0})}
Z^{\rm out}_{lmkn}(E,L_z,Q,0,0,0,0),
\label{eq:changeparameter}
\ee
where
\begin{eqnarray}
\fl
\chi_{lmkn}(t_0,\phi_0,\lambda_{r0},\lambda_{\theta0}) &=&
k \Upsilon_\theta \lambda_{\theta0}
+ n \Upsilon_r \lambda_{r0}
+ m \left[ {\hat \phi}_r(-\lambda_{r0}) + {\hat
    \phi}_\theta(-\lambda_{\theta0}) - \phi_0 \right]
\nn \\ \mbox{} &&
-\omega_{mkn} \left[ {\hat t}_r(-\lambda_{r0}) + {\hat
    t}_\theta(-\lambda_{\theta0}) - t_0 \right].
\label{eq:chilmnkdef}
\end{eqnarray}
This formula can be derived by substituting the expressions (\ref{eq:deltatrformula}),
(\ref{eq:deltatthetaformula}), (\ref{eq:deltaphirformula}) and
(\ref{eq:deltaphithetaformula}) for the functions $\Delta t_r$,
$\Delta t_\theta$, $\Delta \phi_r$ and $\Delta \phi_\theta$ into Eq.\
(\ref{eq:Zlmknformula}), making the changes of variables in the integral
\be
\lambda_r \to {\tilde \lambda}_r = \lambda_r - \lambda_{r0},\ \ \ \ \
\lambda_\theta \to {\tilde \lambda}_\theta = \lambda_\theta -
\lambda_{\theta0},
\ee
and comparing with Eq.\ (\ref{eq:Zlmknformulafiducial}).  Finally we
note that the phase (\ref{eq:chilmnkdef}) and amplitude
(\ref{eq:changeparameter}) are invariant under the transformations
(\ref{eq:t1}), (\ref{eq:t2}), (\ref{eq:t3}), and (\ref{eq:t4}) that
correspond to the re-parameterization $\lambda \to \lambda + \Delta
\lambda$.  This invariance serves as a consistency check of the
formulae, since we expect the invariance on physical grounds.

\section{Expressions for the time derivatives of energy, angular
  momentum, and rest mass}
\label{sec:Edot}

\subsection{Time averages}

Let ${\cal E}$ be one of the three conserved quantities of geodesic
motion, $E$, $L_z$ or $Q$.  For the purpose of evolving the orbit we
would like to compute the quantity
\be
\left< \frac{d{\cal E} }{d t} \right>_t,
\label{eq:desired}
\ee
that is, the average with respect to the Boyer-Lindquist time
coordinate $t$ of the derivative of ${\cal E}$ with respect to $t$.
However, the quantity that is most naturally computed is the
derivative with respect to proper time $\tau$, and the type of average
that is most easily computed is the average with respect to Mino time
$\lambda$.  In this section we therefore rewrite the quantity
(\ref{eq:desired}) in terms of a Mino-time average of $d{\cal
E}/d\tau$.

In the adiabatic limit, we can choose a time interval $\Delta t$ which
is long compared to the orbital timescales but short compared to the
radiation reaction time\footnote{A natural choice for $\Delta t$ is the geometric
mean of the orbital time and the radiation reaction time; this is
the time it takes for the phase difference between the geodesic orbit
and the true orbit to become of order unity.}.
Then, to a good approximation we have
\be
\left< \frac{d {\cal E}} {d t} \right>_t = \frac{\Delta {\cal E}}{\Delta t},
\ee
where $\Delta {\cal E}$ is the change in ${\cal E}$ over this
interval.  Now let $\Delta \lambda$ be the change in Mino time over
the interval.  From Eq.\ (\ref{eq:tmotion1}) we have
\be
\Delta t = \Gamma \Delta \lambda + {\rm oscillatory~terms}.
\ee
Now the oscillatory terms will be bounded as $\Delta t$ is taken
larger and larger, and therefore in the adiabatic limit they will
give a negligible fractional correction to $\Delta t$.  Hence we get
\begin{eqnarray}
\left< \frac{d {\cal E}} {d t} \right>_t &=& \frac{1}{\Gamma}
\frac{\Delta {\cal E}}{\Delta \lambda} \nn \\
\mbox{} &=&
\frac{1}{\Gamma} \left< \frac{d {\cal E}} {d \lambda} \right>_\lambda,
\label{eq:dcalEdt00}
\end{eqnarray}
where the $\lambda$ subscript on the angular brackets means an average
with respect to $\lambda$.
Note that using the definition (\ref{eq:Gammadef}) of $\Gamma$ we
can rewrite this formula as
\begin{equation}
\left< \frac{d {\cal E}} {d t} \right>_t =
\frac{ \left< d {\cal E} / d \lambda \right>_\lambda }
{\left< d t / d \lambda \right>_\lambda }.
\end{equation}
Finally we can use Eq.\ (\ref{eq:Minotime}) to rewrite the Mino-time
derivative in Eq.\ (\ref{eq:dcalEdt00}) in
terms of a proper time derivative.  This gives the final
formula which we will use:
\begin{eqnarray}
\left< \frac{d {\cal E}} {d t} \right>_t &=&
%\frac{1}{\Gamma}
%\frac{\Delta {\cal E}}{\Delta \lambda} \nn \\
%\mbox{} &=&
\frac{1}{\Gamma} \left< \Sigma \, \frac{d {\cal E}} {d \tau} \right>_\lambda.
\label{eq:used}
\end{eqnarray}

\subsection{Formulae for time derivatives of energy and angular momentum}

We now specialize to the cases of energy and angular momentum, ${\cal
E} = E$ or ${\cal E} = L_z$.  For these cases the result for the
time-averaged time derivative is
\begin{eqnarray}
\fl
\left< \frac{d {\cal E}}{dt} \right>_t = - \frac{1}{64 \pi^2 \mu}
\sum_\Lambda
\left\{ \begin{array}{l} \omega_{mkn} \\
                         m   \\
\end{array} \right\}
\frac{\omega_{mkn}}{|\omega_{mkn}|}
\left[
\frac{1}{|\alpha_\Lambda|^2}
|Z^{\rm out}_\Lambda |^2
+\frac{\omega_{mkn} p_{mkn}}{|\omega_{mkn} p_{mkn}|}
\frac{|\tau_\Lambda|^2}{|\beta_\Lambda|^2}
|Z^{\rm down}_\Lambda |^2 \right]. \nn \\
\label{eq:dcalEdt2}
\end{eqnarray}
Here
\be
p_{mkn} \equiv p_{m\omega_{mkn}} = \omega_{mkn} - m \omega_+,
\ee
cf.\ Eq.\ (\ref{eq:kdef}) above.
Also the symbol $\Lambda$ is a shorthand for the set of indices
$lmkn$; we define
\be
\alpha_\Lambda = \alpha_{lmkn} \equiv \alpha_{\omega_{mkn}lm},
\label{eq:alphaLambdadef}
\ee
and similarly for $\beta_\Lambda$ and $\tau_\Lambda$.  The sum over
$\Lambda$ is defined as
\be
\sum_\Lambda = \sum_{k=-\infty}^\infty \sum_{n=-\infty}^\infty
\sum_{l=0}^\infty \sum_{m=-l}^l.
\ee
The expression in curly brackets in Eq.\ (\ref{eq:dcalEdt2})
means either a factor of $\omega_{mkn}$ (for energy) or a factor of
$m$ (for angular momentum).
The first term in the square brackets corresponds to the flux through
future null infinity, and the second term corresponds the flux through
the horizon.  Note that for energy the future null infinity term is always
negative; this makes sense since the expression (\ref{eq:dcalEdt2})
is the rate of change of orbital energy.

The flux expression (\ref{eq:dcalEdt2})
has been derived from the radiation
reaction force.  As shown by Gal'tsov, the same result
is obtained by computing the flux directly using the stress-energy
tensor of the scalar field  \cite{Galtsov:1982}.
%[A special case of this
%result has been independently derived in Ref.\ \cite{Gralla:2005et}]

Note that the final answer (\ref{eq:dcalEdt2}) is independent of
our choices for the normalization constants $\alpha_\Lambda$ and
$\beta_\Lambda$, since the amplitudes $Z^{\rm out}_\Lambda$ and
$Z^{\rm down}_\Lambda$ contain factors of $\alpha_\Lambda$ and
$\beta_\Lambda$ that compensate for the factors that appear explicitly
in Eq.\ (\ref{eq:dcalEdt2}).

\subsection{Derivation}
\label{sec:derivation1}

For energy and angular momentum, we can write the conserved quantity as
the inner product of a Killing vector $\xi^\alpha$ with the 4-velocity
of the particle:
\be
{\cal E} = \xi_\alpha u^\alpha.
\ee
Here ${\vec \xi} = - \partial/\partial t$ for ${\cal E} = E$, and
${\vec \xi} = \partial /\partial \phi$ for ${\cal E} = L_z$.
Taking a time derivative gives
\be
\frac{d {\cal E}}{d \tau} = u^\beta \nabla_\beta ( \xi_\alpha
u^\alpha) = (\nabla_\beta \xi_\alpha) u^\beta u^\alpha + \xi_\alpha
(u^\beta \nabla_\beta u^\alpha) = \xi_\alpha a^\alpha,
\ee
where $a^\alpha$ is the 4-acceleration.  Here we have used the fact
that ${\vec \xi}$ is a Killing vector so $\nabla_{(\alpha}
\xi_{\beta)} =0$.

Next, we use the formula (\ref{eq:selfacc0}) for the
self-acceleration.  This gives
\begin{eqnarray}
\frac{d {\cal E}}{d \tau} &=& \frac{q}{\mu} \left[ \xi^\alpha
\nabla_\alpha \Phi_{\rm rad} + (\xi_\alpha u^\alpha) u^\beta
\nabla_\beta \Phi_{\rm rad} \right] \nn \\
\mbox{} &=&  \frac{q}{\mu} \xi^\alpha
\nabla_\alpha \Phi_{\rm rad} +\frac{q}{\mu} {\cal E} \frac{d \Phi_{\rm
rad}}{d \tau}.
\label{eq:dcalEdtau0}
\end{eqnarray}
Now the second term here is a total time derivative to leading order
in $q$, since ${\cal E}$ is conserved to zeroth order in $q$.  Hence
the change in ${\cal E}$ over an interval from $\tau_1$ to $\tau_2$
associated with the second term will be
\be
\frac{q}{\mu} {\cal E} \left[ \Phi_{\rm rad}(\tau_2) - \Phi_{\rm
rad}(\tau_1) \right].
\ee
This is a term which will oscillate but will not grow secularly with
time (since $\Phi_{\rm rad}$ does not grow secularly with time).
Hence in the adiabatic limit this term will be smaller
than the change due to the first term in Eq.\
(\ref{eq:dcalEdtau0}) by the ratio of the orbital timescale to the
inspiral timescale, which is negligible in the approximation we are using.
Dropping the second term in Eq.\
(\ref{eq:dcalEdtau0}) and substituting
into Eq.\ (\ref{eq:used}) we get
\be
\left< \frac{d {\cal E}} {d t} \right>_t = \frac{q}{\mu \Gamma}
\, \left< \,\Sigma \, \xi^\alpha \nabla_\alpha \Phi_{\rm rad} \right>_\lambda.
\label{eq:dcalEdtau1}
\ee

Next, we can obtain an explicit expression for the radiative field
$\Phi_{\rm rad}$ by substituting the harmonic decomposition
(\ref{eq:harmonicdecomposition}) into Eq.\ (\ref{eq:Phiradformula0}).
This gives
\begin{eqnarray}
\fl
\Phi_{\rm rad}(x) = \frac{1}{32 \pi i} \sum_{k = -\infty}^\infty
\sum_{n=-\infty}^\infty \sum_{l=0}^\infty \sum_{m=-l}^l
\frac{\omega_{mkn}}{|\omega_{mkn}|}
\bigg[
\frac{1}{|\alpha_\Lambda|^2}
Z^{\rm out}_\Lambda \pi^{\rm out}_\Lambda(x)
\nonumber \\
\lo
+\frac{\omega_{mkn} p_{mkn}}{|\omega_{mkn} p_{mkn}|}
\frac{|\tau_\Lambda|^2}{|\beta_\Lambda|^2}
Z^{\rm down}_\Lambda \pi^{\rm down}_\Lambda(x)
\bigg].
\label{eq:Phiradformula0a}
\end{eqnarray}
Here as above $\Lambda$ denotes the set of indices $lmkn$, and we have
defined
\be
\pi^{\rm out,down}_{lmkn} = \pi^{\rm out,down}_{\omega_{mkn}lm}.
\ee
We now substitute this expression into Eq.\
(\ref{eq:dcalEdtau1}), and use the fact that the operator $\xi^\alpha
\nabla_\alpha$ gives a factor of $i \omega$ (for the energy) or $i m$
(for the angular momentum).  We get
\begin{eqnarray}
\fl
\left< \frac{d {\cal E}}{dt} \right>_t = \frac{q}{32 \pi \mu \Gamma}
\sum_\Lambda
\left\{ \begin{array}{l} \omega_{mkn} \\
                         m   \\
\end{array} \right\}
\frac{\omega_{mkn}}{|\omega_{mkn}|}
\bigg[
\frac{1}{|\alpha_\Lambda|^2}
Z^{\rm out}_\Lambda \left< \Sigma \, \pi^{\rm out}_\Lambda \right>_\lambda
\nonumber \\
\lo
+\frac{\omega_{mkn} p_{mkn}}{|\omega_{mkn} p_{mkn}|}
\frac{|\tau_\Lambda|^2}{|\beta_\Lambda|^2}
Z^{\rm down}_\Lambda \left< \Sigma \, \pi^{\rm down}_\Lambda \right>_\lambda
\bigg],
\label{eq:dcalEdt1}
\end{eqnarray}
where the expression in curly brackets means either a factor of
$\omega_{mkn}$ or a factor of $m$.

We now discuss the evaluation of the quantity
\be
\left< \Sigma \, \pi_{\omega_{mkn}lm}^{\rm out} \right>_\lambda
\ee
which appears in Eq.\ (\ref{eq:dcalEdt1}).  This is the time average
of the ``out'' mode function evaluated on the geodesic, and can be easily
evaluated using the harmonic decomposition of Sec.\
\ref{sec:harmonic}.  By comparing the integrands of Eqs.\
(\ref{eq:Zlmout1}) and (\ref{eq:Zoutformula2}) we find that
\be
\fl
\Sigma[z^\alpha(\lambda)] \pi_{\omega_{mkn}lm}^{\rm
out}[z^\alpha(\lambda)] =  - \frac{1}{q}
J_{\omega_{mkn}lm}(\lambda,\lambda)^* e^{-i
\lambda (\Gamma \omega_{mkn} - \Upsilon_\phi m)},
\ee
where $z^\alpha(\lambda) = [ t(\lambda), r(\lambda),
 \theta(\lambda),\phi(\lambda)]$ is the geodesic.
Now using the Fourier series expansion (\ref{eq:doubleFourierseries})
of $J_{\omega lm}$ gives
\be
\fl
\Sigma[z^\alpha(\lambda)] \pi_{\omega_{mkn}lm}^{\rm
out}[z^\alpha(\lambda)] =  - \frac{1}{q} \sum_{k',n'}
J_{\omega_{mkn}lmk'n'}^*e^{-i
\lambda \Gamma (\omega_{mkn} - \omega_{mk'n'})}.
\ee
From this expression it is clear that averaging over $\lambda$ kills
all the terms in the sum except $k'=k$, $n'=n$, and we obtain
\be
\fl
\left< \Sigma[z^\alpha(\lambda)] \pi_{\omega_{mkn}lm}^{\rm
out}[z^\alpha(\lambda)] \right>_\lambda =  - \frac{1}{q}
J_{\omega_{mkn}lmkn}^* = - \frac{\Gamma}{2 \pi q} (Z^{\rm out}_{lmkn})^*.
\label{eq:Zoutuseful}
\ee
For the last equation we have used Eq.\ (\ref{eq:Zlmknans}).
Finally substituting this result, together with a similar equation for
the ``down'' modes, in to Eq.\ (\ref{eq:dcalEdt1}) we obtain the final
result (\ref{eq:dcalEdt2})

\subsection{Time derivative of renormalized rest mass}

Equation (\ref{eq:dmudt2}) for the time derivative of the renormalized rest
mass $\mu(\tau)$ can be immediately integrated with respect to $\tau$ between two
times $\tau_1$ and $\tau_2$.  This yields
\be
\mu(\tau_2) - \mu(\tau_1) = q \Phi_{\rm rad}[z(\tau_2)] - q \Phi_{\rm
  rad}[z(\tau_1)],
\ee
where $x^\alpha = z^\alpha(\tau)$ is the geodesic.
Now since the zeroth-order orbit is bound, the right hand side of this
equation does not contain any secularly growing terms.  Instead it
consists only of oscillatory terms.  Therefore, over long timescales
and in the adiabatic limit (the regime in which we are working in this
paper), the rest mass is conserved:
\be
\frac{d\mu}{d\tau} =0.
\ee

\section{Expression for the time derivative of the Carter constant}
\label{sec:Kdot}

\subsection{Formula for time derivative}

We now turn to the corresponding computation for the Carter constant
$K$ defined by Eq.\ (\ref{eq:Kdef}).  The result is
\begin{eqnarray}
\fl
\left< \frac{dK}{dt} \right>_t = - \frac{1}{32 \pi^2 \mu} \sum_\Lambda
\frac{\omega_{mkn}}{|\omega_{mkn}|}
\left[
\frac{1}{|\alpha_\Lambda|^2}
( {\tilde Z}^{\rm out}_\Lambda)^*
Z^{\rm out}_\Lambda
+\frac{\omega_{mkn} p_{mkn}}{|\omega_{mkn} p_{mkn}|}
\frac{|\tau_\Lambda|^2}{|\beta_\Lambda|^2}
({\tilde Z}^{\rm down}_\Lambda)^* Z^{\rm down}_\Lambda
\right].
\nonumber \\
\label{eq:dKdt}
\end{eqnarray}
This expression has the same structure as the expression
(\ref{eq:dcalEdt2}) for
the time derivatives of the energy and angular momentum, except that
the squared amplitudes $|Z^{\rm out}_\Lambda|^2$ and $|Z^{\rm
down}_\Lambda|^2$ have been replaced with products of the amplitudes
$Z^{\rm out}_\Lambda$ and $Z^{\rm down}_\Lambda$ with two new
amplitudes
${\tilde Z}^{\rm out}_\Lambda$ and ${\tilde Z}^{\rm down}_\Lambda$.
These new amplitudes are defined by the equation
\begin{eqnarray}
\fl
{\tilde Z}^{\rm out,down}_{lmkn} =  \frac{4 \pi q}{\Gamma \Lambda_r
  \Lambda_\theta}
e^{-i m \phi_0} e^{i \omega_{mkn} t_0}
\int_0^{\Lambda_r} d\lambda_r \, \int_0^{\Lambda_\theta}
d\lambda_\theta \ e^{i(k \Upsilon_\theta \lambda_\theta + n \Upsilon_r
  \lambda_r)}
\Theta_{\omega_{mkn}lm}[\theta(\lambda_\theta)]^*
\nn \\ \mbox{}  \times
e^{- i m \Delta \phi_r(\lambda_r)} e^{-i m \Delta \phi_\theta(\lambda_\theta)}
e^{i \omega_{mkn} \Delta t_r(\lambda_r)} e^{i
  \omega_{mkn} \Delta t_\theta(\lambda_\theta)}
\nn \\ \mbox{}
\times \left\{ G_{mkn}[\lambda_r,\lambda_\theta]
R^{\rm out,down}_{\omega_{mkn}lm}[r(\lambda_r)]^*
+ G[\lambda_r,\lambda_\theta]
\frac{d R^{\rm out,down}_{\omega_{mkn}lm}}{dr}[r(\lambda_r)]^*
\right\},
\nonumber \\
\label{eq:tildeZlmknformula}
\end{eqnarray}
where
\be
R^{\rm out,down}_{\omega_{mkn}lm}(r) = \frac{u^{\rm
    out,down}_{\omega_{mkn}lm}(r)}{\varpi(r)},
\ee
\be
G_{mkn}(\lambda_r,\lambda_\theta) \equiv -
\frac{\Sigma}{\Delta} (\varpi^2 E - a L_z)
 (\varpi^2 \omega_{mkn}- a m) +
%\omega_{mkn} r^2 T,
2 i r u_r \Delta,
\label{eq:Gmnkdef}
\ee
\be
G(\lambda_r,\lambda_\theta) = - i \Delta \Sigma u_r.
\label{eq:Gdef1}
\ee
On the right hand sides of Eqs.\ (\ref{eq:Gmnkdef}) and
(\ref{eq:Gdef1}), it is understood that $\Sigma(r,\theta)$,
$\Delta(r)$ and $\varpi(r)$ are evaluated at $r = r(\lambda_r)$ and
$\theta = \theta(\lambda_\theta)$.  Also it is understood that
$u_r$ is given as a function of $r$ by Eq.\
(\ref{eq:urformula}), and hence as a function of $\lambda_r$ by using $r
= r(\lambda_r)$ and by resolving the sign ambiguity in
Eq.\ (\ref{eq:urformula}) using Eq.\ (\ref{eq:psidef}).

Several features of the formulae (\ref{eq:dKdt}) and
(\ref{eq:tildeZlmknformula}) are worth noting:

\begin{itemize}

\item The formula (\ref{eq:tildeZlmknformula}) for the new amplitudes
${\tilde Z}^{\rm out}_{lmkn}$ and ${\tilde Z}^{\rm down}_{lmkn}$.
has a very similar structure to the formula (\ref{eq:Zlmknformula})
defining the amplitudes $Z^{\rm out}_{lmkn}$ and $Z^{\rm
  down}_{lmkn}$.  In particular, the derivation of the formula
(\ref{eq:changeparameter}) for the dependence of the amplitudes on
the parameters $t_0$, $\phi_0$, $\lambda_{r0}$ and $\lambda_{\theta0}$
of the geodesic (via an overall phase) carries through as before.
This means that the time derivative of the Carter constant is
independent of these parameters, as expected, since the phase from the untilded
amplitudes cancels the phase from the tilded amplitudes in Eq.\
(\ref{eq:dKdt}).

\item Although we have no general definition for ``flux of Carter'',
  presumably the first term in Eq.\ (\ref{eq:dKdt}) corresponds to
  something like the flux of Carter to future null infinity, and the
  second term to the flux of Carter down the black hole horizon.

\item As was the case for the original amplitudes, to
evaluate the double integral (\ref{eq:tildeZlmknformula}) it is
convenient to use the identity (\ref{eq:averageidentity}) to
express the integral in terms of the variables $\chi$ and $\psi$
instead of $\lambda_\theta$ and $\lambda_r$.

\end{itemize}

\subsection{Derivation}

Taking a time derivative of the expression (\ref{eq:Kdef}) for $K$ gives
\begin{eqnarray}
\frac{d K}{d \tau} &=& u^\gamma \nabla_\gamma (K_{\alpha\beta} u^\alpha u^\beta) =
\nabla_{(\gamma} K_{\alpha\beta)} u^\gamma u^\alpha u^\beta + 2 K_{\alpha\beta} u^\alpha u^\gamma \nabla_\gamma u^\beta \nn \\
\mbox{} &=& 2 K_{\alpha\beta} u^\alpha a^\beta.
\end{eqnarray}
Here we have used the Killing tensor equation $\nabla_{(\gamma} K_{\alpha\beta)} =0$.
Now substituting the expression (\ref{eq:selfacc0}) for the
self-acceleration gives
\begin{eqnarray}
\frac{d K}{d \tau} &=& \frac{2 q}{\mu} \left[ K_{\alpha\beta} u^\alpha
  u^\beta u^\gamma \nabla_\gamma \Phi_{\rm rad} + K_{\alpha\beta}
  u^\alpha \nabla^\beta \Phi_{\rm rad} \right]
\nn \\
\mbox{} &=&  \frac{2 q}{\mu} K \frac{d \Phi_{\rm rad}}{d \tau}
+\frac{2 q}{\mu} K^{\alpha\beta} u_{\alpha} \nabla_\beta \Phi_{\rm rad}.
\label{eq:dKdtau0}
\end{eqnarray}
Now the first term here is a total time derivative to leading order
in $q$, since $K$ is conserved to zeroth order in $q$.  Hence
this term can be neglected for the reason explained in Sec.\
\ref{sec:derivation1}.  Dropping the first term and substituting
into Eq.\ (\ref{eq:used}) we get
\be
\left< \frac{d K} {d t} \right>_t = \frac{2 q}{\mu \Gamma}
\, \left< \,\Sigma \, K^{\alpha\beta} u_\alpha \nabla_\beta \Phi_{\rm rad} \right>_\lambda.
\label{eq:dKdtau1}
\ee

Next, we use the explicit expression (\ref{eq:Phiradformula0a})
for the radiation field $\Phi_{\rm rad}$.
This gives
\begin{eqnarray}
\fl
\left< \frac{d K}{dt} \right>_t = \frac{q}{16 \pi i \mu \Gamma}
\sum_\Lambda
\frac{\omega_{mkn}}{|\omega_{mkn}|}
\bigg[
\frac{1}{|\alpha_\Lambda|^2}
Z^{\rm out}_\Lambda \left< \Sigma K^{\alpha\beta} u_\alpha
\nabla_\beta \, \pi^{\rm out}_\Lambda \right>_\lambda
\nonumber \\
\lo
+\frac{\omega_{mkn} p_{mkn}}{|\omega_{mkn} p_{mkn}|}
\frac{|\tau_\Lambda|^2}{|\beta_\Lambda|^2}
Z^{\rm down}_\Lambda \left< \Sigma K^{\alpha\beta} u_\alpha
\nabla_\beta \, \pi^{\rm down}_\Lambda \right>_\lambda
\bigg].
\label{eq:dKdt1}
\end{eqnarray}
We now define the amplitudes
\be
{\tilde Z}^{\rm out}_{lmkn} = \frac{2 \pi q}{\Gamma i} \left< \Sigma
K^{\alpha\beta} u_\alpha \nabla_\beta ( \pi^{\rm out}_{\omega_{mkn}lm}
)^* \right>_\lambda
\label{eq:tildeZoutdef}
\ee
and
\be
{\tilde Z}^{\rm down}_{lmkn} = \frac{2 \pi q}{\Gamma i} \left< \Sigma
K^{\alpha\beta} u_\alpha \nabla_\beta ( \pi^{\rm down}_{\omega_{mkn}lm}
)^* \right>_\lambda.
\label{eq:tildeZdowndef}
\ee
Substituting these definitions into Eq.\ (\ref{eq:dKdt1}) gives the
formula (\ref{eq:dKdt}).  Therefore it remains only to derive the
formula (\ref{eq:tildeZlmknformula}) for the amplitudes ${\tilde
  Z}_\Lambda^{\rm out}$ and ${\tilde Z}_\Lambda^{\rm down}$.
We will derive the formula for
${\tilde Z}_\Lambda^{\rm out}$; the derivation of the formula for
${\tilde Z}_\Lambda^{\rm down}$ is similar.

We start by simplifying the differential operator $K^{\alpha\beta} u_\alpha
\nabla_\beta$ which appears in the definitions (\ref{eq:tildeZoutdef})
and (\ref{eq:tildeZdowndef}).  Using the expression
(\ref{eq:Kalphabeta}) for the Killing tensor we can write this as
\be
K^{\alpha\beta} u_\alpha \nabla_\beta
 = \Sigma (l^\alpha u_\alpha) n^\beta \nabla_\beta
 + \Sigma (n^\alpha u_\alpha) l^\beta \nabla_\beta + r^2
 \frac{d}{d\tau}.
\ee
Using the definitions
(\ref{eq:vecldef}) and (\ref{eq:vecndef}) of ${\vec l}$ and ${\vec
n}$, the definitions (\ref{eq:Edef}) and (\ref{eq:Lzdef}) of $E$ and
$L_z$, and the fact that $\partial_t$ and $\partial_\phi$ reduce to $i
\omega_{mkn}$ and $-i m$ when acting on $(\pi^{\rm
  out}_{\omega_{mkn}lm})^*$ gives
\begin{eqnarray}
\fl
K^{\alpha\beta} u_\alpha \nabla_\beta
 &=& r^2 \frac{d}{d\tau} - \frac{i}{2 \Delta}( -\varpi^2 E + a L_z +
 u_r \Delta )
(-\varpi^2 \omega_{mkn} + a m - i \Delta \partial_r)
\nn \\ \mbox{} &&
 - \frac{i}{2 \Delta}( -\varpi^2 E + a L_z - u_r \Delta )
(-\varpi^2 \omega_{mkn} + a m + i \Delta \partial_r)  \\
\mbox{} &=& r^2 \frac{d}{d\tau} - \frac{i}{\Delta}( -\varpi^2 E + a
 L_z) (-\varpi^2 \omega_{mkn} + a m) - \Delta u_r \partial_r.
\label{eq:K3}
\end{eqnarray}

Consider now the contribution of the first term in Eq.\
(\ref{eq:K3}) to the expression (\ref{eq:tildeZoutdef})
for ${\tilde Z}^{\rm out}_{lmkn}$.  Using the relation (\ref{eq:Minotime})
between $d\tau$ and $d\lambda$ we can write this as
\begin{equation}
\frac{2 \pi q}{\Gamma i} \left< \Sigma r^2 \frac{d}{d\tau}
( \pi^{\rm out}_{\omega_{mkn}lm}
)^* \right>_\lambda =
\frac{2 \pi q}{\Gamma i} \left<  r^2 \frac{d}{d\lambda}
( \pi^{\rm out}_{\omega_{mkn}lm}
)^* \right>_\lambda.
\end{equation}
We can integrate this by parts with respect to $\lambda$; the boundary
term which is generated can be neglected for the reason explained in
Sec.\ \ref{sec:derivation1}.  This gives
\begin{equation}
- \frac{2 \pi q}{\Gamma i} \left< \frac{d r^2}{d\lambda}
( \pi^{\rm out}_{\omega_{mkn}lm}
)^* \right>_\lambda =
\frac{2 \pi q}{\Gamma} \left<  2 i r u_r \Delta
( \pi^{\rm out}_{\omega_{mkn}lm}
)^* \right>_\lambda,
\end{equation}
where we have used $dr/d\lambda = \Sigma u^r = \Delta u_r$ from Eqs.\
(\ref{eq:Minotime}) and (\ref{eq:kerrmetric}).
Combining this with the result obtained from substituting the
second and third terms in Eq.\ (\ref{eq:K3}) into
Eq. (\ref{eq:tildeZoutdef}) gives
\be
\fl
{\tilde Z}^{\rm out}_{lmkn} = \frac{2 \pi q}{\Gamma } \left<
G_{mkn}(\lambda,\lambda) ( \pi^{\rm out}_{\omega_{mkn}lm} )^*
+ G(\lambda,\lambda) \partial_r ( \pi^{\rm out}_{\omega_{mkn}lm} )^*
\right>_\lambda,
\label{eq:tildeZoutformula1}
\ee
where the functions $G_{mkn}(\lambda_r,\lambda_\theta)$ and
$G(\lambda_r,\lambda_\theta)$ are
[cf.\ Eqs.\ (\ref{eq:Gmnkdef}) and (\ref{eq:Gdef1}) above]
\be
G_{mkn}(\lambda_r,\lambda_\theta) \equiv - \frac{\Sigma}{\Delta}
(\varpi^2 E -a L_z)
 (\varpi^2 \omega_{mkn} - a m) + 2 i r u_r \Delta,
\label{eq:Gmnkdef2}
\ee
\be
G(\lambda_r,\lambda_\theta) = - i \Delta \Sigma u_r.
\label{eq:Gdef2}
\ee
As explained above, it is understood that the functions $\Sigma(r,\theta)$,
$\Delta(r)$ and $\varpi(r)$ on the right hand sides of Eqs.\
(\ref{eq:Gmnkdef2}) and (\ref{eq:Gdef2})
are evaluated at $r = r(\lambda_r)$ and
$\theta = \theta(\lambda_\theta)$.  Also it is understood that
$u_r$ is given as a function of $r$ by Eq.\
(\ref{eq:urformula}), and hence as a function of $\lambda_r$ by using $r
= r(\lambda_r)$ and by resolving the sign ambiguity in
Eq.\ (\ref{eq:urformula}) using Eq.\ (\ref{eq:psidef}).

The average over $\lambda$ in Eq.\ (\ref{eq:tildeZoutformula1}) can be
evaluated using the same techniques as in Secs.\ \ref{sec:harmonic}
and \ref{sec:Edot} above.  Using the definition (\ref{eq:pioutdef}) of the
``out'' mode function and the definition (\ref{eq:omegamkndef}) of
$\omega_{mkn}$,
the quantity inside the angular brackets in
Eq.\ (\ref{eq:tildeZoutformula1}) can be written as
\be
{\tilde J}_{mkn}(\lambda,\lambda),
\ee
where
\begin{eqnarray}
\fl
{\tilde J}_{mkn}(\lambda_r,\lambda_\theta)
=  2
e^{-i m \phi_0} e^{i \omega_{mkn} t_0}
\, e^{i(k \Upsilon_\theta \lambda_\theta + n \Upsilon_r
  \lambda_r)}
\Theta_{\omega_{mkn}lm}[\theta(\lambda_\theta)]^*
\nn \\ \mbox{}  \times
e^{- i m \Delta \phi_r(\lambda_r)} e^{-i m \Delta \phi_\theta(\lambda_\theta)}
e^{i \omega_{mkn} \Delta t_r(\lambda_r)} e^{i
  \omega_{mkn} \Delta t_\theta(\lambda_\theta)}
\nn \\ \mbox{}
\times \left\{ G_{mkn}[\lambda_r,\lambda_\theta]
R^{\rm out,down}_{\omega_{mkn}lm}[r(\lambda_r)]^*
+ G[\lambda_r,\lambda_\theta]
\frac{d R^{\rm out,down}_{\omega_{mkn}lm}}{dr}[r(\lambda_r)]^*
\right\}.
\nonumber \\
\label{eq:tildeJformula}
\end{eqnarray}
This function is biperiodic:
\begin{eqnarray}
{\tilde J}_{mkn}(\lambda_r + \Lambda_r, \lambda_\theta) &=&
{\tilde J}_{mkn}(\lambda_r, \lambda_\theta) \nn \\
\mbox{}
{\tilde J}_{mkn}(\lambda_r , \lambda_\theta + \Lambda_\theta) &=&
{\tilde J}_{mkn}(\lambda_r, \lambda_\theta).
\end{eqnarray}
Hence it can be expanded as a double Fourier series, and the average
over $\lambda$ of ${\tilde J}_{mkn}(\lambda,\lambda)$ is
\be
\left< {\tilde J}_{mkn}(\lambda,\lambda) \right>_\lambda
=\frac{1}{\Lambda_r \Lambda_\theta}  \int_0^{\Lambda_r} d\lambda_r
\int_0^{\Lambda_\theta} d\lambda_\theta {\tilde
  J}_{mkn}(\lambda_r,\lambda_\theta).
\label{eq:Jav}
\ee
Substituting Eq.\ (\ref{eq:tildeJformula}) into Eq.\ (\ref{eq:Jav})
and back into Eq.\ (\ref{eq:tildeZoutformula1}) now gives the formula
(\ref{eq:tildeZlmknformula}).

%\section{Evolution of orbits and asymptotic scalar waveform}

\section{Circular orbits}
\label{sec:circular}

\subsection{Known formula for time derivative of the Carter constant}

In this section we study the prediction of the formula (\ref{eq:dKdt})
for the time derivative of the Carter constant, for the special case
of circular orbits (orbits of constant Boyer-Lindquist radius $r$).  It is known that
circular orbits remain circular while evolving under the influence of radiation
reaction in the adiabatic regime \cite{Kennefick:1995za,Ryan:1995xi,Thesis:Mino}.
Also for circular orbits it is possible to relate the Carter constant
$K$ to the energy $E$ and angular momentum $L_z$ of the orbit:
specializing Eq.\ (\ref{eq:Kformula1}) to zero radial velocity $u_r
=0$ yields
\begin{equation}
K = \frac{\left(\varpi^2 E - aL_z\right)^2}{\Delta} - r^2\;.
\end{equation}
In order for this relation to be preserved under slow evolution of
$K$, $E$ and $L_z$, we must have
\be
\left< \frac{dK}{dt}\right>_t =
 \frac{2(\varpi^2 E - a L_z)}{\Delta}\left[\varpi^2 \left<
   \frac{dE}{dt} \right>_t - a
\left< \frac{d L_z}{dt} \right>_t \right]\;.
\label{eq:circtocirccheck}
\ee
We now verify that our expression (\ref{eq:dKdt}) for $\left< dK/dt
\right>_t$ satisfies the condition (\ref{eq:circtocirccheck}).

\subsection{Verification that our result reproduces this formula}

For circular orbits, the following simplifications apply to the
harmonic decomposition of the geodesic source derived in Sec.\
\ref{sec:harmonic}.  First, the functions $\Delta t_r(\lambda)$,
${\hat t}_r(\lambda)$, $\Delta \phi_r(\lambda)$ and ${\hat
  \phi}_r(\lambda)$ defined by Eqs.\ (\ref{eq:deltatrformula}),
(\ref{eq:hattrdef}), (\ref{eq:deltaphirformula}) and
(\ref{eq:hatphirdef}) vanish identically.  Second,
the biperiodic function $J_{\omega lm}(\lambda_r,\lambda_\theta)$ defined in Eq.\
(\ref{eq:Jomegalmdef}) is therefore independent of $\lambda_r$.  It follows that
the double Fourier series (\ref{eq:doubleFourierseries})
is replaced by the single Fourier series
\be
J_{\omega lm}(\lambda_\theta) =
\sum_{k=-\infty}^\infty
J_{\omega lmk} e^{-i k \Upsilon_\theta \lambda_\theta},
\label{eq:singleFourierseries}
\ee
where the coefficients $J_{\omega lmk}$
are given by
\be
J_{\omega lmk} = \frac{1}{\Lambda_\theta}
\int_0^{\Lambda_\theta} d\lambda_\theta
\,
e^{i k \Upsilon_\theta \lambda_\theta}
J_{\omega lm}(\lambda_\theta).
\label{eq:Jomegalmkdef}
\ee
The effect of this change on the subsequent formulae for fluxes and
amplitudes in Secs.\ \ref{sec:Edot} and \ref{sec:Kdot}
can be summarized as: (i) remove the sums over $n$; (ii)
remove the indices $n$ from the amplitudes $Z_{lmkn}$; (iii) remove
the averaging operation
\[
\frac{1}{\lambda_r} \int_0^{\Lambda_r} d\lambda_r
\]
from the formulae for the amplitudes; and (iv) evaluate the
integrands in the formulae for the amplitudes at $n=0$.
Also the formula (\ref{eq:omegamkndef})
for the frequency $\omega_{mkn}$ is replaced by
\be
\omega_{mk} = m \Omega_\phi + k \Omega_\theta.
\ee

With these simplifications, the formula (\ref{eq:dcalEdt2}) for the time
derivatives of energy and angular momentum becomes
\begin{eqnarray}
\fl
\left< \frac{d {\cal E}}{dt} \right>_t = - \frac{1}{64 \pi^2 \mu}
\sum_{lmk}
\left\{ \begin{array}{l} \omega_{mk} \\
                         m   \\
\end{array} \right\}
\frac{\omega_{mk}}{|\omega_{mk}|}
\bigg[
\frac{1}{|\alpha_{lmk}|^2}
|Z^{\rm out}_{lmk} |^2
\nonumber \\
\lo
+\frac{\omega_{mk} p_{mk}}{|\omega_{mk} p_{mk}|}
\frac{|\tau_{lmk}|^2}{|\beta_{lmk}|^2}
|Z^{\rm down}_{lmk} |^2 \bigg],
\label{eq:dcalEdt2circ}
\end{eqnarray}
where ${\cal E} = E$ or $L_z$, and the expression in curly brackets
means either $\omega_{mk}$ (for $E$) or $m$ (for $L_z$).  Similarly
the expression (\ref{eq:dKdt}) for the time derivative of the Carter constant
becomes
\begin{eqnarray}
\fl
\left< \frac{dK}{dt} \right>_t = - \frac{1}{32 \pi^2 \mu} \sum_{lmk}
\frac{\omega_{mk}}{|\omega_{mk}|}
\bigg[
\frac{1}{|\alpha_{lmk}|^2}
( {\tilde Z}^{\rm out}_{lmk})^*
Z^{\rm out}_{lmk}
\nonumber \\
\lo
+\frac{\omega_{mk} p_{mk}}{|\omega_{mk} p_{mk}|}
\frac{|\tau_{lmk}|^2}{|\beta_{lmk}|^2}
({\tilde Z}^{\rm down}_{lmk})^* Z^{\rm down}_{lmk}
\bigg].
\label{eq:dKdtcirc}
\end{eqnarray}
Comparing Eqs.\ (\ref{eq:dcalEdt2circ}) and (\ref{eq:dKdtcirc}), we
see that the condition (\ref{eq:circtocirccheck})
will be satisfied as long as the amplitudes ${\tilde Z}_{lmk}$ satisfy
\be
{\tilde Z}_{lmk}^{{\rm out},{\rm down}} =
\frac{\left(\varpi^2 E- a
L_z\right)}{\Delta}\left(\varpi^2\omega_{mk} - m a\right)
Z_{lmk}^{{\rm out},{\rm down}}\;.
\label{eq:amplitudeidentity}
\ee
Therefore it suffices to derive the identity
(\ref{eq:amplitudeidentity}).  We will derive this identity for the
``out'' modes; the ``down'' derivation is identical.

The particular form of the expressions for the amplitudes that are
most useful here are Eqs.\ (\ref{eq:Zoutuseful}) and
(\ref{eq:tildeZoutformula1}):
\be
Z^{\rm out}_{lmk} = - \frac{2 \pi q}{\Gamma} \left< \Sigma \left(
\pi_{lmk}^{\rm out} \right)^* \right>_\lambda,
\label{eq:Zoutuseful1}
\ee
and
\be
{\tilde Z}^{\rm out}_{lmk} =  \frac{2 \pi q}{\Gamma} \left<
G_{mk}(\lambda,\lambda) \left( \pi_{lmk}^{\rm out} \right)^*
+ G(\lambda,\lambda) \partial_r \left( \pi_{lmk}^{\rm out} \right)^*
\right>_\lambda.
\label{eq:tildeZoutformula2}
\ee
Here we have defined $\pi_{lmk}^{\rm out} = \pi_{\omega_{mk}lm}^{\rm
  out}$.  It follows from Eq.\ (\ref{eq:Gdef2}) together with
$u_r=0$ that the function $G(\lambda,\lambda)$ vanishes, so the
formula for ${\tilde Z}_{lmk}^{\rm out}$ simplifies to
\be
{\tilde Z}^{\rm out}_{lmk} =  \frac{2 \pi q}{\Gamma} \left<
G_{mk}(\lambda,\lambda) \left( \pi_{lmk}^{\rm out} \right)^*
\right>_\lambda.
\label{eq:tildeZoutformula3}
\ee
We now substitute into Eq.\ (\ref{eq:tildeZoutformula3})
the expression (\ref{eq:Gmnkdef2}) for $G_{mk}$.  The second term in
Eq.\ (\ref{eq:Gmnkdef2}) vanishes since $u_r=0$, and of all the
factors in the first term, only $\Sigma$ depends on $\lambda$; the
remaining factors are constants and can be pulled out of the average
over $\lambda$.  This yields
\be
{\tilde Z}_{lmk}^{\rm out} = - \frac{2 \pi q}{\Gamma} \frac{\left( \varpi^2
E - a L_z \right)}{\Delta} \left( \varpi^2 \omega_{mk} - a m \right) \left<
\Sigma \left( \pi_{lmk}^{\rm out} \right)^* \right>_\lambda.
\ee
Comparing this with the formula (\ref{eq:Zoutuseful1}) for $Z_{lmk}^{\rm out}$ now
yields the identity (\ref{eq:amplitudeidentity}).

\ack
This research was supported in part by NSF grants PHY-0140209
and PHY-0244424 and by NASA Grant NAGW-12906.  We thank the anonymous
referees for several helpful comments.

\appendix

\section{Accuracy of adiabatic waveforms}
\label{sec:accuracy}

In this appendix we estimate the accuracy of the adiabatic waveforms,
expanding on the brief treatment given in Ref.\ \cite{dfh05}.
Specifically we estimate the magnitude of
the correction to the
waveform's phase that corresponds to the term $\phi_2(t)$ in Eq.\
(\ref{eq:emri_phase}).
We can roughly estimate this term by using post-Newtonian expressions
for the waveform in which terms corresponding to $\phi_2(t)$ are readily identified.
While the post-Newtonian approximation is not strictly valid in the
highly relativistic regime near the horizon of interest here, it
suffices to give some indication of the accuracy of the adiabatic
waveforms.  In this appendix we specialize for simplicity to circular,
equatorial orbits.  We also specialize to gravitational radiation
reaction, unlike the body of the paper which dealt with the scalar case.
Our conclusion is that adiabatic waveforms will likely be sufficiently accurate
for signal detection.

The procedure we use is as follows.  We focus attention on the phase
$\Psi(f)$ of the Fourier transform of the dominant, $l=2$ piece of the
gravitational waveform.  Here $f$ is the gravitational wave frequency.
Now a change to the phase function of the form
$\Psi(f) \to \Psi(f) + \Psi_0 + 2 \pi f \Delta t$ is not observable;
the constant $\Delta t$ corresponds to a change in the time of arrival
of the signal.  Therefore we focus on the observable quantity $d^2
\Psi/df^2$, and we write this as
\be
\frac{d^2 \Psi}{d f^2}(f) = G(f,m_1,m_2).
\label{eq:Gadef}
\ee
Here $G$ is a function, $m_1$ is the mass of the black hole, and $m_2$
is the mass of the inspiralling compact object\footnote{We also
define $M = m_1 + m_2$, the total mass, and $\mu = m_1 m_2/M$, the
reduced mass.   Throughout the body of the paper we referred to the mass
of the inspiralling object as $\mu$ and the mass of the black hole as
$M$; this is valid to leading order in $\mu/M$.  In this
appendix however we need to be more accurate, and hence we revert to
using $m_1$ and $m_2$ for the two masses.}.  The specific form of $G$ obtained from
post-Newtonian theory is discussed below.

We next expand the function $G$ as a power
series in $m_2$, at fixed $f$ and $m_1$.  The leading order term scales as
$m_2^{-1}$, and we obtain
\be
G(f,m_1,m_2) = \frac{G_0(f,m_1)}{m_2} + G_1(f,m_1) + G_2(f,m_1) m_2 + \ldots
\label{eq:Gexpand}
\ee
The first term in this expression corresponds to the term $\Phi_1(t)$
in Eq.\ (\ref{eq:emri_phase}); it gives the leading-order, adiabatic
waveform.  The error incurred from using adiabatic waveforms is
therefore
\be
\fl
\Delta G(f,m_1,m_2) = G(f,m_1,m_2) - \frac{G_0(f,m_1)}{m_2} =
G_1(f,m_1) + G_2(f,m_1) m_2 + \ldots
\ee
We want to estimate the effects of this phase error.

It is useful to split the phase error into two terms:
\be
\Delta G(f,m_1,m_2) = \Delta G_1(f,m_1,m_2) + \Delta G_2(f,m_1,m_2),
\label{eq:DeltaGsplit}
\ee
where
\be
\Delta G_1(f,m_1,m_2) = G(f,m_1,m_2) - \frac{G_0(f,m_1 + m_2)}{m_2}
\label{eq:DeltaG1def}
\ee
and
\be
\Delta G_2(f,m_1,m_2) = \frac{G_0(f,m_1 + m_2)}{m_2} - \frac{G_0(f,m_1)}{m_2}.
\ee
The phase error $\Delta G_2$ corresponds to the error that would be
caused by using an incorrect value of the mass of the black hole.  The
effect of this error on the data analysis would be to cause a
systematic error in the inferred best fit value of the black hole
mass.  However, the fractional error would be of order $\mu/M \ll 1$.
This error will not have any effect on the ability of adiabatic
waveforms to detect signals when used as search templates.
Therefore, we will neglect this error, and focus on the remaining
error term $\Delta G_1$.

We now compute explicitly the phase error $\Delta \Psi_1(f)$ that
corresponds to the term $\Delta G_1$ in Eq.\ (\ref{eq:DeltaGsplit}).
We use the post-3.5-Newtonian expression for $\Psi(f)$, which
can be computed from the orbital energy $E(f)$ and gravitational wave
luminosity ${\dot E}(f)$
via the equation \cite{Poisson:1995vs}
\be
\frac{d^2 \Psi}{df^2} = - 2 \pi \frac{dE/df(f)}{{\dot E}(f)}.
\ee
For the energy $E(f)$ we use the expression given in Eq.\ (50) of Ref.\
\cite{Blanchet:1999pm}, and for the luminosity ${\dot E}$ we use Eq.\
(12.9) of Ref.\ \cite{Blanchet:2001aw} \footnote{With parameter values $\lambda
= -1987/3080$ from Ref.\ \protect{\cite{Blanchet:2003gy}} and $\theta =
-11831/9240$ from Refs. \protect{\cite{Blanchet:2004bb,Blanchet:2004ek}}.}.
This gives
\be
\Psi(f) = 2 \pi f t_c + \phi_c + \frac{3 M}{128 \mu y^5} {\cal F}(f),
\label{eq:Psi0}
\ee
where $t_c$ and $\phi_c$ are constants, $y = (\pi M f)^{1/3}$, and
\begin{eqnarray}
\fl
{\cal F} = 1 + \frac{3715\,y^2}{756} - 16\,\pi \,y^3 + \frac{15293365\,y^4}{508032} +
  \frac{11583231236531\,y^6}{4694215680} - \frac{6848\,\gamma\,y^6}{21}
\nonumber \\
- \frac{640\,{\pi }^2\,y^6}{3}
+ \frac{77096675\,\pi \,y^7}{254016} + \frac{55\,y^2\,\mu }{9\,M} +
  \frac{27145\,y^4\,\mu }{504\,M}
\nonumber \\
- \frac{15737765635\,y^6\,\mu }{3048192\,M} +
  \frac{2255\,{\pi }^2\,y^6\,\mu }{12\,M} + \frac{1014115\,\pi
  \,y^7\,\mu }{3024\,M} +   \frac{3085\,y^4\,{\mu }^2}{72\,M^2}
\nonumber \\
+ \frac{76055\,y^6\,{\mu }^2}{1728\,M^2} -
  \frac{36865\,\pi \,y^7\,{\mu }^2}{378\,M^2} -
  \frac{127825\,y^6\,{\mu }^3}{1296\,M^3}
+ \frac{a y^3}{3} \left(113 \nu^2 + \frac{75 \mu}{M} \right)
\nonumber \\
-  \frac{13696\,y^6\,\log (2)}{21} + \frac{38645\,\pi \,y^5\,\log (y)}{252} -
  \frac{6848\,y^6\,\log (y)}{21}
\nonumber \\
+ \frac{5\,\pi \,y^5\,\mu \,\log (y)}{M}.
\label{eq:fullphase}
\end{eqnarray}
Here $\gamma$ is Euler's constant, $a$ is the dimensionless spin parameter of the
black hole, and $\nu = m_1/M = 1/2 + \sqrt{1/4 -\mu/M}$.
We have also added to Eq.\ (\ref{eq:fullphase}) the spin-dependent
terms up to post-2-Newtonian order, taken from Eq.\ (1) of Ref.\
\cite{Buonanno:2002fy}, specialized to a non-spinning
particle on a circular, equatorial orbit.

We now insert the expression (\ref{eq:Psi0}) for $\Psi(f)$ into Eqs.\
(\ref{eq:Gadef}) and (\ref{eq:DeltaG1def}), and integrate twice with
respect to $f$ to obtain $\Delta \Psi_1(f)$.  The result is
\begin{eqnarray}
\fl
\Delta \Psi_1(f) =
\frac{3}{128 y^5} \bigg[
 \frac{55\,y^2}{9} +
  \frac{27145\,y^4 }{504} - \frac{15737765635\,y^6 }{3048192} +
  \frac{2255\,{\pi }^2\,y^6 }{12} + \frac{1014115\,\pi
  \,y^7 }{3024}
\nonumber \\
  +\frac{3085\,y^4\,{\mu }}{72\,M}
+ \frac{76055\,y^6\,{\mu }}{1728\,M} -
  \frac{36865\,\pi \,y^7\,{\mu }}{378\,M} -
  \frac{127825\,y^6\,{\mu }^2}{1296\,M^2}
\nonumber \\
+ 5\,\pi \,y^5 \,\log (y)
+ \frac{a y^3}{3} (75 + 113 q)
\bigg],
\label{eq:DeltaPsians}
\end{eqnarray}
where $q = (\nu^2-1) M/\mu = -2 + O(\mu/M)$.
Now as discussed above, a change to the phase function of the form
$\Psi(f) \to \Psi(f) + \Psi_0 + 2 \pi f \Delta t$ is not observable.
This freedom allows us to set $\Psi(f_0) = \Psi'(f_0)
= 0$
%for all templates
where $f_0$ is a fixed frequency,
which amounts to
replacing our expression $\Delta \Psi_1(f)$ for the phase error with
\be
\delta \Psi_1(f) = \Delta \Psi_1(f) - \Delta\Psi_1(f_0) - \Delta
\Psi_1'(f_0)(f-f_0).
\ee

\begin{figure}
\begin{center}
\epsfig{file=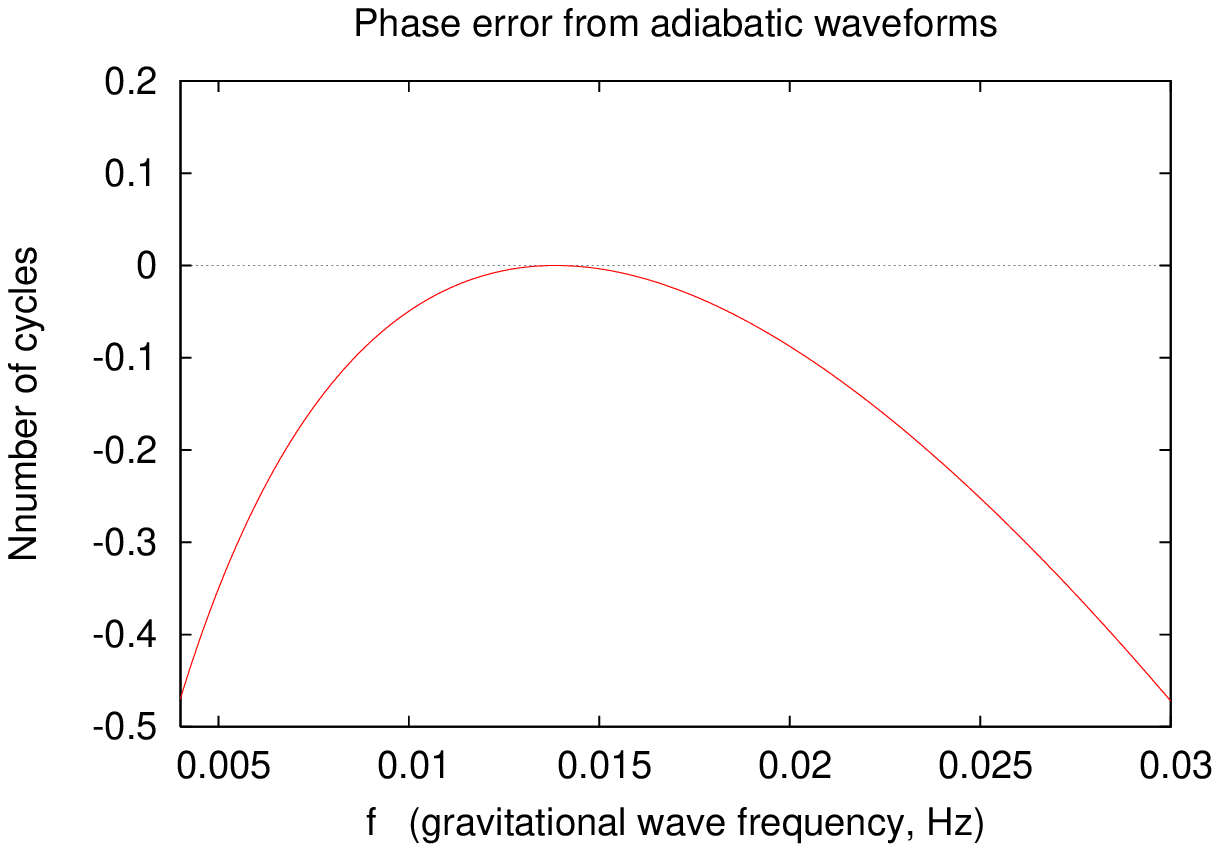,angle=0,width=9cm}
\caption{This plot shows the error in the phase of the Fourier
transform of the waveform that results from using adiabatic
waveforms, estimated using post-3.5-Newtonian waveforms,
for a $10 M_\odot$ compact object spiralling into a $10^6
M_\odot$ black hole with spin parameter $a = 0.999$.  For this system
the last year of inspiral extends from $f = 0.004$ Hz to $f = 0.03$ Hz
\cite{Finn:2000sy}.}
\label{fig:phase}
\end{center}
\end{figure}

We now evaluate the phase error $\delta \Psi_1(f)$ for
some typical sources.  First, we take $\mu = 1 M_\odot$, $M = 10^6
M_\odot$, $a =0.999$.
For this case the last year of inspiral extends from $f = 10^{-2}$ Hz
to $\sim 3 \times 10^{-2}$ Hz \cite{Finn:2000sy}; the maximum
value of $|\delta \Psi_1|$ is $0.19$ cycles if we
take $f_0 = 0.018$ Hz.
As a second example with a higher mass ratio, we take $\mu = 10
M_\odot$, $M = 10^6 M_\odot$, $a = 0.999$.
The last year of inspiral for this case extends
from $f = 0.004$ Hz to $f = 0.03$ Hz \cite{Finn:2000sy};
the maximum
value of $|\delta \Psi_1|$ is $0.47$ cycles if we
take $f_0 = 0.0138$ Hz (see Fig.\ \ref{fig:phase})\footnote{If the
spin-dependent term in Eq.\ (\protect{\ref{eq:DeltaPsians}}) is
omitted, the errors are about $\sim 50\%$ larger.}.  For signals from
intermediate mass black holes that may be detectable by LIGO, the phase errors are
yet smaller.

These phase errors are sufficiently large that adiabatic waveforms
cannot be used as data-analysis templates; the template will not quite
stay in phase with the signal over the entire $\sim M/\mu \sim 10^5$ cycles of inspiral.
However, the requirements on detection templates are much less
stringent: phase coherence is only needed for $\sim 3$ weeks,
rather than a year \cite{Gair:2004iv}.  In addition, in the matched
filtering search process phase error
will tend to be compensated for by small systematic errors in
the best-fit mass parameters.  Because the adiabatic waveforms are
{\it almost} good enough for data-analysis templates (phase errors
$\sim1$ cycle in the worst cases), adiabatic waveforms will likely
be accurate enough for detection templates both for ground-based and
space-based detectors.

\section{Typos in Gal'tsov}
\label{sec:typos}

This appendix lists some of the typos in the paper by Gal'tsov
\cite{Galtsov:1982}:

\begin{itemize}

\item The discussion of the range of summation of the indices $l$, $m$
  given after Eq. (2.5) is incorrect.  It should be summation $l$,$m$
  with  $l \ge |s|$ and $|m| \le l$ rather than $l \le |s|$ and $|m|
  \le 1$.

\item In the definition of the variable ${}_s u$ given just before
  Eq.\ (2.7), the factor $(\tau^2 + a^2)^{-1/2}$ should be $(r^2 +
  a^2)^{-1/2}$.

\item In the first of the four equations in Eqs.\ (2.9), the term
$r^2 e^{-i \omega r^*}$ should be $r^s e^{-i \omega r^*}$.

\item The right hand side of Eq. (2.25) should be divided by
$-4\pi$, as already discussed in Sec.\ \ref{sec:retardedformula} above.

\item In Eq. (2.30), the quantity ${\bf f}_\Lambda^{\rm out}$ should
  be replaced by its complex conjugate ${\bar {\bf f}}_\Lambda^{\rm
  out}$.

\item In the first line of Eq.\ (3.2), the quantity
  $Z(\theta',\varphi')$ should be replaced by its complex conjugate
$Z(\theta',\varphi')^*$.

\item In the second of Eqs.\ (3.3), the right hand side should read
\[
\kappa_s \tau_s {\bar \tau}_s (w k/|w k|) ( {}_sv^{\rm up} + {\bar
  \sigma}_{-s} \, {}_sv^{\rm in})
\]
rather than
\[
\kappa_s \tau_s {\bar \tau}_s (w k/|w k|) ( {}_sv^{\rm up}) + {\bar
  \sigma}_{-s} \, {}_sv^{\rm in},
\]
i.e. the closing bracket is in the wrong place.

\item In the second term on the first line of Eq.\ (3.6), the first
  factor of ${}_s {\bar v}^{\rm up}(r)$ should be omitted.

\item In Eq. (3.9), the argument of the last factor should be $x'$
  rather than $x x'$.

\item Gal'tsov's equation (4.13) for the rate of change of energy or angular
momentum of the orbit is correct only for sources which are smooth
functions of frequency.  However, for bound
geodesics, the inner products $(\pi,J)$ are sums of delta functions
in frequency, so it does not make sense to square these inner products
as Gal'tsov does.  The correct version of this equation is
our Eq.\ (\ref{eq:dcalEdt2}) above.

\end{itemize}

\section*{References}

%\bibliographystyle{hplain1}
%\bibliography{scalar}

\end{document}